\documentclass[useAMS,usenatbib]{mn2e}
\usepackage{graphicx}
\title[The SEDs of the revised 200 mJy sample]{The SEDs of the revised 200 mJy sample.}
\author[S.~Ant\'on, I.W.A.~Browne, M.~J~.M~.~March\~a, M. Bondi, A. Polatidis]{S.~Ant\'on$^{1,2}$
\thanks{e-mail:Sonia.Anton@oal.ul.pt}, I.W.A. Browne$^1$, M~.J.~M.~March\~a$^2$, M. Bondi$^3$, 
A. Polatidis$^4$ \\
$^1$~Jodrell~Bank~Observatory, University of Manchester, U.K.\\
$^2$~CAAUL, Observat\'orio Astron\'omico de Lisboa, Tapada de Ajuda, 1349-018, Lisboa, Portugal.\\
$^3$~Istituto di Radioastronomia del CNR via Gobetti 101, I-40129, Bologna, Italy\\
$^4$~Max Planck Institut fur Radioastronomie, Auf dem Hugel 69, Bonn, Germany}     
             
\begin{document}

\date{}

% \pagerange{\pageref{firstpage}--\pageref{lastpage}} \pubyear{2002}

\maketitle

\label{firstpage}

\begin{abstract}
We address the question of why low-luminosity radio sources with
similar flat radio spectra show a range of optical activity. The
investigation is based on the spectral energy distributions (SEDs) of
objects from the 200 mJy sample. We gathered new data from the VLA at
43 GHz, from SCUBA in the JCMT at 2000, 1350 and 850 $\mu$m and from 
the ISOPHOT instrument on {\it ISO} at 170, 90, 60 and 25 $\mu$m.
There is considerable diversity amongst the SEDs of the objects: 
there are objects with steep broadband spectra  between centimetre and millimetre
bands (14\% of the sample), there are those with flat broadband spectra over most of the
spectral range (48\% of the sample) and there are those which show pronounced
sub-mm/infrared excesses (27\% of the sample). Some objects of the first
group have two sided radio morphology indicating that their pc-scale emission
is not dominated by beamed jet emission. Amongst the objects that have smooth
broadband spectra from the radio to the infrared there are passive elliptical
galaxies as well as the expected BL Lacs objects. The most pronounced 
sub-mm/infrared excesses are shown by the broad emission line objects. 
\end{abstract}


\section{Introduction}
Active galaxies are distinguished by the broad range of the
electromagnetic spectrum over which they radiate and by the fact that
most of the luminosity comes from the compact nuclear region of the galaxy.
Considering the emission in the radio domain, AGNs come, broadly, in
two flavours: those with weak radio emission, the ``radio-quiet'', and
those with strong radio emission, the ``radio-loud''. The radio-loud
objects are generally divided further into core-dominated objects and
lobe-dominated objects. The former generally have flat radio spectra
and it is upon this sub-set of objects we concentrate in this paper.
At optical wavelengths, amongst the high-luminosity flat-spectrum AGN,
we find objects with strong optical emission lines, the quasars. At
lower luminosities we find objects with very weak or absent emission
lines and polarised and variable continuum, the BL Lacs,
Seyfert-like objects with strong emission lines, and objects with no
sign of optical activity, the passive elliptical galaxies.
Our primary objective is to investigate the origin of the panoply of
optical properties displayed by objects with very similar radio
properties, in particular the flat-spectrum, core-dominated,
low-luminosity radio sources.  The investigation is based on the 200
mJy sample (\citealt{marchatese}, \citealt{mar95}, hereinafter M96). The sample
 has been slightly
revised as a result of better available data (\citealt{antontese}) and
the revised sample is presented here. All the sources are
core-dominated when observed with the VLA with 0.2~arcsec resolution
at 8.4 GHz and have flat radio spectra between 1.4 and 5 GHz.
Their ``radio homogeneity'', however, disappears at
the optical wavelengths: about 30\%\ of the objects show optical
properties of BL Lacs, but in the remaining $\sim$~70\%\ of the
objects the optical properties split into broad and narrow emission
line objects and apparently inactive elliptical galaxies (M96). 
The range of optical activity amongst the 200 mJy
objects begs the question: why in some objects does the non-thermal
component, clearly present in the radio, disappear at optical
frequencies?  The origin of the split in the optical properties must
either arise from differences in the AGN emission and/or differences
in the host galaxy properties. In order to try to identify that origin
we have embarked in a multi-wavelength study of the 200 mJy
sample. This work includes the study of the large-scale properties
such as the structure of the host galaxies and the small scale
properties, in particular the analysis of the spectral energy
distributions of the AGNs. In March\~a et al. (in prep) we present
the surface photometry of approximately half of the sample, and we
find that the large-scale properties of the sources are very
homogeneous and also similar to those found in radio galaxies in
general. In this paper we present the SEDs of the 200 mJy sample, and
we pose the following questions:\\ 
\noindent $\bullet$ Is the lack of optical activity in some objects due to
extinction?\\ 
\noindent $\bullet$ Do all low luminosity flat spectrum radio sources
 have beamed BL Lac emission?  \\ 
Extinction could hide core optical non-thermal emission. Radio
core-dominated objects are expected to be viewed down the radio axis and thus,
according to unification models, should be least likely to be affected
by extinction. These models, however, only take into account central
obscuring regions, but more extended dusty regions may also be
present and prevent us from detecting the AGN emission. The
far-infrared band offers the best chance to trace dust emission, and
for this reason new infrared data were gathered with the {\it Infrared Space 
Observatory} ({\it ISO}).\\
In what concerns the second question, the fact that an object has a 
flat radio-spectrum does not necessarily prove that the emission originates in 
relativistic jets and is beamed. If the emission comes from self-absorbed lobes, 
as for example in Compact Symmetric Objects (CSOs), then the flat radio spectrum 
synchrotron emission is not 
expected  to extend to higher frequency bands. To investigate this, 
new sub-mm data were gathered with SCUBA. The main astronomical objective of these
observations is to try
and look how far the synchrotron component extends into the
mm/sub-millimetre part of the spectrum. We use VLBI observations (Bondi et al, in prep.)
to distinguish between beamed core-jet sources from intrinsically
compact and young CSOs.\\
The framework in which we will discuss the SEDs of the objects is the 
following:\\ 
\noindent 1. The centimetre wavelength properties of the 200 mJy objects
(flat radio spectra plus compact radio structure) are similar to other
objects that have dominant synchrotron cores at centimetre wavelengths.
Thus we assume that their core radio emission is synchrotron in origin.\\
\noindent 2. This synchrotron emission may extend into the sub-mm, 
infrared and even the optical part of the spectrum. In this latter case one
would call the object a BL Lac. \\
\noindent 3. Cold dust may contribute to the sub-mm emission. \\
\noindent 4. If the optical synchrotron photons are not detected 
because they are absorbed by dust, an excess of infrared emission (due to 
reprocessed energy) showing above the extrapolation of the non-thermal 
broadband spectra might be present.\\
 \noindent This paper is organised as follows. In 
Section ~\ref{section-obs} the properties of the 200 mJy sample are 
summarised, the new observations and data reduction are presented. In 
Section ~\ref{section-results} the shape of the SEDs are 
discussed. The important points  of this work are summarised in 
Section ~\ref{section-conclusions}. Throughout the paper we assume 
H$_o$=50 kms$^{-1}$Mpc$^{-1}$ and $q_o$=0.

\begin{table}
\caption{200 mJy selection criteria}
\label{200mjy}
\begin{tabular}{@{}ccccc}
\hline
S$_{\mbox{\tiny 5GHz}}$ & $\alpha_{1.4-5 GHz}$ & R  &
 $\delta_{\tiny 1950}$ & $|b|$\\ 
$\geq$ 200 mJy & $\geq -$ 0.5 & $\leq$ 17 mag & 
$\geq$ 20$^o$ &$\geq$ 12$^o$  \\
\hline 
\end{tabular}
\medskip

The spectral index $\alpha$ ($S\propto\nu^{\alpha}$) is defined between
 1.4 GHz (NVSS catalogue) and 5 GHz (GB6 catalogue), R magnitudes are
 from APM, corrected by the calibration presented in \citet{antontese}.
\end{table}

\section{The sample, the observations and the data reduction}
\label{section-obs}
The 200 mJy sample is a radio-selected sample, which comprises all
the JVAS (Patnaik et al., 1992; Browne et al., 1998; Wilkinson et al., 1998) 
sources with core-dominated morphology when observed with the VLA at 8.4 GHz in 
A-configuration and which obey the criteria presented in Table~\ref{200mjy}. 
When the sample was first selected \citep{marchatese}, the R magnitudes had been  
estimated through ocular 
examination of the Palomar Observatory Sky Survey (POSS) Red
plates. This method is fairly good for the bright objects, but some 
fuzzy objects could have been missed, due to an underestimate
of their magnitude. Meanwhile the POSS plates have 
been digitised by the Automatic Plate Measuring (APM) machine, permitting one 
to obtain better R magnitudes. This motivated us to re-select the 200 mJy sample, 
following the criteria in \citet{marchatese}, but this time the list of objects was
constrained by their APM magnitudes. In the process we developed a calibration 
algorithm that improves the APM magnitudes for  blended or fuzzy appearance 
objects (Ant\'on, 2000). These improved magnitudes were used in the 
re-selection of the 200 mJy sample.

\begin{table*}
\caption{The table contains the following information: 1st column 1959.0 IAU name,
2nd and third columns J2000.0  right ascension and declination respectively, 4th column
R magnitudes from APM corrected by the calibration presented in Ant\'on (2000), 5th column 
redshift, 6th column optical-type classification, 7th column radio
morphology as shown by VLBI resolution, 8th column the shape of the broadband SED}
\begin{tabular}{lccrrcll}
\hline \hline
\multicolumn{1}{c}{Name} & \multicolumn{1}{c}{{\it ra}} &
\multicolumn{1}{c}{$\delta$} & 
\multicolumn{1}{c}{R} & \multicolumn{1}{c}{z} & Optical & VLBI & SED \\
\multicolumn{1}{c}{B1950} & \multicolumn{1}{c}{J2000.0} &
\multicolumn{1}{c}{J2000.0} &
\multicolumn{1}{c}{mag} & \multicolumn{1}{c}{} & type & type & type \\ 
\hline \hline
0035+227 & 00 38 08.10 &  +23 03 28.44 & 15.6     & 0.096 & PEG & 2-S$^{2}$    & II  \\ 
0046+316 & 00 48 47.14 &  +31 57 25.09 & 13.0     & 0.015 & Sy2 & c+j$^{17}$    & II  \\ 
0055+300 & 00 57 48.89 &  +30 21 08.84 & 12.5     & 0.015 & PEG & c+j$^{20}$     & II  \\ 
0109+224 & 01 12 05.82 &  +22 44 38.80 & 16.1     &       & BLL & c+j$^{18}$    & III \\
0116+319 & 01 19 35.00 &  +32 10 50.01 & 11.3     & 0.060 & PEG & 2-S$^{3}$     & II  \\ 
0125+487 & 01 28 08.06 &  +49 01 05.98 & 16.5     & 0.067 & Sy1 & c+j$^{1}$     & III \\ 
0149+710 & 01 53 25.85 &  +71 15 06.47 & 15.0     & 0.022 & PEG & c+j$^{1}$     & II \\
0210+515 & 02 14 17.93 &  +51 44 51.97 & 16.0     & 0.049 & BLL & c+j$^{1}$     & III \\ 
0251+393 & 02 54 42.63 &  +39 31 34.71 & 16.0     & 0.289 & Sy1 & c+j$^{8}$     & III \\ 
0309+411 & 03 13 01.96 &  +41 20 01.19 & 17.0     & 0.134 & Sy1 & c+j$^{8}$     & III \\ 
0316+41  & 03 19 48.16 &  +41 30 42.10 & 12.7     & 0.017 & Sy1 & 2-S$^{21}$    & II  \\
0321+340 & 03 24 41.16 &  +34 10 45.80 & 15.5     & 0.061 & Sy1 & c+j$^{10}$    & III \\ 
0651+410 & 06 55 10.02 &  +41 00 10.15 & 13.8     & 0.021 & PEG & c+j$^{2}$   & II   \\ 
0651+428 & 06 54 43.53 &  +42 47 58.73 & 15.0     & 0.126 & BLC & c+j$^{1}$     & III \\ 
0652+426 & 06 56 10.55 &  +42 37 01.34 & 14.3     & 0.059 & PEG &               &I    \\
0716+714 & 07 21 53.45 &  +71 20 36.36 & 16.0     &       & BLL & c+j$^{4}$    &III  \\
0719+255 & 07 22 49.04 &  +25 28 33.92 & 14.4     &       & UNK &               & I   \\
0729+562 & 07 33 28.61 &  +56 05 41.73 & 16.7     & 0.104 & PEG &               & I   \\ 
0733+597 & 07 37 30.09 &  +59 41 03.19 & 13.7     & 0.041 & PEG & ? $^{16}$      &I    \\ 
0806+350 & 08 09 38.89 &  +34 55 37.26 & 15.5     & 0.082 & BLL & c+j$^{2}$   & III  \\
0836+290 & 08 39 15.85 &  +28 50 39.34 & 14.9     & 0.079 & PEG & c+j$^{19}$     & ?   \\   
0848+686 & 08 53 18.90 &  +68 28 19.01 & 14.0     & 0.039 & PEG & c+j$^{2}$   &I  \\ 
0902+468 & 09 06 15.54 &  +46 36 19.02 & 16.7     & 0.084 & PEG & 2-S$^{2}$   &I  \\ 
0912+297 & 09 15 52.40 &  +29 33 23.98 & 16.6     &       & BLL & c+j$^{18}$    &III  \\
0925+504 & 09 29 15.36 &  +50 13 34.39 & 16.7     &       & BLL & c+j$^{8}$     &?  \\ 
1027+749 & 10 31 22.02 &  +74 41 58.33 & 15.5     & 0.123 & Sy1 & c+j$^{10}$    &?  \\
1055+567 & 10 58 37.73 &  +56 28 11.18 & 16.1     & 0.144 & BLL & c+j$^{1}$     &III  \\
1101+384 & 11 04 27.31 &  +38 12 31.79 & 12.6     & 0.031 & BLL & c+j$^{12}$    &III  \\
1123+203 & 11 25 58.74 &  +20 05 54.38 & 16.7     & 0.133 & BLL & c+j$^{10}$    & III \\ 
1133+704 & 11 36 26.41 &  +70 09 27.30 & 13.9     & 0.046 & BLL & c+j$^{9}$    &III  \\
1144+352 & 11 47 22.13 &  +35 01 07.53 & 15.0     & 0.063 & PEG & c+j$^{8}$     & II \\ 
1146+596 & 11 48 50.36 &  +59 24 56.36 & 12.0     & 0.009 & PEG & 2-S$^{15}$     & II  \\
1147+245 & 11 50 19.21 &  +24 17 53.85 & 17.0     &       & BLL & c+j$^{5}$    &  III \\ 
1215+303 & 12 17 52.08 &  +30 07 00.62 & 15.9     & 0.130 & BLL & c+j$^{1}$     &III  \\
1217+295 & 12 20 06.82 &  +29 16 50.72 & 13.7     & 0.002 & Sy1 & 2-S$^{2}$   &  II \\ 
1219+285 & 12 21 31.69 &  +28 13 58.50 & 15.8     & 0.102 & BLL & c+j$^{6}$    & III \\ 
1241+735 & 12 43 11.22 &  +73 15 59.26 & 14.2     & 0.075 & PEG & c+j$^{1}$     &III  \\ 
1245+676 & 12 47 33.33 &  +67 23 16.46 & 16.6     & 0.103 & PEG & 2-S$^{2}$   &I  \\ 
1246+586 & 12 48 18.84 &  +58 20 27.94 & 15.8     &       & BLL & c+j$^{16}$     &III  \\
1254+571 & 12 56 14.23 &  +56 52 25.24 & 12.8     & 0.041 & Sy1 & c+j+d$^{14}$  & II  \\  
1404+286 & 14 07 00.39 &  +28 27 14.69 & 14.7     & 0.075 & Sy1 & 2-S$^{13}$     & II  \\ 
1418+546 & 14 19 46.60 &  +54 23 14.79 & 15.5     & 0.151 & BLL & c+j$^{5}$    &III  \\ 
1421+511 & 14 23 14.19 &  +50 55 37.2  & 16.6     & 0.274 & Sy1 & u  $^{10}$    &III  \\
1424+240 & 14 27 00.39 &  +23 48 00.04 & 16.5     &       & BLL & c+j$^{10}$    & III \\ 
1429+400 & 14 31 20.54 &  +39 52 40.78 & 16.7     & 1.213 & Sy1 & c+j$^{10}$    & ?   \\ 
1532+236 & 15 34 57.22 &  +23 30 11.61 & 11.7     & 0.018 & Sy2 & SNR           & II  \\ 
1551+239 & 15 53 43.59 &  +23 48 25.48 & 14.8     & 0.115 & PEG &               &III  \\ 
1558+595 & 15 59 01.70 &  +59 24 21.85 & 12.9     & 0.057 & PEG & 2-S$^{1}$     & II  \\ 
1645+292 & 16 47 26.88 &  +29 09 49.60 & 16.2     & 0.132 & BLC &               &I \\ 
1646+499 & 16 47 34.91 &  +49 50 00.58 & 16.5     & 0.045 & Sy1 & c+j$^{1}$     &III  \\
1652+398 & 16 53 52.22 &  +39 45 36.61 & 11.5     & 0.03  & BLL & c+j$^{11}$      &III  \\
1658+302 & 17 00 45.23 &  +30 08 12.90 & 12.8     & 0.036 & PEG &               & III \\ 
1703+223 & 17 05 29.33 &  +22 16 07.59 & 12.3     & 0.049 & PEG & c+j$^{2}$      & I  \\ 
1744+260 & 17 46 48.28 &  +26 03 20.35 & 17.0     & 0.147 & Sy2 & (c+j$^{22}$)   & III \\
\hline \hline
\end{tabular}
\label{200table}
\end{table*}

\begin{table*}
\centerline{Table ~\ref{200table} continued}
\begin{tabular}{lccrrcll}
\hline \hline
\multicolumn{1}{c}{Name} & \multicolumn{1}{c}{{\it ra}} &
\multicolumn{1}{c}{$\delta$} & 
\multicolumn{1}{c}{R} & \multicolumn{1}{c}{z} & Optical & VLBI & SED \\
\multicolumn{1}{c}{B1950} & \multicolumn{1}{c}{J2000.0} &
\multicolumn{1}{c}{J2000.0} &
\multicolumn{1}{c}{mag} & \multicolumn{1}{c}{} & type & type & type \\ 
\hline \hline
1755+626 & 17 55 48.44 &  +62 36 44.12 & 11.0     & 0.024 & PEG & ? $^{2}$   &II  \\
1807+698 & 18 06 50.68 &  +69 49 28.11 & 15.1     & 0.046 & BLL & c+j$^{7}$     & III \\
1959+650 & 19 59 59.85 &  +65 08 54.67 & 16.5     & 0.047 & BLL & c+j$^{1}$    &III  \\
2116+81  & 21 14 01.18 &  +82 04 48.35 & 15.5     & 0.084 & Sy1 & c+j$^{15}$     &III  \\ 
2202+363 & 22 04 21.10 &  +36 32 37.09 & 16.0     & 0.073 & PEG & u  $^{1}$     & ?   \\ 
2217+259 & 22 19 49.74 &  +26 13 27.96 & 13.1     & 0.085 & Sy1 &              &II \\
2227+306 & 22 29 34.11 &  +30 57 10.29 & 16.8     & 0.322 & Sy1 &              & ?  \\ 
2319+317 & 23 21 54.95 &  +32 04 07.59 & 16.0     & 0.001 & UNK &              & ?  \\ 
2320+203 & 23 23 20.34 &  +20 35 23.52 & 12.8     & 0.038 & PEG & c+j$^{2}$     &III  \\ 
2337+268 & 23 40 00.84 &  +27 08 01.37 & 13.9     & 0.032 & UNK & c+j$^{2}$     &II  \\
\hline \hline
\end{tabular}
\medskip

{\it Keys}: 1. Optical classification: {\bf BLL}=BL Lacs, {\bf BLC}=BL Lac Candidates,
{\bf Syf1}=Seyfert-1 type, {\bf Syf2}=Seyfert-2 type, {\bf PEG}=Passive Elliptical 
Galaxies; UNK = Unknown spectrum. 2. Radio morphology classification:  {\bf c+j}=core-jet; {\bf c}=core, 
{\bf c+j+d} Core+jet+disk,{\bf 2-S } 2-sided, {\bf u} unresolved. SED classification: 
 {\bf I} steep single power-law broadband spectra, {\bf II}  broadband spectra with one or more bumps, 
{\bf III} broken power-law broadband spectra.\\
{\it References}
 1. Bondi et al. (2001)
 2. Bondi et al (in preparation)
 3. Conway (1996)
 4. Gabuzda et al. (1998)
 5. Gabuzda et al. (1996a)
 6. Gabuzda et al (1996b)
 7. Gabuzda et al. (1989)
 8.  Henstock et al. (1995)
 9.  Kollgaard et al. (1996)
 10. NRAO VLBA calibrators list
 11. Pearson et al. (1993)
 12. Piner et al. (1999)
 13. Stanghellini et al. (2001)
 14. Taylor et al. (1999)
 15. Taylor et al. (1996)
 16. Taylor et al (1994)
 17. Ulvestad et al (1999)
 18. USNO Astrometric VLBI database
 19. Venturi et al (1995)
 20. Venturi et al.(1993)
 21. Vermeulen et al. (1994)   
 22. This object shows core-jet morphology at MERLIN resolution (Augusto et al., 1998). 
\end{table*}
\noindent With the re-selection of the 200 mJy sample, and in comparison with
the list of objects presented in M96, seven new objects entered the 
sample, and one object was subtracted. The later is the  object 2214+201. 
We gathered a spectrum of the object,
and the spectrum is identical to that of a M star (Ant\'on, 2000). We also obtained
 an I image with the Nordic Optical  Telescope. Its  radial profile is that of a 
point source and no residuals were found after a PSF subtraction 
\citep[for details see][]{antontese}. We conclude that the detected optical object
is most probably a star in front of the radio source. In this case we do not know 
the R magnitude of the optical counterpart of the radio object, and for this reason 
the object has been subtracted from the sample. The 200 mJy sample comprises 64 
objects, which are listed in Table~\ref{200table}.\\
M96 obtained optical spectra and optical polarisation
for a large number of 200 mJy objects. Based on the equivalent width (EW) of 
the strongest emission line and the 4000\AA\ break contrast (C; see 
section~\ref{section-optical}) the objects may be classified in four broad 
categories:\\
\noindent $\bullet$ {\bf BL Lac objects (BLL)}. These are objects with featureless spectra 
that obey the ``classical'' BL Lac definition: C~$\leq$~0.25 and 
EW~$\leq$~5~\AA.\\
\noindent $\bullet$ {\bf BL Lac Candidates (BLC)} 
These are objects with break contrast 0.25~$<$C~$<$~0.4 and EW~$<$~40\AA.\\
\noindent $\bullet$ {\bf Seyfert-like objects}
These are the objects with EW~$>$~60 \AA.
In this case the value of the 4000\AA\ break does not give a good
estimate of the non-thermal continuum, as in most cases
there will be contamination from the accretion disk emission, in
particular in the case of broad emission line objects. The objects
are divided into Seyfert1-like {\bf (Sy1)}, and into Seyfert2-like  {\bf (Sy2)} depending
whether their emission lines are broad or narrow, respectively. \\
\noindent $\bullet$ {\bf Passive Elliptical Galaxies (PEGs)} 
These are objects with C~$\geq$~0.40 and weak emission 
lines EW~$<$~30-40 \AA. 

\subsection{New data and data reduction}

\subsubsection{VLA observations -- 43 GHz data}

As part of a  program to measure flux densities of potential 
high-frequency 
calibration sources, 16 objects of the 200~mJy sample were observed
with the VLA in its D-configuration at 43~GHz. Each observation
lasted $\sim$1~min. The calibration and data reduction were performed
in AIPS in the standard manner with 3C286 being used as the primary
flux density calibrator. 

\subsubsection{SCUBA data}
\label{section-scuba}

\begin{table*}
\caption{{\it ISO} and SCUBA data. For each source (IAU name in the first column), 
the following columns refer to 25, 60, 90 and 170$\mu$m {\it ISO} flux densities and 
uncertainties (see text), and to the 850, 1350 and 2000 $\mu$m 
SCUBA flux densities and errors (see text).}
\begin{tabular}{lrrrrrrrrrrrrrr}
 \hline \hline 
\multicolumn{1}{c}{Name} & 
\multicolumn{2}{c}{25$\mu$m} & 
\multicolumn{2}{c}{60$\mu$m} &
\multicolumn{2}{c}{90$\mu$m} & 
\multicolumn{2}{c}{170$\mu$m} & 
\multicolumn{2}{c}{850$\mu$m} & 
\multicolumn{2}{c}{1350$\mu$m} &
\multicolumn{2}{c}{2000$\mu$m} \\
\multicolumn{1}{c}{} & 
\multicolumn{1}{c}{F} & 
\multicolumn{1}{c}{$\sigma$} & 
\multicolumn{1}{c}{F} &
\multicolumn{1}{c}{$\sigma$} & 
\multicolumn{1}{c}{F} & 
\multicolumn{1}{c}{$\sigma$} & 
\multicolumn{1}{c}{F} & 
\multicolumn{1}{c}{$\sigma$} & 
\multicolumn{1}{c}{F} & 
\multicolumn{1}{c}{eF$_{850}$} & 
\multicolumn{1}{c}{F} &
\multicolumn{1}{c}{eF$_{1350}$} & 
\multicolumn{1}{c}{F} &
\multicolumn{1}{c}{eF$_{2000}$} \\
\hline
0046+316 &    260 & 57  & 1139   & 330 & 1108   &  302& 1451   & 300 &        &       &        &       &        &      \\
0116+319 &   $<$9 &     & $<$372 &     & $<$165 &     &$<$264  &     &        &       &        &       &        &      \\
0125+487 &    19  &  3  &        &     &        &     &        &     &        &       &        &       &        &      \\
1133+704 & $<$51  &     & $<$345 &     & 66     &   21&$<$123  &     &   69.9 & 13.8  &   56.6 & 13.1  &   78.8 & 13.7 \\ 
1144+352 &    11  &  3  & $<$156 &     & 158    &   40&   363  &  27 &   86.1 & 12.7  &   84.4 & 12.6  &  149.2 & 12.8 \\
1146+596 &        &     &        &     &        &     &        &     &   62.9 & 13.3  &   57.6 & 14.6  &   94.9 & 18.3 \\ 
1147+245 &        &     &        &     &        &     &        &     &  344.9 & 23.9  &  337.5 & 40.5  &  558.0 & 32.6 \\ 
1215+303 &        &     &        &     &        &     &        &     &  153.8 & 29.6  &  172.1 & 21.8  &  198.8 & 24.2 \\ 
1217+295 &        &     &        &     &        &     &        &     &   44.6 & 11.0  &   50.4 &  9.6  &   37.8 &  7.3 \\ 
1219+285 &        &     &        &     &        &     &        &     &  623.9 & 51.0  &  635.8 & 74.0  &  736.0 & 41.2 \\ 
1241+735 &        &     &        &     &        &     &        &     &   39.7 &  8.9  &   45.8 &  9.6  &   40.6 &  8.8 \\ 
1245+676 &        &     &        &     &        &     &        &     &   11.4 &  3.6  &  $<$ 13&       &   19.4 &  2.8 \\ 
1254+571 &        &     &        &     &        &     &        &     &   78.0 & 24.5  &   49.8 & 12.1  &   28.1 &  6.7 \\ 
1404+286 &        &     &        &     &        &     &        &     &   18.5 &  5.2  &  $<$ 16&       &   41.1 &  9.9 \\ 
1418+546 &        &     &        &     &        &     &        &     &  170.9 & 15.6  &  234.2 & 28.2  &  338.5 & 16.6 \\ 
1421+511 &        &     &        &     &        &     &        &     &   24.6 &  4.4  &   48.3 & 10.1  &   69.9 &  7.7 \\ 
1424+240 &        &     &        &     &        &     &        &     &   47.1 &  8.1  &  151.7 & 17.9  &  140.1 & 10.8 \\ 
1532+236 &        &     &        &     &        &     &        &     &  455.7 & 46.9  &  271.0 & 32.3  &   88.1 &  8.8 \\ 
1551+239 &        &     &        &     &        &     &        &     &   36.5 &  4.6  &   89.8 & 12.8  &  125.0 &  7.4 \\ 
1558+595 &  $<$24 &     & $<$111 &     & $<$219 &     &   306  &  86 & $<$ 28 &       &  $<$ 19&       &  $<$ 30&      \\ 
1645+292 &        &     &        &     &        &     &        &     &   20.1 &  6.0  &   24.6 &  5.3  &   26.4 &  7.5 \\ 
1646+499 &    11  &  3  & $<$126 &     & $<$108 &     &   221  &  21 &        &       &  205.5 & 24.9  &  274.4 & 13.1 \\ 
1652+398 &        &     &        &     &        &     &        &     &        &       &  452.3 & 52.9  &  678.0 & 26.6 \\ 
1658+302 &        &     &        &     &        &     &        &     &   30.8 &  4.3  &   35.7 &  6.1  &   38.1 &  6.9 \\ 
1703+223 &        &     &        &     &        &     &        &     &   15.2 &  3.5  &        &       &        &      \\
\hline \hline
\end{tabular}
\label{tablescubaflux}
\end{table*}

\noindent We gathered 850, 1350, 2000 $\mu$m observations at
JCMT\footnotemark\footnotetext{James Clerk Maxwell Telescope} using
SCUBA\footnotemark\footnotetext{Submillimetre Common-User Bolometer
Array}. A set of 22 objects was observed between May and July of
1998. The observations were carried by a service observer in a
fall-back program, during 32 hours of allocated time. The objects
observed were chosen from the 200 mJy list by the observer to fit
into the available time and available RA range.  
Being fall-back observations
they were taken during non-optimum weather conditions.  The
observations were performed in standard SCUBA photometric mode, which
involves the conventional technique of chopping and nodding in order
to remove the dominant sky background - the chopper throw was 60
arcseconds.
\noindent The data were processed with {\sc SURF} 
(SCUBA User Reduction Facility)
package. The reduction process was based on the techniques described in
\citet{stevens97} and consisted of nod compensation, extinction correction and calibration.
The calibrators observed during our observations were: irc+10216, 16293-2422 
and  Uranus. Starlink package {\rm  FLUXES} 
supplies the flux densities of the planets for the time and day of 
observation, corrected for the SCUBA beam-size. \\
\noindent The results of the SCUBA observations are presented in 
Table~\ref{tablescubaflux}: flux densities are in mJy and the error on the flux density, 
eF, includes the calibration uncertainties and the instrumental flux 
uncertainties, added in quadrature. 
\noindent Table~\ref{tablescubaflux} shows that the SCUBA observations
were very successful: amongst 22 objects, only one (1558+595) was not 
detected at any of the 3 SCUBA wavelengths. There are few cases of 
non-detections at some wavelengths and for these upper limits are
quoted as 3$\sigma$.

\subsubsection{ISOPHOT Observations}
\label{section-iso}

\noindent A set of seven sources from the 200 mJy sample was observed
at 25, 60, 90 and 170 $\mu$m wavelengths by {\it ISO} with the imaging photo-polarimeter 
ISOPHOT.
The list of objects corresponds to the number
that could be observed in the total observing time allocated, and obeying :
z $\leq$ 0.07, $|b|> 12^o$, did not belong to GTO programs and
their coordinates were permitted by the satellite constraints.
The main idea was to compare the SEDs of BL Lac objects with their relatives, 
and for this aim we selected objects of different optical type:
three PEGs, one Seyfert2-like object, two Seyfert1-like object and one BL Lac.
The observations 
were performed with P2 detector to obtain
data at 25 $\mu$m, C100 detector to obtain data at 60 and 90 $\mu$m,  and C200
detector to obtain data at 170 $\mu$m; all  with the same 
integration time on-source and on-background.
Observations at  25 $\mu$m were taken with triangular chopper mode with a 
throw of 60 arcsec between positions. The 60, 90 and 170 $\mu$m observations were 
performed in rectangular chopper mode, the chopper throw was 180 arcsec between
positions.\\
\noindent The data were reduced using 
PIA\footnotemark\footnotetext{PIA is a joint development by the 
ESA Astrophysics Division and the ISOPHOT consortium.}
(PHT-Interactive Analysis).
\noindent Table~\ref{tablescubaflux} presents the results: the flux densities 
are the mean of the flux densities (obtained from the several chopper measurements),
and $\sigma$ is the dispersion of the flux densities about the 
mean\footnotemark\footnotetext{The values of $\sigma$ are the dispersion of 
the flux densities about the mean; they are not 
the dispersion on the mean  $\sigma$/$\sqrt{\mbox{N}}$. 
In a Gaussian distribution the relevant quantity is 
$\sigma$/$\sqrt{\mbox{N}}$, a quantity that contains the number
of measurements used.  However, in our case the data are not 
statistically well behaved, and $\sigma$ gives a conservative estimate of
the significance of the measurement}. 
 The systematic error associated with
the photometric calibration is estimated to be $\approx$ 30\%.
The error estimate includes drift and chopper 
frequency errors (the main sources of uncertainty), and the power calibration
 error, that contributing $\approx$ 10\% of the total error.
 Considering that there were some
problems affecting the ISOPHOT detection system, the most relevant point 
is the detection of the source. 
The dispersion of the flux densities (presented in 
Table~\ref{tablescubaflux}) is a measure of the statistical significance of 
the detection.
\noindent A detection means that the mean flux density is higher than
3$\sigma$, otherwise the measurement is considered as a non-detection
and an upper limit of 3$\sigma$ for the flux density is recorded.  At
25 $\mu$m four out of the seven objects were detected, at 60 $\mu$m
only one source out of six was detected\footnote{The source 0125+487
was only observed at 25 $\mu$m}, at 90 $\mu$m three out of six were
detected and at 170 $\mu$m four out of six. Most of the detections
look ``physically reasonable'' and when {\it IRAS} observations are
available for comparison, the {\it ISO}  and {\it IRAS} detections are consistent.

\subsubsection{Optical data}
\label{section-optical}

One of the main goals is to compare the properties of the 200 mJy
objects at different bands, hence the effort to produce their SEDs.
At optical wavelengths much of the light of the low redshift 200 mJy objects 
is from the host galaxy, however, we are primarily interested in studying 
the AGN properties,
which means that we have to estimate the strength of the extra emission
contribution to the optical flux. In order to do this we make use of a
common feature of the spectra of early-type galaxies usually referred
to as the 4000\AA\ break. This feature which refers to a flux
discontinuity occurring across 4000\AA\ is commonly quantified by the
break contrast (C) defined as:
\begin{eqnarray}
C & = & \frac{F_{\nu}^{+} - F_{\nu}^{-}}{F_{\nu}^{+}}
\label{cdefinition}
\end{eqnarray}

where $F_{\nu}^+$ and $F_{\nu}^-$ are the red-ward and blue-ward
flux densities of the break, and they are the
the mean flux density in the 4050--4250 \AA\ range and 3750--3950 \AA\ range,
respectively. If it happens that besides the thermal
emission there is an extra source of continuum, then that depression
will be decreased, and the measure of contrast as defined in
Eq. \ref{cdefinition} can then be used to estimate that extra
contribution.
In a study carried out by \cite{dressler87} on a large sample of
early-type galaxies, the authors concluded that the galaxies showed
a very narrow range of contrasts ($<C^{G}>$ = 0.49$\pm 0.1$),  with most of 
them with contrast above 0.4. This led M96 to suggest that optically dull 
galaxies with measured
$C<0.4$ corresponded to sources where there was an extra
contribution due to the AGN ($F^{AGN}$). This quantity can be
obtained from the observed spectrum via:
\begin{eqnarray}
F^{AGN} & = & \frac{<C^{G}> - C}{<C^{G}>} \; F_o
\label{opticalflux}
\end{eqnarray}
where $F_o$ is total optical flux density. 
Making use of the spectra in M96 we have obtained an
estimate of the contribution of the AGN to the optical flux. 
The working assumption is that the extra continuum 
represents the nuclear component. The origin of the nuclear 
emission might be thermal from accretion disk emission, or non-thermal. 
The approach here represents an effort to minimise as much as possible the 
contribution of starlight, which for some objects is clearly the major source
of emission in the optical band. For those objects whose contrasts are C$>$ 0.45, C=0.45 was adopted in
order to have a quantitative upper limit to the nuclear emission.
The flux calibration of the spectra in M96 is
reliable at a 10\%\ level (Ant\'on 2000).  This level of accuracy is
sufficient for the SED analysis.

\section{Discussion of the SEDs}
\label{section-results}

In Figure ~\ref{seds} we present the SEDs for all the objects in the
200~mJy sample. To supplement our own data we have plotted and  used in the 
following analysis relevant
data found in the NASA Extragalactic database (NED) and in the
literature. In particular we include measurements from {\it IRAS}, from
{\it 2MASS} and from the {\it ROSAT} All-Sky-Survey \citep{Voges99}. We see a
considerable diversity both in the spectral coverage and in the
overall shapes of the SEDs. If we discount those objects with very
poor spectral coverage in the sub-mm and infrared parts of the
spectrum, we can tentatively identify three broad spectral types of
objects: (a) those with initially
steep radio power-law spectra ($\alpha < 0.4$) and perhaps steepening onto the
sub-mm/IR/optical bands, (b) the objects that show an infrared bump that we
interpret as dust emission and (c) the broken power-law objects that 
show flat broadband spectra up to the millimetre band, and sometimes beyond,
before steepening. We also list these classifications in 
Table \ref{200table}, as class I, II and III, respectively. 
Nearly all the known BL Lac objects fall into the third, broken
power-law, category. It is interesting that, with the exception of
1418+546, any flux variability is insufficient to show up in plots of
this crude flux resolution. It should also be noted that a few PEGs
(e.g. 0055+300) look similar to BL Lacs when we discount the near
infrared/optical emission which is predominantly starlight, not
AGN emission. There are more PEGs where just the sub-mm emission
appears as a smooth continuation of the radio spectrum. We discuss
this further below. A typical steep spectrum object is 0035+227 and
one with a lumpy spectrum 1404+286. We now look at the different spectral 
ranges individually. 

\subsection{From the radio to the submillimetre}
\begin{figure}
\centerline{
\includegraphics[angle=-90,width=6.5cm]{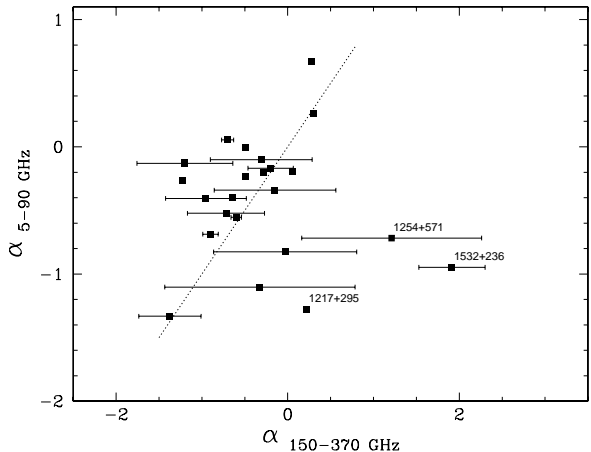}}
\caption{The spectral indices at the sub-mm band  vs the
spectral indices at the radio band (see text). 
The dashed line represents $\alpha_{\rm{sub-mm}}=\alpha_{\rm {radio}}$.}
\label{colorcolorsmm}
\end{figure}
One of the main goals is to establish if the synchrotron
emission extends up to the sub-mm band. We look to see
whether, statistically, the sub-mm emission can be described as a
smooth extension of what we already know to be synchrotron emission,
i.e. the radio emission, or if there is a discontinuity 
indicating additional thermal emission from cold dust.
From Figure ~\ref{seds} it is noticeable that the majority of the
objects show a spectrum consistent with a power-law continuing from the radio to the sub-mm
wavelengths, suggesting that the synchrotron emission does indeed
extend up to the sub-mm band.  Assuming that the emission may be
represented by a power-law (F $\sim\nu^{\alpha}$), the spectral
indices $\alpha_{\rm {radio}}$ and $\alpha_{\rm{sub-mm}}$ were computed through a
weighted least-square fit to the (available) data between 5~$\le$$\nu$$<$ $90$~GHz 
and 150~$\le$$\nu$$\le$~370~GHz
respectively. Errors are not quoted for $\alpha_{\rm {radio}}$ because the
data were not gathered simultaneously or with instruments of the same
angular resolution; it is very difficult to quantify the effects of
variability and resolution. For the same reason, errors of
$\alpha_{\rm{sub-mm}}$ are not quoted if the data were taken in different
epochs. In Figure ~\ref{colorcolorsmm} the dashed line represents
$\alpha_{\rm{sub-mm}}=\alpha_{\rm {radio}}$.
The large majority of the 200 mJy objects are grouped on or just to
the left of the $\alpha_{\rm{sub-mm}}$ = $\alpha_{\rm {radio}}$ line, indicating
either the same or a slightly steeper spectral index at sub-mm
wavelengths compared to the radio. This is what is expected for
synchrotron emission with a power-law energy spectrum modified at the
highest energies by synchrotron losses.
The objects 1217+295, 1254+571, 1532+236 depart from the general trend and 
are discussed individually in Section~\ref{individual}.
Considering the sub-mm
properties of the different optical-classes, we find
the following\footnotemark\footnotetext{Unfortunately the numbers per class for
Seyfert2-like and BLC are so small that they had to be excluded from
the above analysis.}:\\
\noindent {\bf BL Lac objects} -- The mean sub-mm spectral index is
$\overline{\alpha}_{\rm{sub-mm}}$=$-$0.25$\pm 0.37$ which is consistent with
$-$0.48$\pm$0.22 obtained by Gear et al (1994) for a sample of 30 BL
Lacs. Not surprisingly, all the objects have flat radio spectra and
flat sub-mm spectra, in good agreement with previous studies (e.g.,
\citealt{kna90}, henceforth K90, \citealt{bloom94}, \citealt{gear94}).\\
\noindent {\bf Seyfert1-like objects} -- In this group we find diverse
SEDs, with objects showing a falling spectrum and others showing a 
rising spectrum. The $\alpha_{\rm{sub-mm}}$ distribution
is broad,  the  mean sub-mm spectral index is 
$\overline{\alpha}_{\rm{sub-mm}}=  -0.43\pm 0.87$.\\ 
\noindent {\bf PEG objects} -- These objects have flattish sub-mm spectra 
and their mean spectral index is $\overline{\alpha}_{\rm{sub-mm}} \sim -0.58\pm0.39$.\\ 

\noindent In summary, we do not find a clear split of properties between
the radio and sub-mm band that could account for the split of properties
detected in the optical regime. Broadly, the SEDs of the objects 
up to the  submillimetre regime (and for the list of the objects presented
in Table 3) are similar.

\subsection{The infrared emission}

\begin{figure}
\centerline{
\includegraphics[width=8cm]{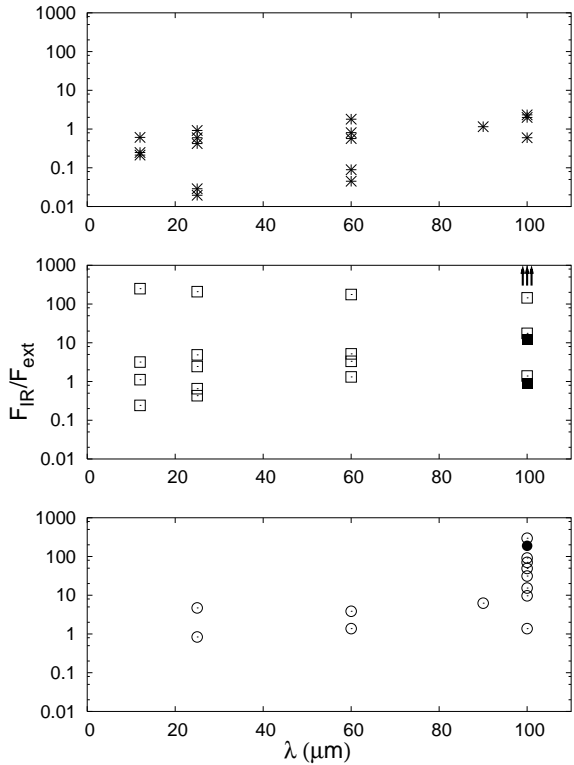}}
\caption{The ratio between the flux density detected at 12, 25, 60, 90 and 100 $\mu$m, 
F$_{\rm{IR}}$, and the flux density extrapolated from the  radio/sub-mm band ,
F$_{\rm{ext}}$, for different objects. Stars represent BL Lacs (top panel), square symbols 
represent Seyfert-like objects (middle panel) and circles represent PEGs (lower panel).
An object may not have been
detected at the all  wavelengths. In the case of 3 objects, the data are from 170$\mu$m 
and they are represented by filled symbols at 100$\mu$m. The Seyfert-like objects
 1217+295, 1254+571 (Mrk 231) and 1532+236 (Arp 220)  
are the only objects with F$_{\rm{IR}}$/F$_{\rm{ext}}$$>>$ 1000, being represented by 
arrows (see middle plot).}
\label{meaninfrared}
\end{figure}

\begin{figure}
\includegraphics[width=8cm]{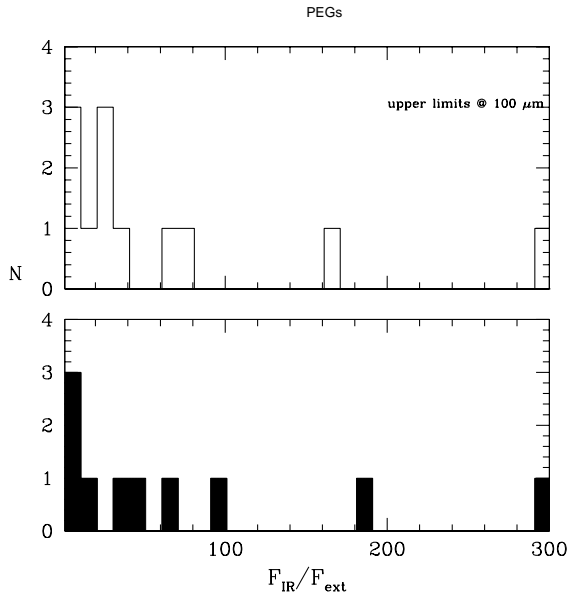}
\caption{{\bf Bottom}  ratio between the flux density detected at the largest infrared 
wavelength available (90, or 100, or 170 $\mu$m), and the flux density extrapolated from 
the  radio/sub-mm band at that wavelength for the PEGS {\bf Top} 
 ratio between an upper limit of the flux density at 100 $\mu$m,  and the flux density 
extrapolated from  the  radio/sub-mm band at 100$\mu$m for the PEGS.}
\label{meaninfraredhisto}
\end{figure}

The far- and mid-infrared properties of some of the 200 mJy objects are
analysed using our own {\it ISO} data and archival {\it IRAS} data. The main goal
is a) to investigate if, for a given optical magnitude of the host
galaxies, the 200 mJy objects have an excess of infrared emission
compared to normal radio-quiet elliptical galaxies and, b) to
investigate the nature of the infrared excess; i.e. is it mainly
synchrotron emission or is it dust emission?  \\
K90 compared the {\it IRAS} infrared
luminosities of normal E/S0 galaxies with radio galaxies and found
that the latter have stronger infrared emission relative to the
optical emission than do normal E/S0 galaxies. 
In order to investigate how much of the infrared emission of the
200 mJy objects can be attributed to the ``normal'' emission of the
host galaxy, we computed the 
median value of the ratio of the luminosity at 90$\mu$m/170$\mu$m ({\it ISO} data) 
or 100 $\mu$m ({\it IRAS}
data) with the luminosity in the B band,  L$_{\rm{IR}}$/L$_{\rm B}$, and compared it with
K90 findings for normal galaxies. We chose the far-infrared fluxes because they 
correspond to the wavelengths where detections are above 50\% in both our and K90 
case.
We found that the median  L$_{\rm{IR}}$/L$_{\rm B}$ is 1.93
for the detected 200 mJy objects \footnotemark\footnotetext{The 
mean value is $<$L$_{\rm{IR}}$/L$_{\rm B}$$>$= 5.0 with a r.m.s. = 7.97},
which is one order of magnitude higher than  L$_{\rm{IR}}$/L$_{\rm B}$= 0.12 found by K90 
for radio elliptical galaxies. 
Comparing with L$_{\rm {IR}}$/L$_{\rm B}$= 0.041 for  E/SO/SOa galaxies from K90 we find  that our 
objects are almost two orders of magnitude brighter in the infrared than normal galaxies.
Thus, even if part of the infrared emission of the 200 mJy objects is
due to the host galaxy, there needs to be another infrared component
to account for the total infrared emission. We now investigate the origin of 
the ``extra'' infrared emission.\\
Figure~\ref{meaninfrared} shows
the ratio between the detected flux densities in the infrared
band with estimates of the non-thermal flux densities extrapolated from the
radio/sub-mm band, F$_{\rm{IR}}$/F$_{\rm {ext}}$, at each of the {\it IRAS}
wavelengths and divided into PEGs, Seyfert1-like objects and BL Lac
objects.  Note that the extrapolated non-thermal emission represents
an upper estimate since it assumes that the radio/sub-mm spectrum
continues as power-law to higher frequencies. Figure~\ref{meaninfrared}
shows that there are objects with F$_{\rm{IR}}$ similar to F$_{\rm{ext}}$,  
and this is what one would expect for a smooth synchrotron spectrum, and
objects in which the infrared emission
is clearly much stronger than the extrapolated non-thermal emission, and
the dominant emission process must be thermal, i.e. dust
emission\footnotemark\footnotetext{Not withstanding the previous
discussion it has been suggested that non-thermal flares could account
for infrared excess in some objects \citep{vanbemmel00}. But, since flares
are a short lived phenomenon, it would be very peculiar if all sources
showing infrared excess (over the extrapolated radio emission) were
observed at a moment of a flare.}. As far as the properties of the objects as 
a group are concerned, the properties of BL Lacs, Seyfert-like objects and PEGs 
split in the infrared band:\\

\noindent $\bullet$ The BL Lacs have F$_{\rm{IR}}$/F$_{\rm {ext}}$~$<$~2. They are the 
only group that as a whole shows smooth broadband spectra from the radio up to the infrared.\\

\begin{figure}
\center{
\includegraphics[width=7cm]{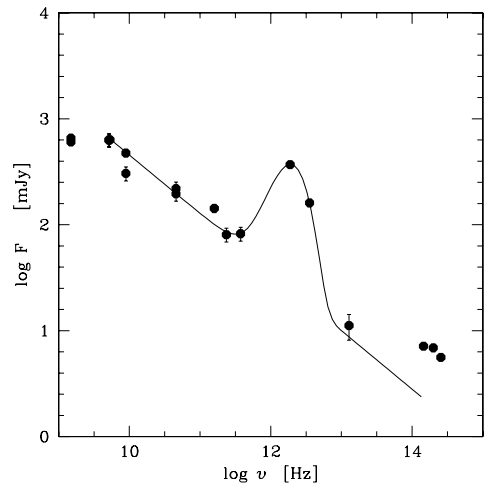}}
\caption{Broadband spectrum of 1144+352, in units of log F$_\nu$ vs log $\nu$. Data points include
data from the literature and the ISO and SCUBA data presented here. The spectrum was fitted
with a  power-law component  plus a greybody component (see text).}
\label{1144ir}
\end{figure}

\noindent  $\bullet$ The PEG objects show a
range of F$_{\rm{IR}}$/F$_{\rm{ext}}$, with a fraction showing sub-mm/infrared excesses
 particularly important 
at relatively large wavelengths (see Fig. 2), thus suggesting the presence of cold dust. 
 We have infrared data for 12 out of 22 PEGs. In order to 
investigate whether the remaining fraction of PEGs might have a big far-infrared bump,
we obtained from the IRAS Scan Processing and Integration tool upper limits of the flux density 
at the location of the non-detected PEG. Figure \ref{meaninfraredhisto} shows 
the distribution of  F$_{\rm{IR}}$/F$_{\rm{ext}}$ \footnotemark\footnotetext{F$_{\rm{ext}}$ as 
before and F$_{\rm{IR}}$ being the flux density at 100 micron, or 90/170 if observed
with ISO} of PEGs: in the lower plot F$_{\rm{IR}}$ represents the detected flux density,
whereas in the upper plot  F$_{\rm{IR}}$ represents an upper limit of the flux density 
at 100 $\mu$m. The distributions are similar. With the available information, the only 
conclusion is that objects with strong infrared bumps as those shown by objects like Arp 220 are
unlikely to be frequent amongst the non-detected PEGs.\\
It is possible to estimate the dust mass of the objects with {\it ISO} and/or
{\it IRAS} data and for which there is a clear visual evidence in the SEDs for dust
emission. For some of the objects dust mass estimates were found in the literature, e.g.
we found from K90: 
M$_{0055+300}^{\rm d}$= 8.5$\times$$10^6$M$_\odot$ and M$_{0116+319}^{\rm d}$=1.4$\times$$10^8$M$_\odot$. 
For the objects with new data, the radio to infrared  broadband spectra were fitted assuming
a power-law component plus a greybody component,
F$_\nu$~=~F$_{\rm pl}$~+~F$_{\rm gb}$,   F$_{\rm pl} \sim \nu^\alpha$ and
F$_{\rm gb} \sim$ $\nu$ I($\nu$,T) (Hildebrand, 1977).  Figure ~\ref{1144ir} shows an example of 
the fitted model for 1144+352. The dust temperature can be estimated and the mass 
of the dust responsible for the far-infrared emission was computed according to K90:\\
\noindent ${\rm M^d}$~=~ 4.8 ${\rm d_L^2}$ (e$^{\frac{144}{\rm T_d}} - 1$) F$_{100}$ ${{\rm M}_\odot}$,
where ${\rm d_L}$=$\frac{zc}{H_o}(1+\frac{z}{2})$ is the luminosity distance (in Mpc)
and F$_{100}$ is the thermal emission at 100 $\mu$m (in Jy). We obtained:\\
\begin{tabular}{@{}ccc}
(M$^{\rm d}$/M$_\odot)_{\mbox{\tiny 1144+352}}$=7$\times$$10^7$ &&   ${\rm T^d}$=22 \\
(M$^{\rm d}$/M$_\odot)_{\mbox{\tiny 1146+596}}$=3$\times$$10^5$ &&   ${\rm T^d}$=36 \\
(M$^{\rm d}$/M$_\odot)_{\mbox{\tiny 1558+595}}$=1$\times$$10^8$ &&   ${\rm T^d}$=22 \\
\end{tabular}

\noindent From IRAS Scan Processing and Integration tool 
we obtained the flux densities at 100$\mu$m for a list of 5 objects. No new data was gathered for 
these objects, and dust mass estimates are based on published data; they are:\\
\begin{tabular}{@{}ccc}
(M$^{\rm d}$/M$_\odot)_{\mbox{\tiny 0035+227}}$=8$\times$$10^7$ && ${\rm T^d}$=37 \\
(M$^{\rm d}$/M$_\odot)_{\mbox{\tiny 0149+710}}$=9$\times$$10^6$ && ${\rm T^d}$=37 \\
(M$^{\rm d}$/M$_\odot)_{\mbox{\tiny 0651+410}}$=7$\times$$10^7$ && ${\rm T^d}$=21 \\
(M$^{\rm d}$/M$_\odot)_{\mbox{\tiny 1755+626}}$=5$\times$$10^6$ && ${\rm T^d}$=29 \\
\end{tabular}

\noindent These results indicate the existence of cold 
dusty regions (20$<$T$<$50 K) with dust  masses similar to those found in typical early-type  
galaxies, i.e. M$^{\rm d}\sim$$10^5-~10^7$M$_\odot$ (e.g. Wiklind 1995). One way dust might be 
also detectable is by its effect on the optical emission lines.
Serote Roos \& Gon\c calves (2004) have concluded that this is not the
case for these objects. They have examined the spectroscopic properties of 90\% of the PEGs of
the 200 mJy sample and found that the dust content  associated with the emitting regions is small 
($E(B-V)<0.25$). Thus, even if there is good evidence for the presence of dusty regions in PEGs,
extinction is not the main cause of the low level of activity in these objects.\\

\noindent $\bullet$ Seyfert1-like objects tend to have lower
sub-mm/infrared excesses than PEGs (see Figure~\ref{meaninfrared}), 
eventhough there are some exceptions, with 
objects having  F$_{\rm{IR}}$/F$_{\rm{ext}}$~$>>$~100. This set comprises 1217+295, 
1254+571 (Mrk 231), 1532+236 (Arp 220), all objects that are known to have large quantities of 
dust (see notes on individual sources in Appendix \ref{individual}).  We note that some Seyfert-like 
objects show  F$_{\rm{IR}}$/F$_{\rm{ext}}$ similar to those shown by BL Lac objects. In those
cases the emission is probably non-thermal in origin.

\subsection{The SEDs  and radio morphology}

In this section we look for any connection between the shapes of the
SEDs and the type of radio-structure of the sources. We have a heterogeneous 
selection of VLA observations ranging in resolution from 0.2 to 45~arcsec 
and we have VLBI observations down to a resolution of $\sim$2~mas.
All objects were
initially selected to have compact cores which are the dominant
features in the 0.2~arcsec resolution VLA 8.4~GHz maps. In addition to
these 8.4~GHz VLA observations, all objects are in the NVSS Catalogue \citep{condon98}, 
which gives information on any radio structure larger than
about an arc-minute. A particularly important question is whether or not the compact radio
emission comes from a nuclear jet and this is why we focus in the VLBI
data here. There are VLBI observations of 58 of the 64 sources in the sample
and in Table \ref{200table} we present the current best classification of the structures
in terms of one-sided core-jet (C-J), two sided (2-S), unresolved single
component (U), or unclassified when the morphology was too complex or
ill-defined. In the 2-S category we include all the objects with jet and/or
lobe emission on both sides of the core or core candidate. This is a 
somewhat simplistic classification scheme, but our main intention here is 
just to separate objects in
which the pc-scale emission is clearly dominated by a beamed jet component
(C-J) from the others (2-S). The 2-S sources contain also some Compact
Symmetric Objects and CSO candidates which are intrinsically small and
young objects. \\
There are 9 objects that are classified as 2-S. Most of them (6) are
classified as PEGs, 4 have an infrared bump in
the SED suggesting the presence of dust emission, and the remaining 2 have a
steep power-law spectrum. In these the
dominant radio emission might {\it not} come from the core but from
compact self-absorbed lobes. In that case, we would not expect the synchrotron
emission from these lobes to be detectable at millimetre or optical
wavelengths. This expectation is supported by the SEDs. \\
There are 40 objects that are classify as having a core-jet radio
structure and it is in this sub-set of objects that we 
look  how far the synchrotron core emission extends up in the
electromagnetic spectrum. We would expect many of these to have broken
power-law SEDs. Twenty eight of the 40 sources do, while the rest are
divided between types I, II and unclassified SEDs.
It is of little surprise that 18 of the 28 core-jet radio objects which
have broken power-law SEDs, are classified as BL Lacs. In addition,
one more is a BL Lac candidate. The remaining nine are
broad-line (6) and  narrow-line (1) objects and PEGs (2). 
We find the high proportion of
broad-line objects with broken power law a somewhat surprising result since in these objects
we might expect the optical and near-infrared emission to be dominated
by accretion disk or dust re-processed emission, respectively. Two of
the six have already been noted as unusual objects having ``hybrid''
properties intermediate between BL Lac objects and Seyfert galaxies;
they possess both a broad-line region and polarised optical continuum
emission (March\~a et al, 1996; Jackson and March\~a, 1999). Perhaps, 
some of the other four objects might be similar hybrids.
We note that none of the PEGS look like infrared/optical
synchrotron emitters -- they would not be classified as PEGS if they
did. However, there are core-jet objects, like 0055+300 and
1144+352, whose radio-to-submm spectra suggest
synchrotron emission but whose SEDs in the infrared are distorted by
additional dust emission.

\section{Conclusions}
\label{section-conclusions}

The main goal of this work was to investigate the
origin of the diverse optical activity amongst similar radio-loud 
low-luminosity flat-spectrum objects. The investigation was based on the
200 mJy sample, a radio-selected sample showing a range of optical
activity. To achieve that goal, multi-wavelength observations were 
obtained (millimetre data with VLA, sub-mm data with SCUBA and infrared data 
with {\it ISO}). Adding  data from the literature allowed us to produce the SEDs of 
64 objects. The SEDs were analysed with two main goals: 
to characterize the non-thermal synchrotron component and to search for
dust emission. Our findings may be summarised as follows:
\begin{enumerate}
\item There is considerable diversity amongst the SEDs but objects fall broadly into 
three groups: there are objects with steep spectra between 1.4 GHz and 43 GHz (14\%
of the sample), there are those with flat spectra over most of the
spectral range (48\% of the sample) and there are those which show pronounced 
sub-mm/infrared excesses (27\% of the sample).
\item Sub-mm observations indicate that 
a major fraction of the 200 mJy objects have smooth and flattish synchrotron 
spectra up to the sub-mm band. This includes BL Lacs, BL Lac
Candidates, and some Seyfert1-like and PEG objects. In only few cases the 
dominant sub-mm emission mechanism is clearly thermal dust emission  (for
example, Arp 220 and Mrk 231). 
\item The broadband spectra between
radio and infrared wavelengths suggest that different
objects have different dominant emission mechanisms in the infrared band.
Some objects show a smooth broadband spectrum  which can be explained 
by synchrotron emission -- all BL Lac and some broad-emission line objects are in
this group.  Others show moderate infrared excess over 
the extrapolated radio/submillimetre emission, suggesting that both
synchrotron emission and dust emission contribute to the infrared -- this
group includes broad and narrow emission line objects and PEGs.
The far-infrared bump present in the majority of the PEGs indicates
the presence of cold dust, which mass we estimate to be similar to 
that found in typical early-type galaxies. Even though there is
dust emission the analysis of Serote Roos \& Gon\c calves (2004) suggests
that it is not strongly affecting the line emission regions. Thus
extinction is unlikely to account for the low level of observed 
activity on PEGs.
\item Correlating the broadband SED shapes with the radio morphology
(through high resolution radio images available in the literature)
allow us to conclude that the sources with two sided morphologies nearly all
have falling spectra between 1.4 GHz and 43 GHz. Not surprisingly, all
the sources showing core-jet radio morphology have flat broadband
spectra often extending well into the sub-mm/infrared range. The vast
majority of these are spectroscopically classified as BL Lacs or Seyfert-like
objects. In a subsequent paper we will discuss in more detail the SEDs of those
objects that are synchrotron emitters, independently of the optical 
classification. We will
compare their spectra with those of blazars, in general, and suggest
that there may be many more objects  with low
frequency peaks in their synchrotron spectra than has been previously
thought.

\end{enumerate}

\section{Acknowledgments}
We thank our anonymous referee for very constructive comments.
SA acknowledges all the support during the visit to the ISOPHOT Data Centre in Heidelberg.
SA  acknowledges the financial support from the European Commission, 
TMR Programme, Research  Network Contract ERBFMRXCT96-0034 ``CERES'', and from the
Portuguese Funda\c c\~ao para a Ci\^encia e Tecnologia through the grant SFRH/BPD/5692/2001. 
M.J.M. M. acknowledges the financial support from 
the Portuguese Funda\c c\~ao para a Ci\^encia e Tecnologia through the grant SFRH/BPD/3610/2000 
This research has made use of the NASA/IPAC Extragalactic
Database (NED) which is operated by the Jet Propulsion Laboratory,
California Institute of Technology, under contract with the National
Aeronautics and Space Administration. "This research has made use of the NASA/ IPAC Infrared Science Archive,
 which is operated by the Jet Propulsion Laboratory, California Institute of Technology, under contract with 
the National Aeronautics and Space Administration."

\appendix

\section{Notes on individual sources}
\label{individual}

$\bullet$ 1146+596 -- Its SED suggests that the non-thermal cut-off
might happen in the sub-millimetre regime. This is a core-dominated source at
JVAS resolution (0.2$''$), however VLBI monitoring  at 5 GHz revealed  twin
relativistic jets in a scale of  miliarcseconds \citep{taylor98}. We note that the absence 
of large-scale radio emission in NVSS map might indicate that this is a young radio 
source. \\
$\bullet$ 1217+295 (NGC 4278) is a well known radio galaxy.  The  
broadband spectrum between 5 and 43 GHz is relatively steep,
 and the sub-mm emission is likely to be thermal, something
that is consistent with the detection of large-scale dusty regions reported
by \citet{carollo97}.\\
$\bullet$ 1245+676 --   NVSS map reveals a large Mpc FRII morphology, but  the
VLBA map shows that the core has a symmetric structure with two mini-lobes
\citep{bruyn91}. This is a so called 'double-double' source.\\
$\bullet$ 1254+571 (Mrk 231) and 1532+236 (Arp 220) -- They have a strong 
inverted sub-mm spectra -- both objects are  Ultra-luminous Infrared galaxies
(e.g. Sanders et al. 1988). The $\alpha_{\rm{sub-mm}}$ of 1532+236 
is 1.91$\pm$ 0.39 and the $\alpha_{\rm{sub-mm}}$ of 1254+571 is 1.21$\pm$1.04.\\
$\bullet$ 1404+286 -- This is a Seyfert1-like object. The broadband spectrum
is inverted in the radio and steep between the radio and the sub-mm band, with 
$\alpha_{\rm {radio}}\sim\alpha_{\rm{sub-mm}}$. 1404+286 is one of the best examples
 of  Gigahertz Peak Sources (GPS), the peak occurring at 5 GHz. \\

\begin{figure*}

\includegraphics[width=5.5cm]{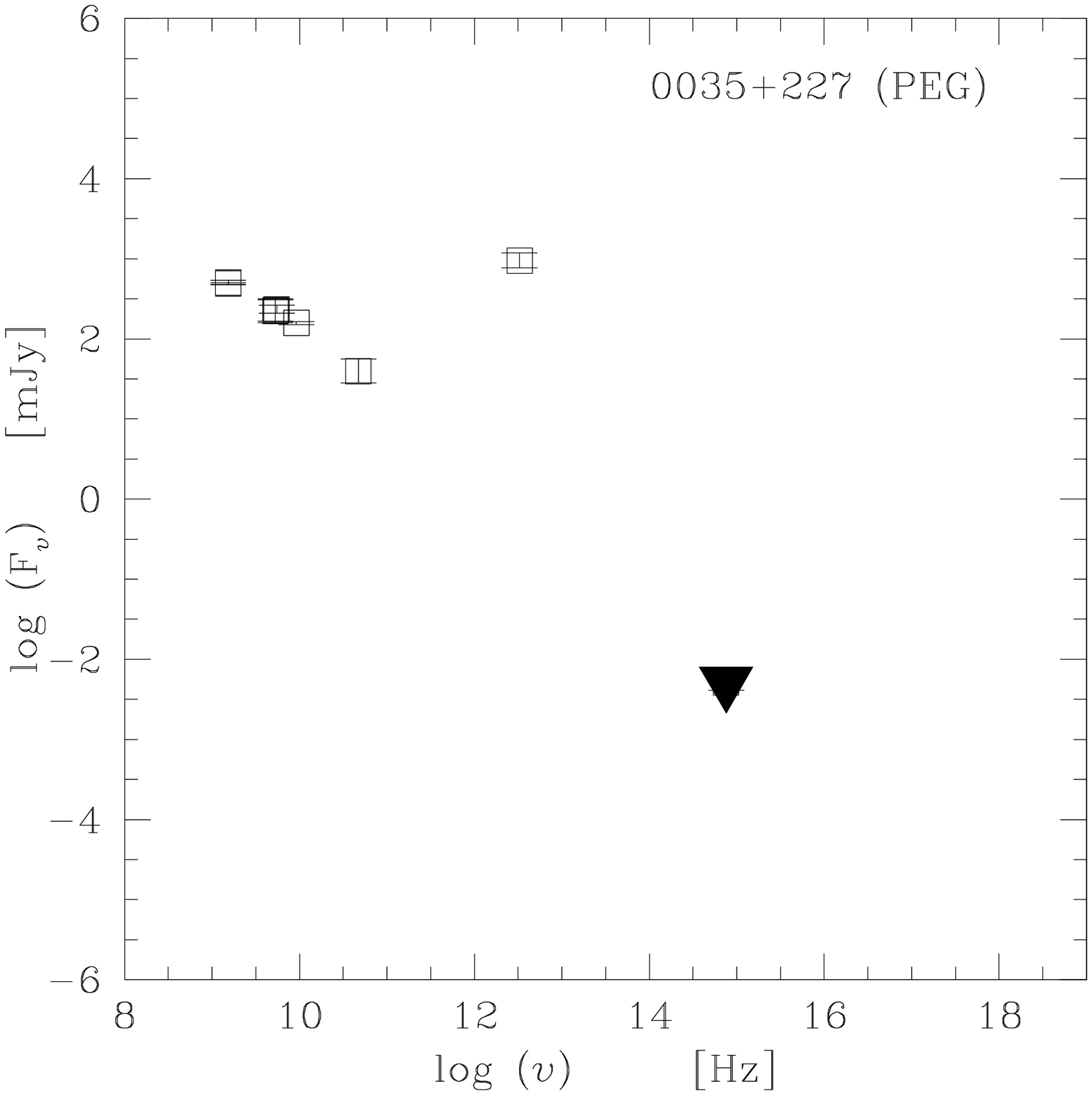}
\includegraphics*[width=5.5cm]{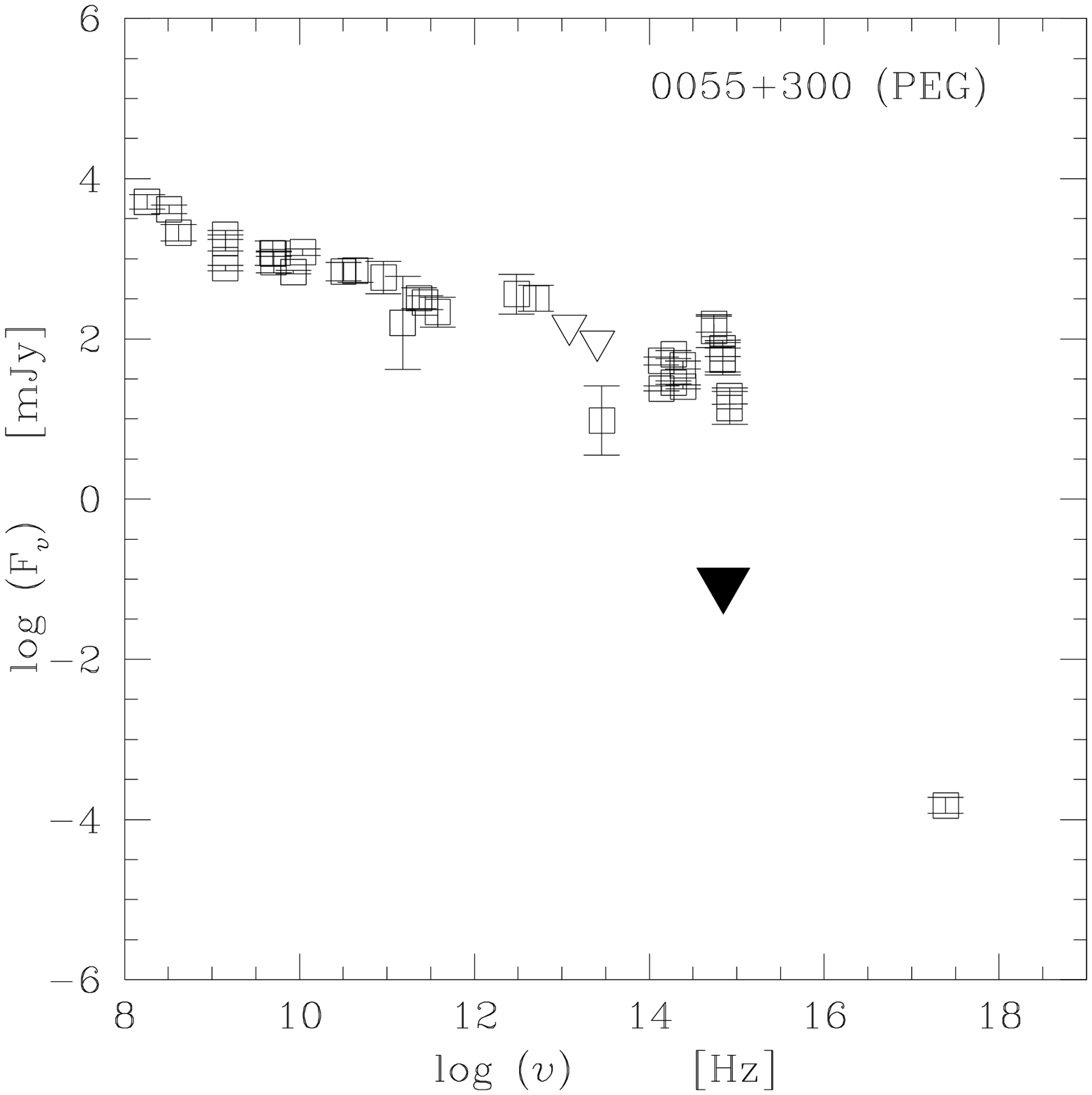}
\includegraphics*[width=5.5cm]{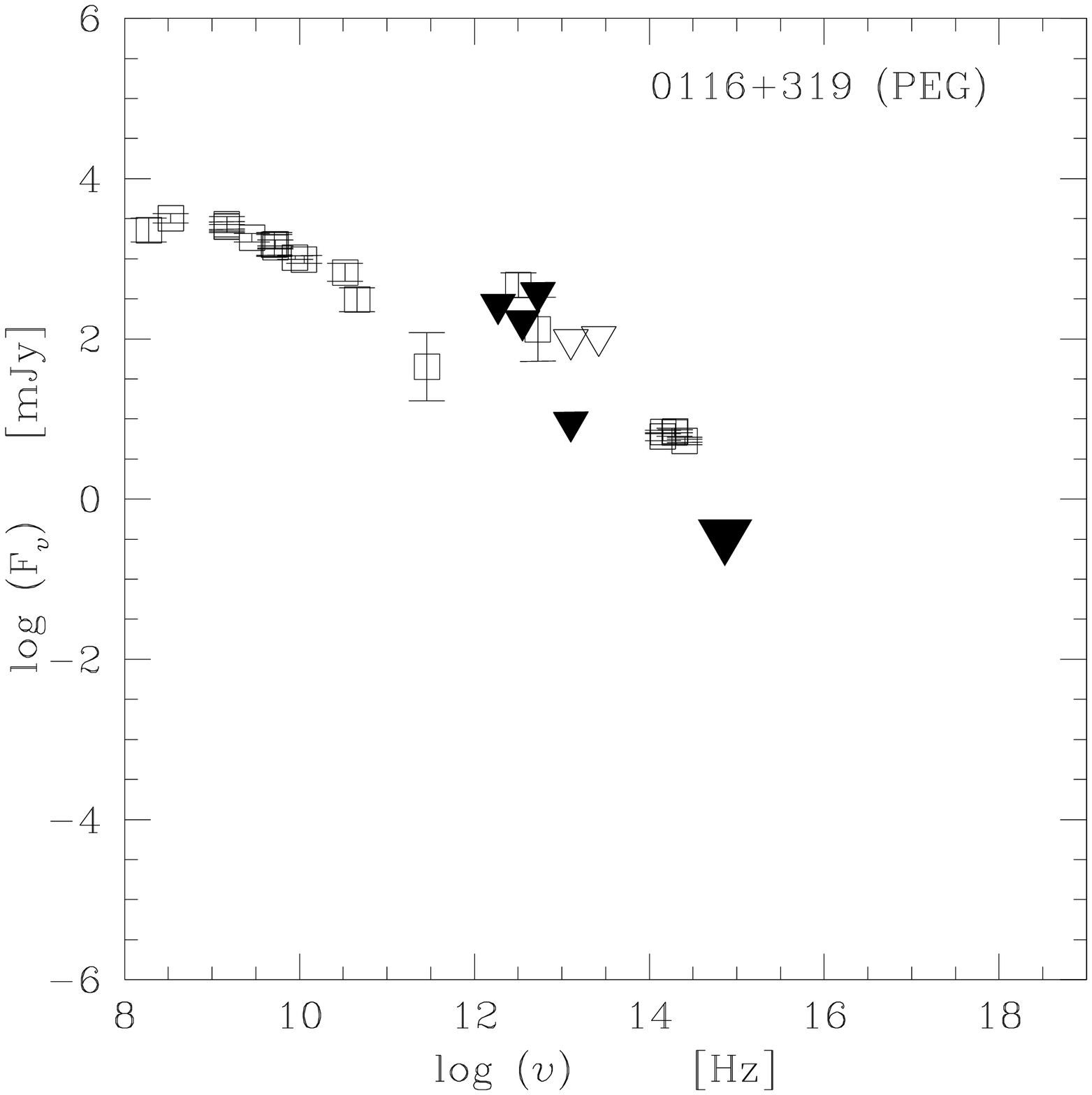}

\includegraphics[width=5.5cm]{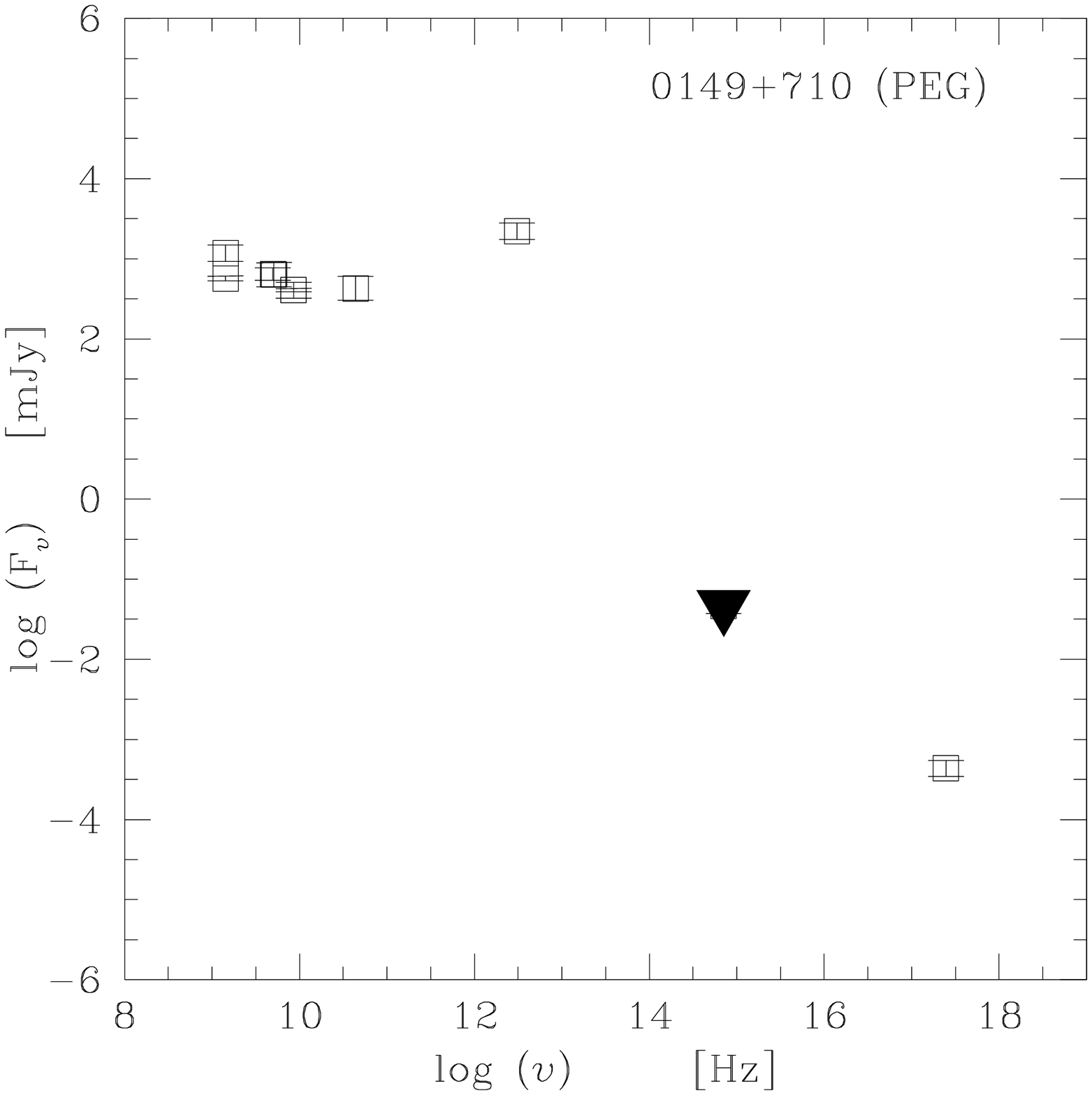}
\includegraphics*[width=5.5cm]{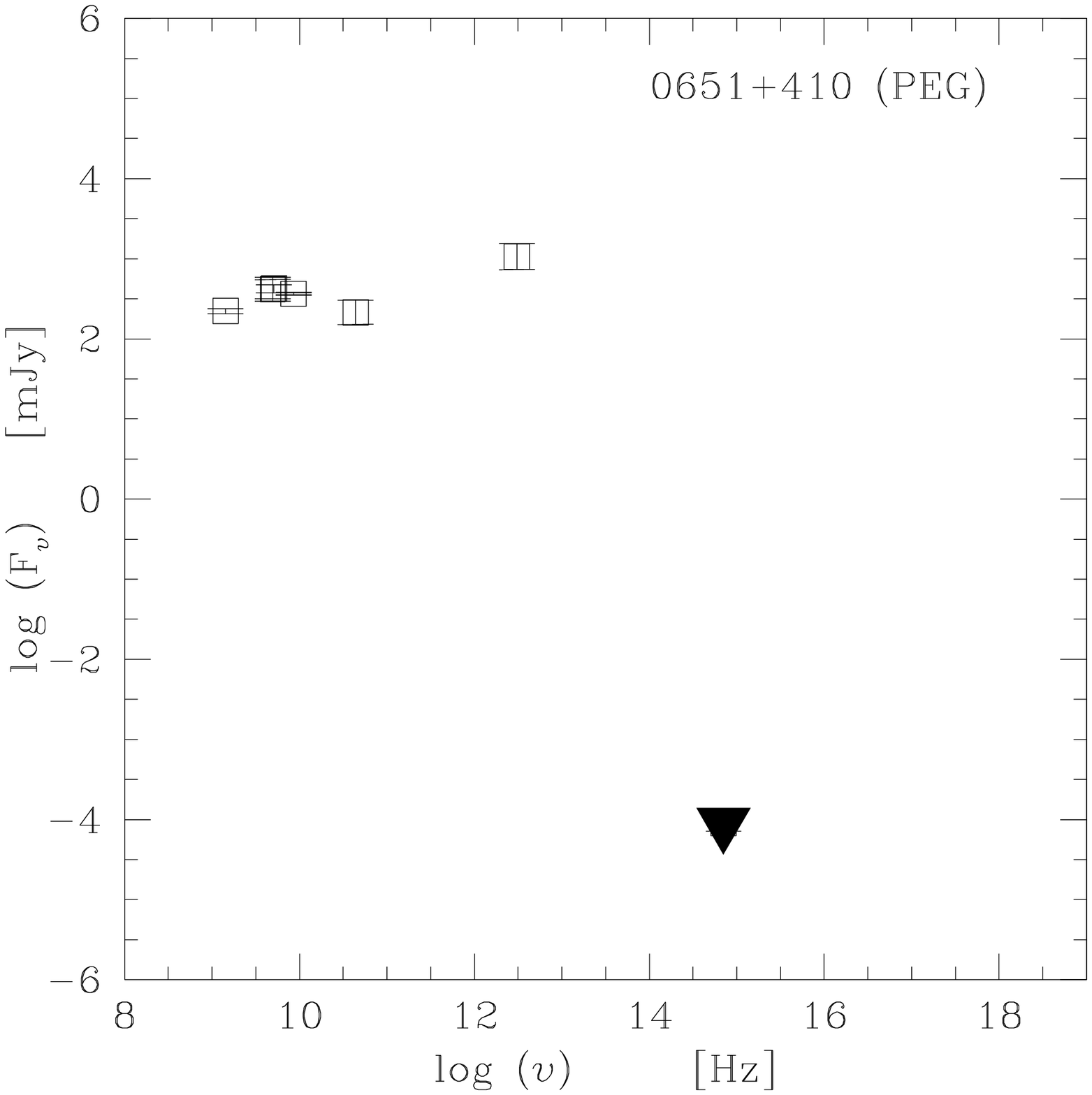}
\includegraphics*[width=5.5cm]{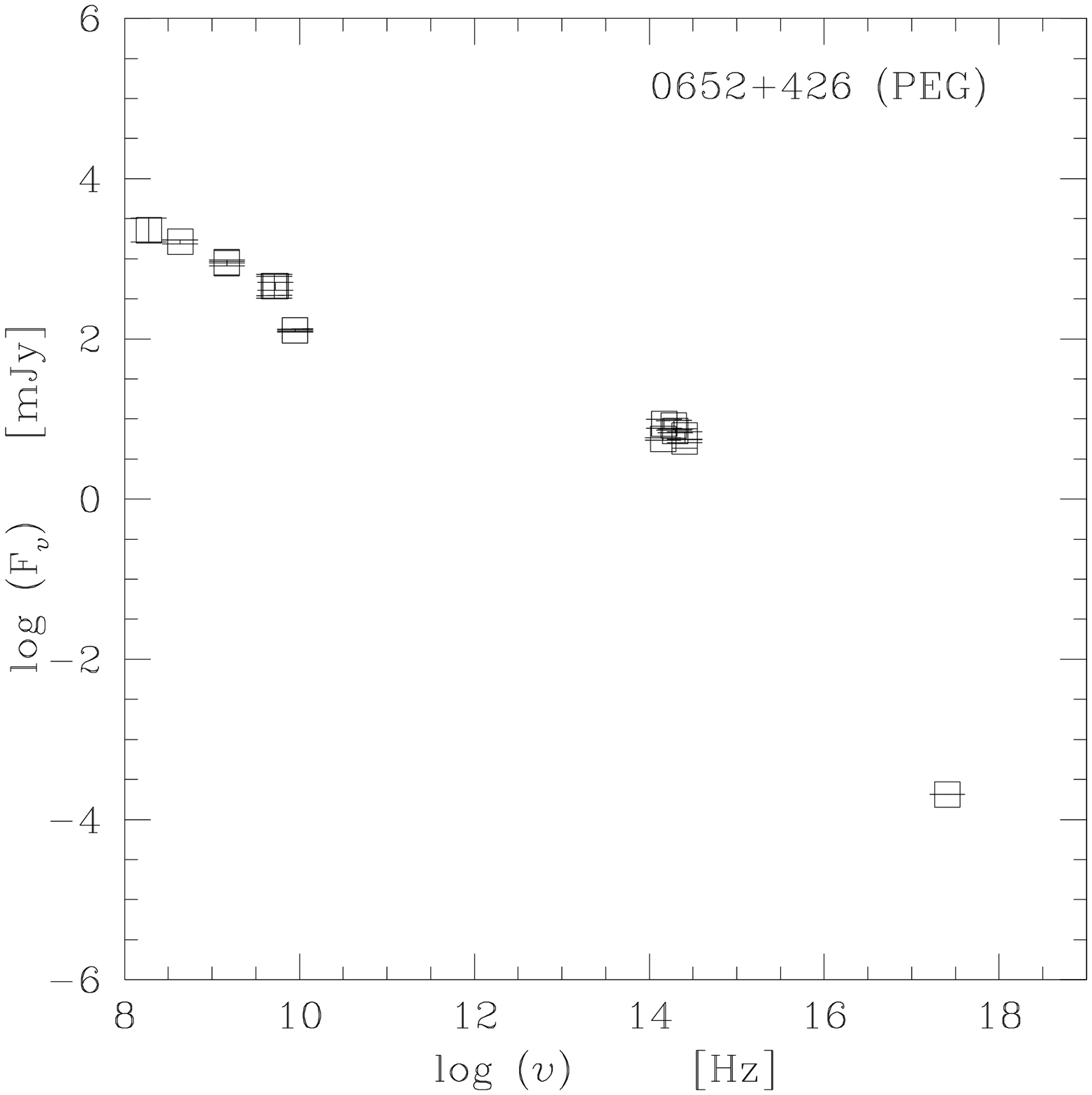}

\includegraphics[width=5.5cm]{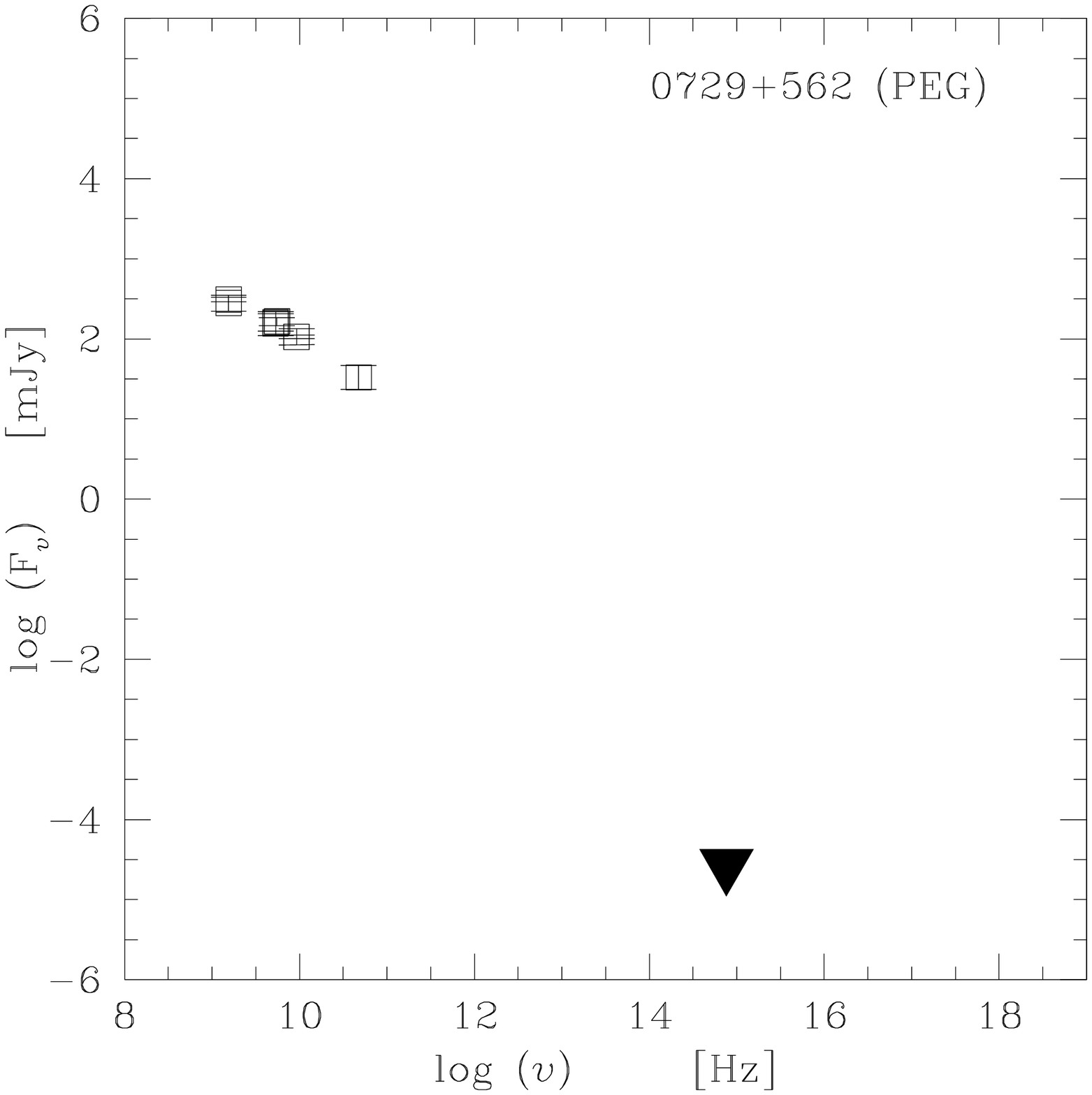}
\includegraphics*[width=5.5cm]{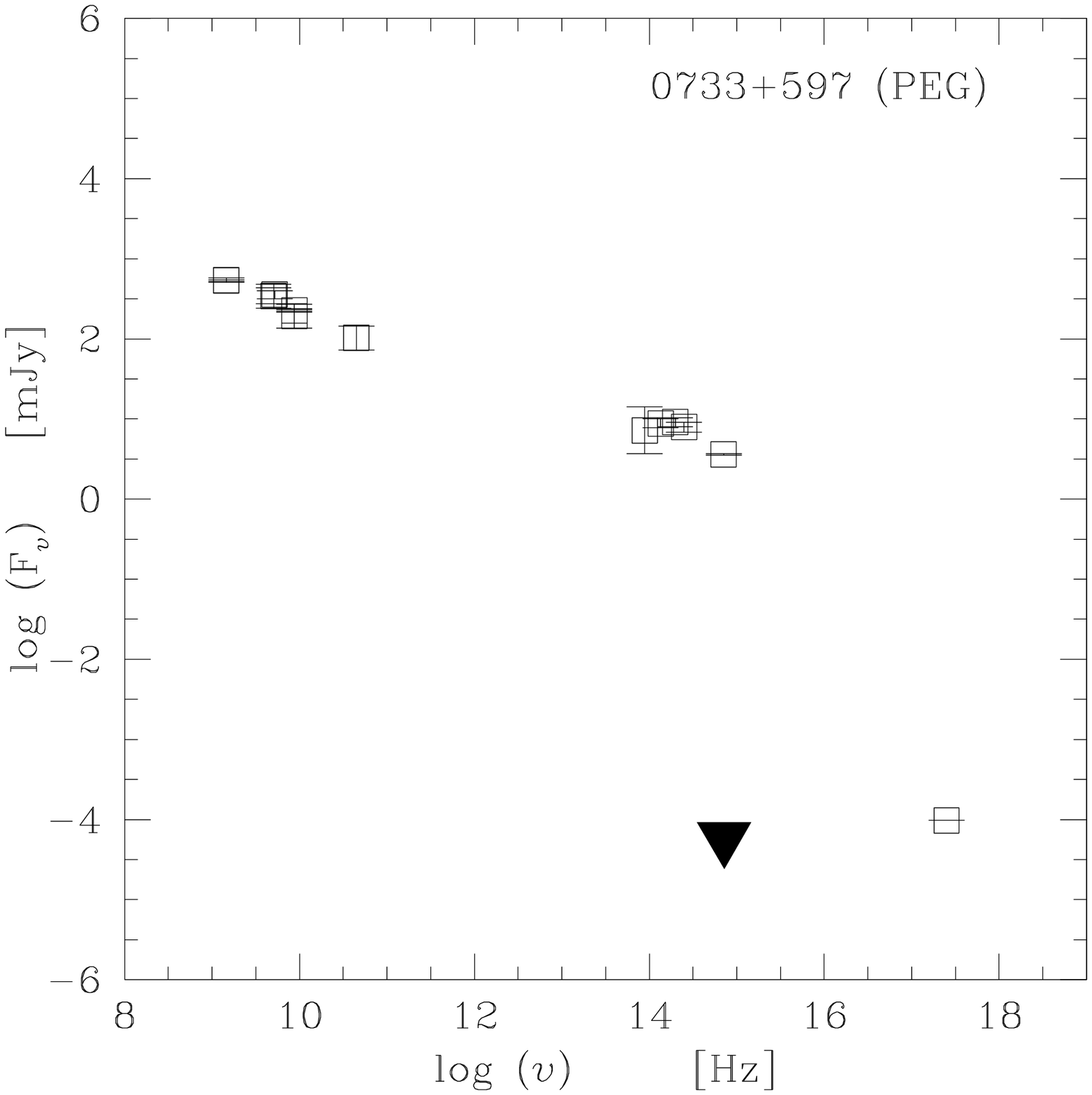}
\includegraphics*[width=5.5cm]{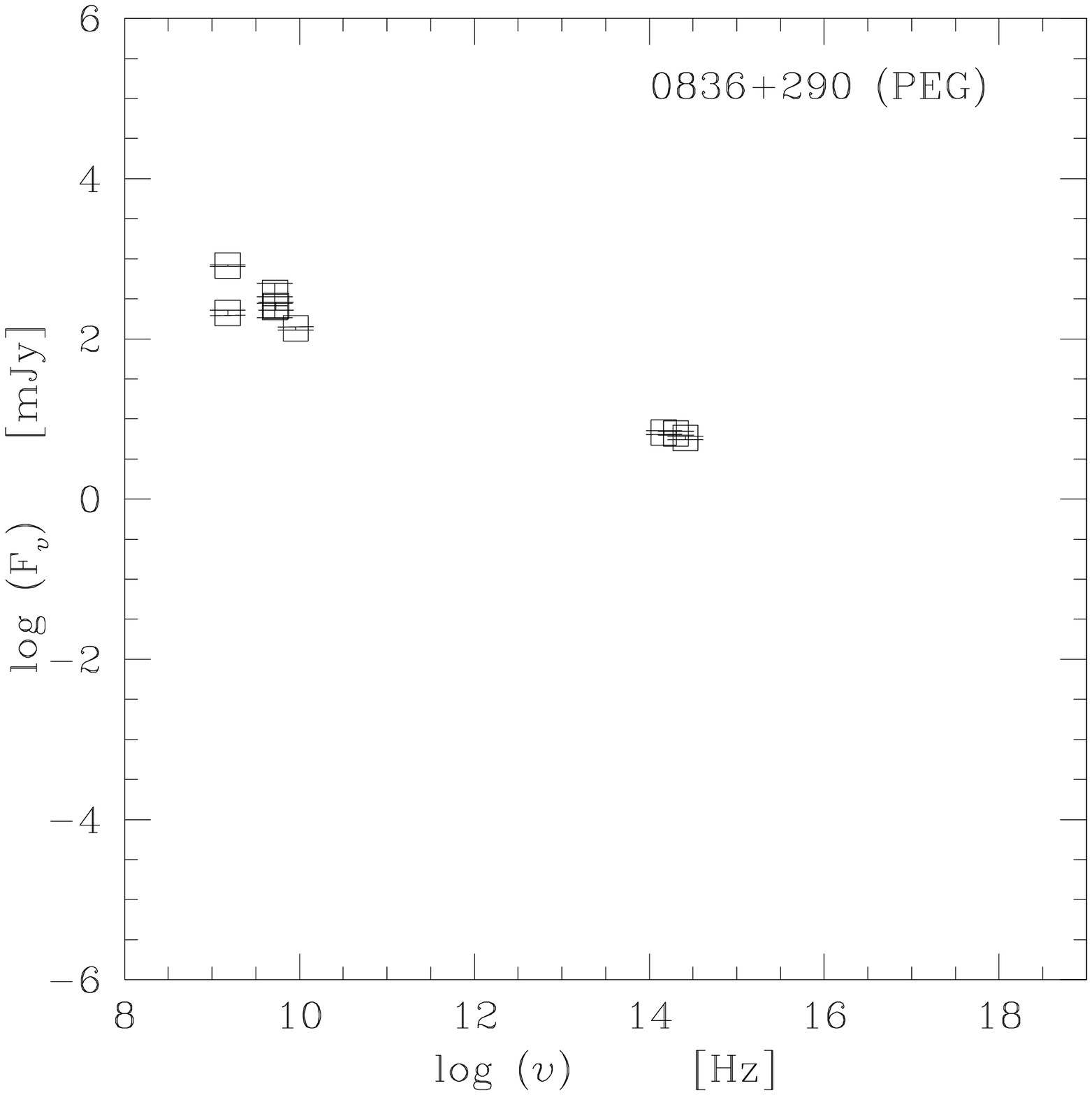}

\includegraphics[width=5.5cm]{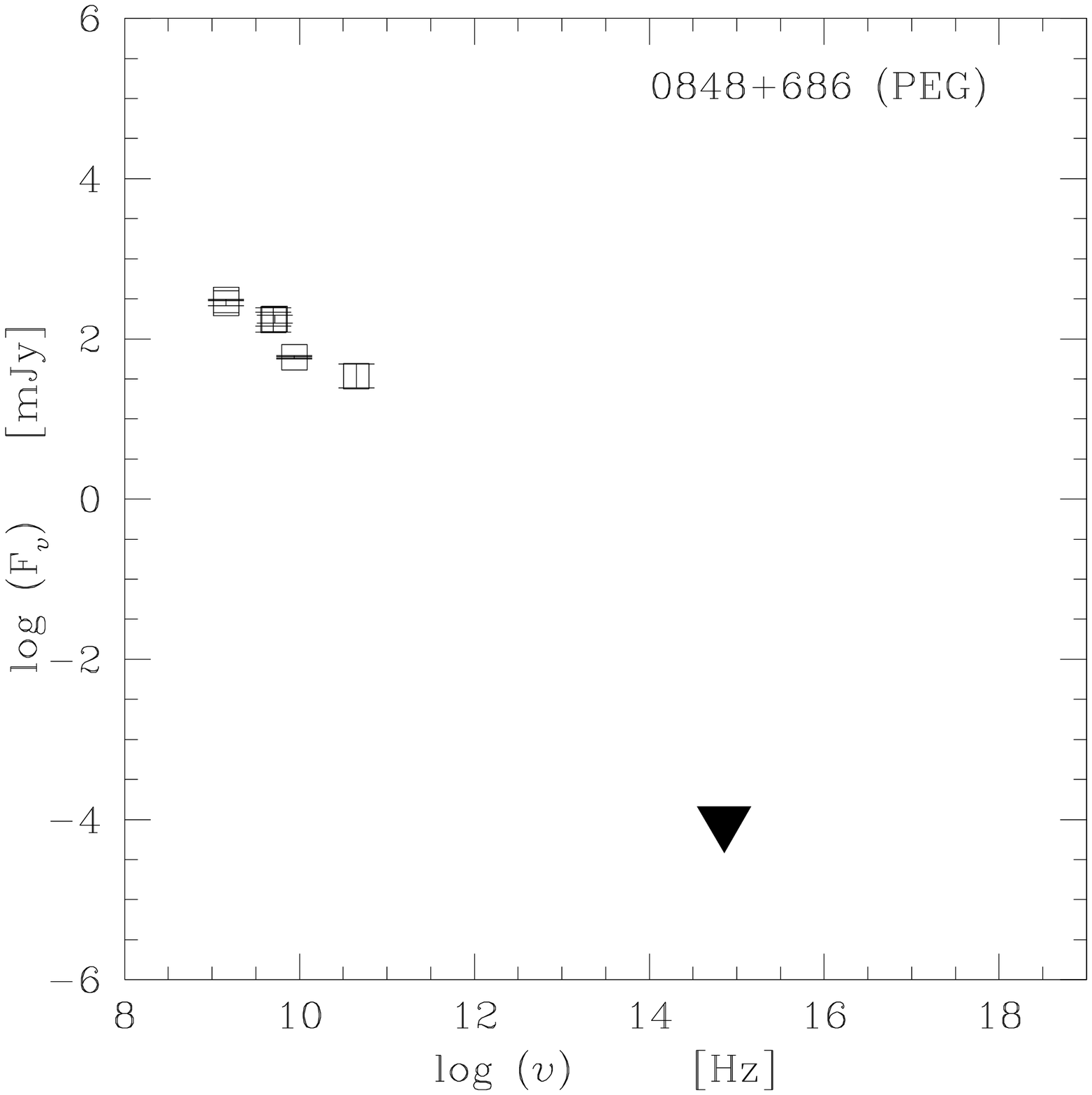}
\includegraphics*[width=5.5cm]{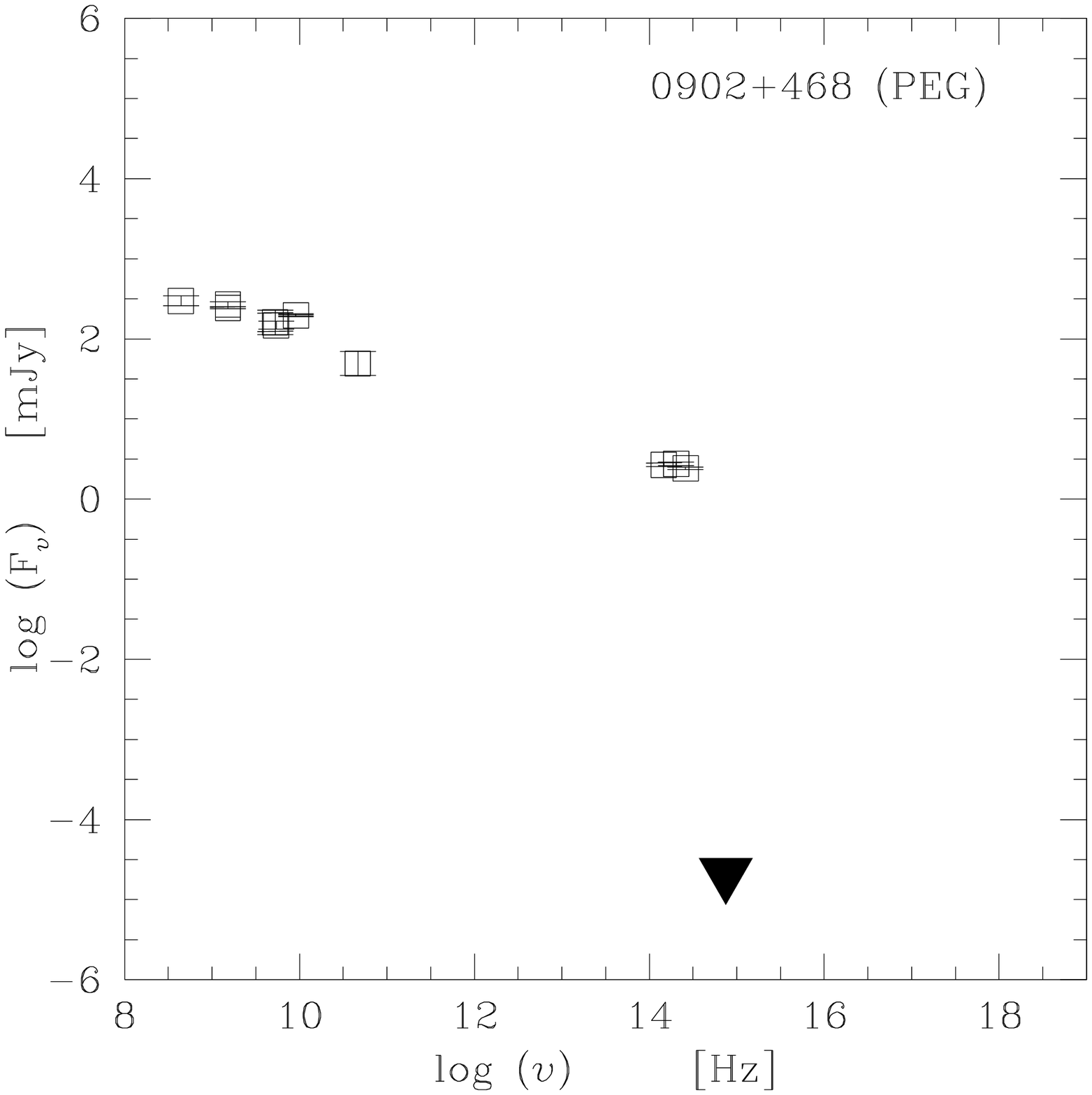}
\includegraphics*[width=5.5cm]{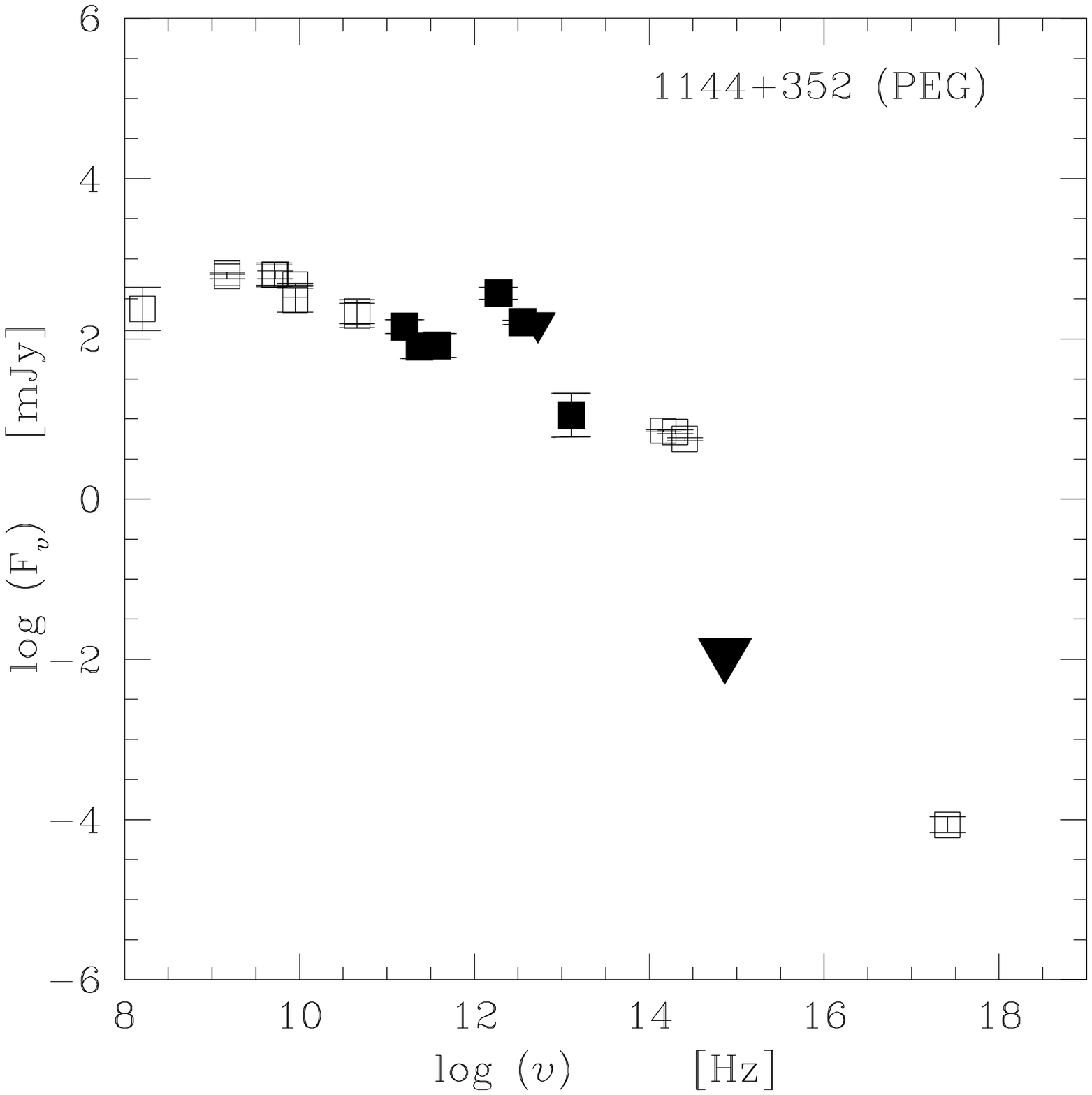}

\end{figure*}

\begin{figure*}

\includegraphics[width=5.5cm]{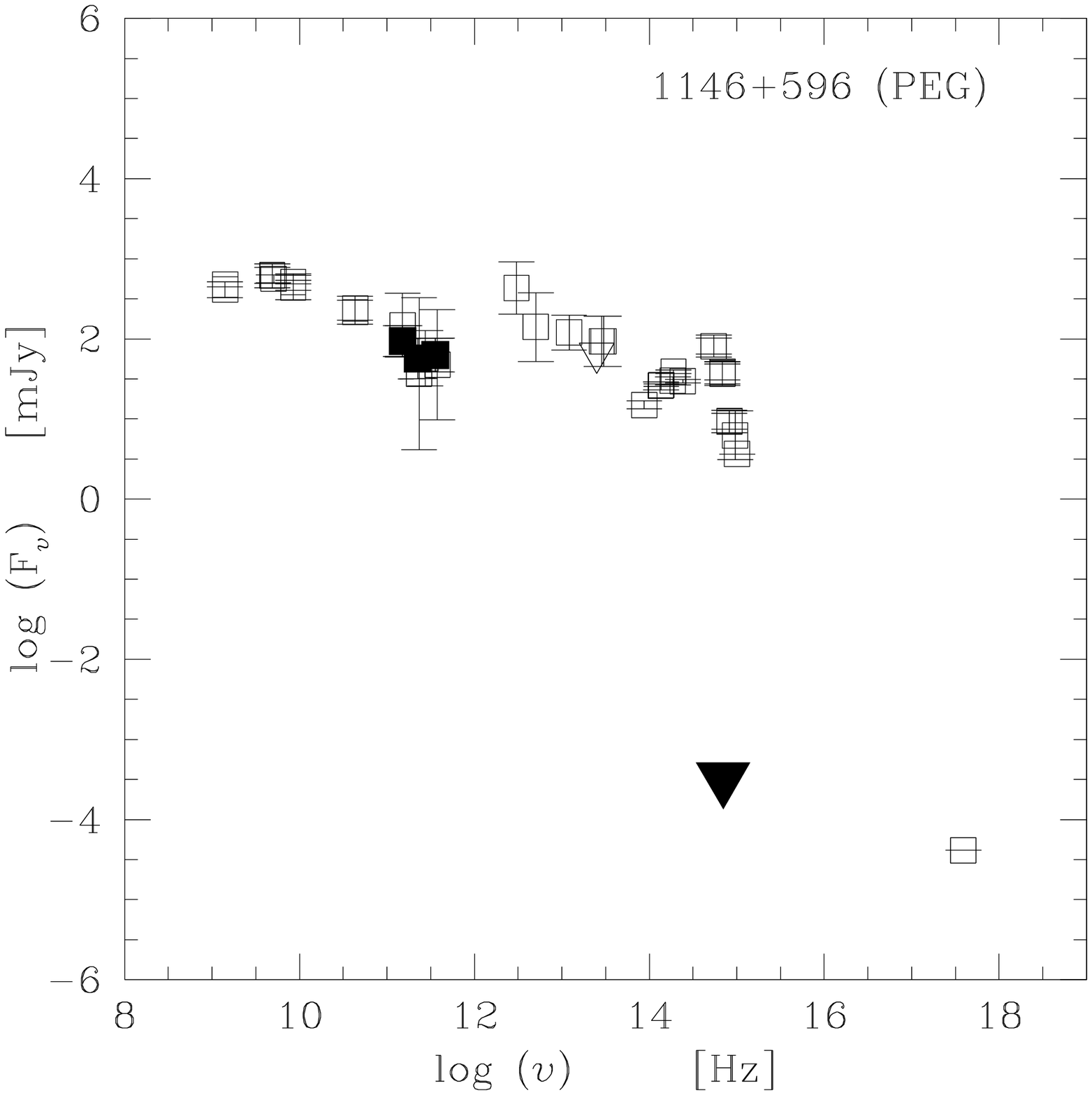}
\includegraphics*[width=5.5cm]{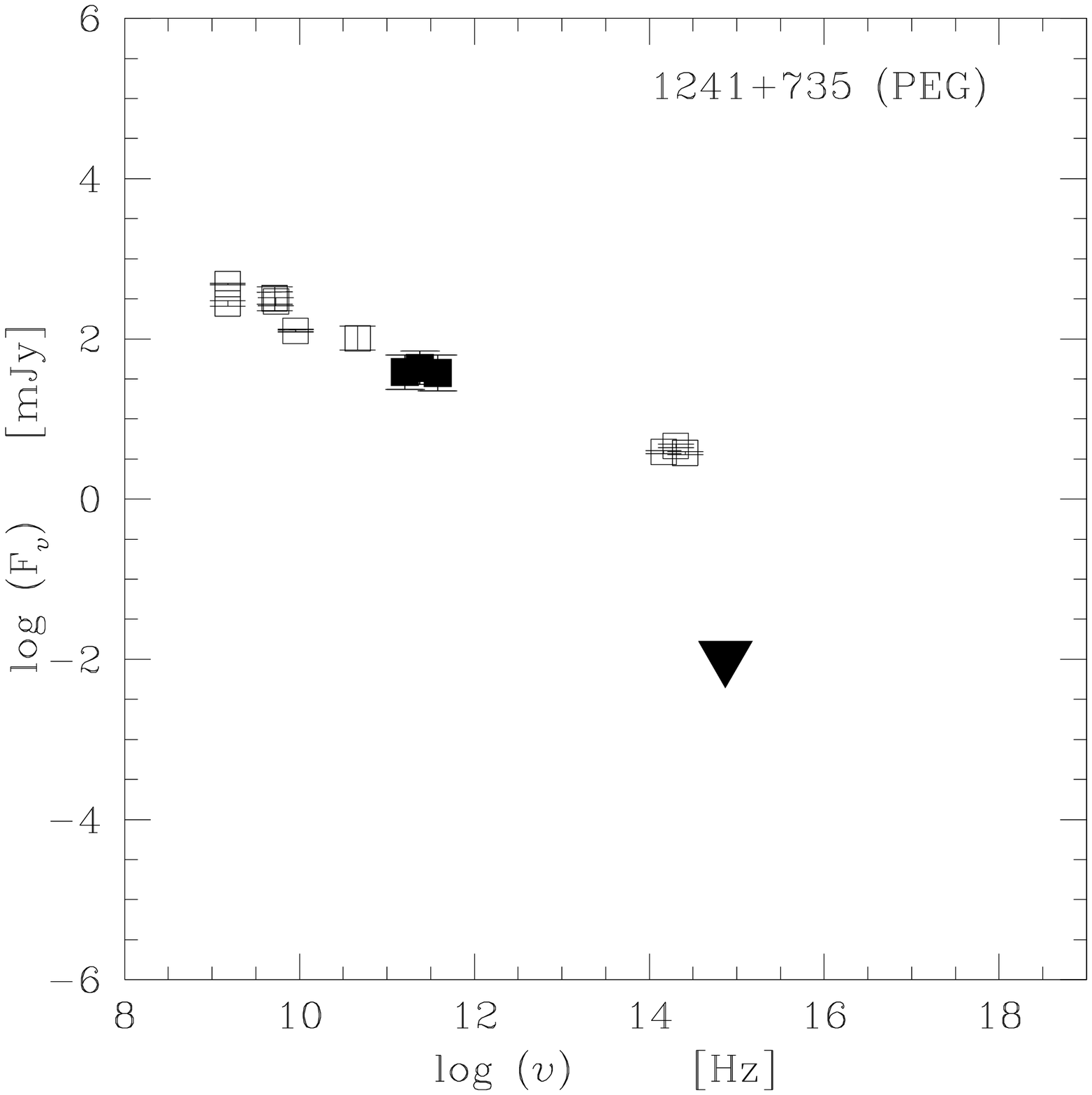}
\includegraphics*[width=5.5cm]{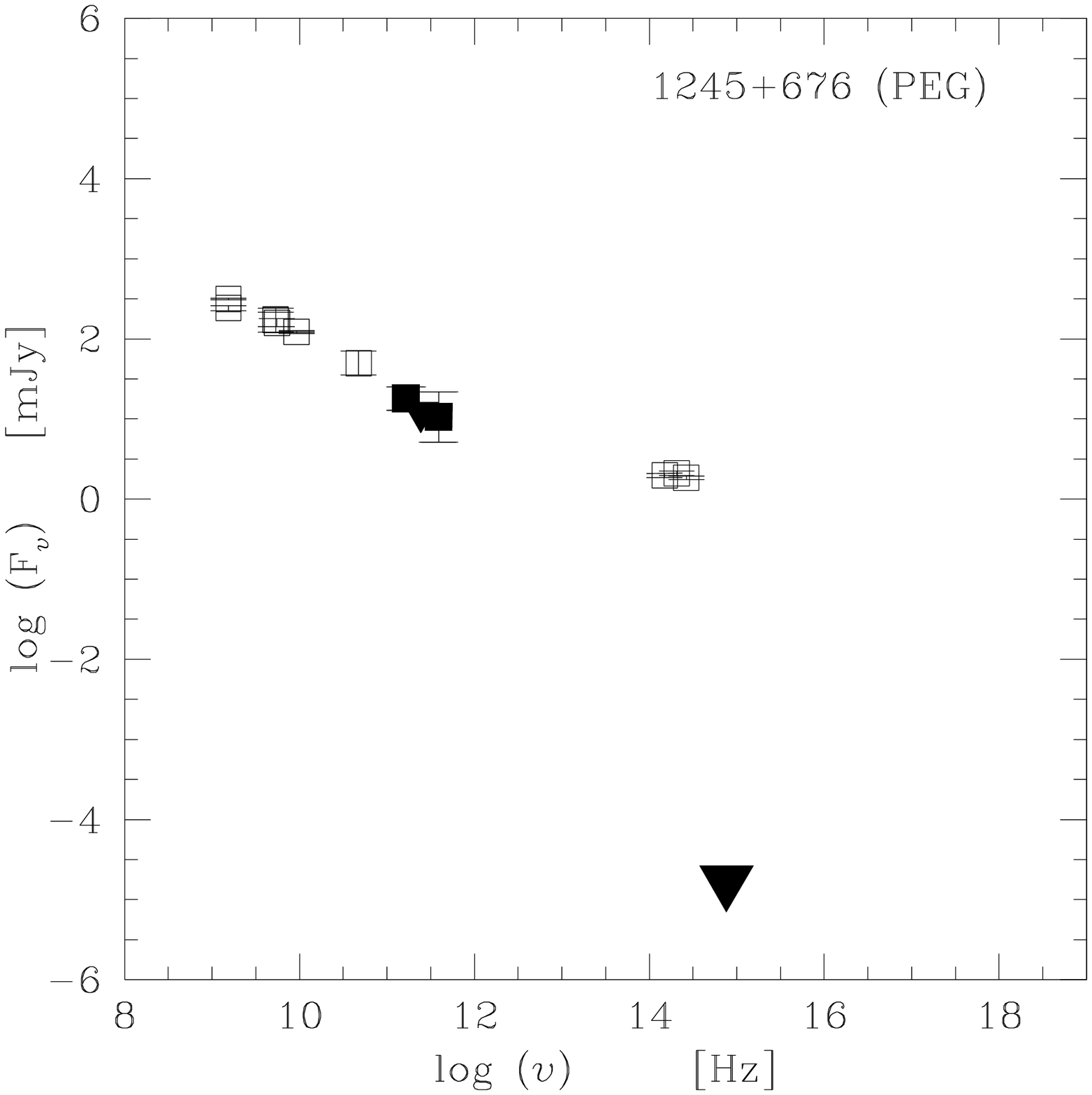}

\includegraphics[width=5.5cm]{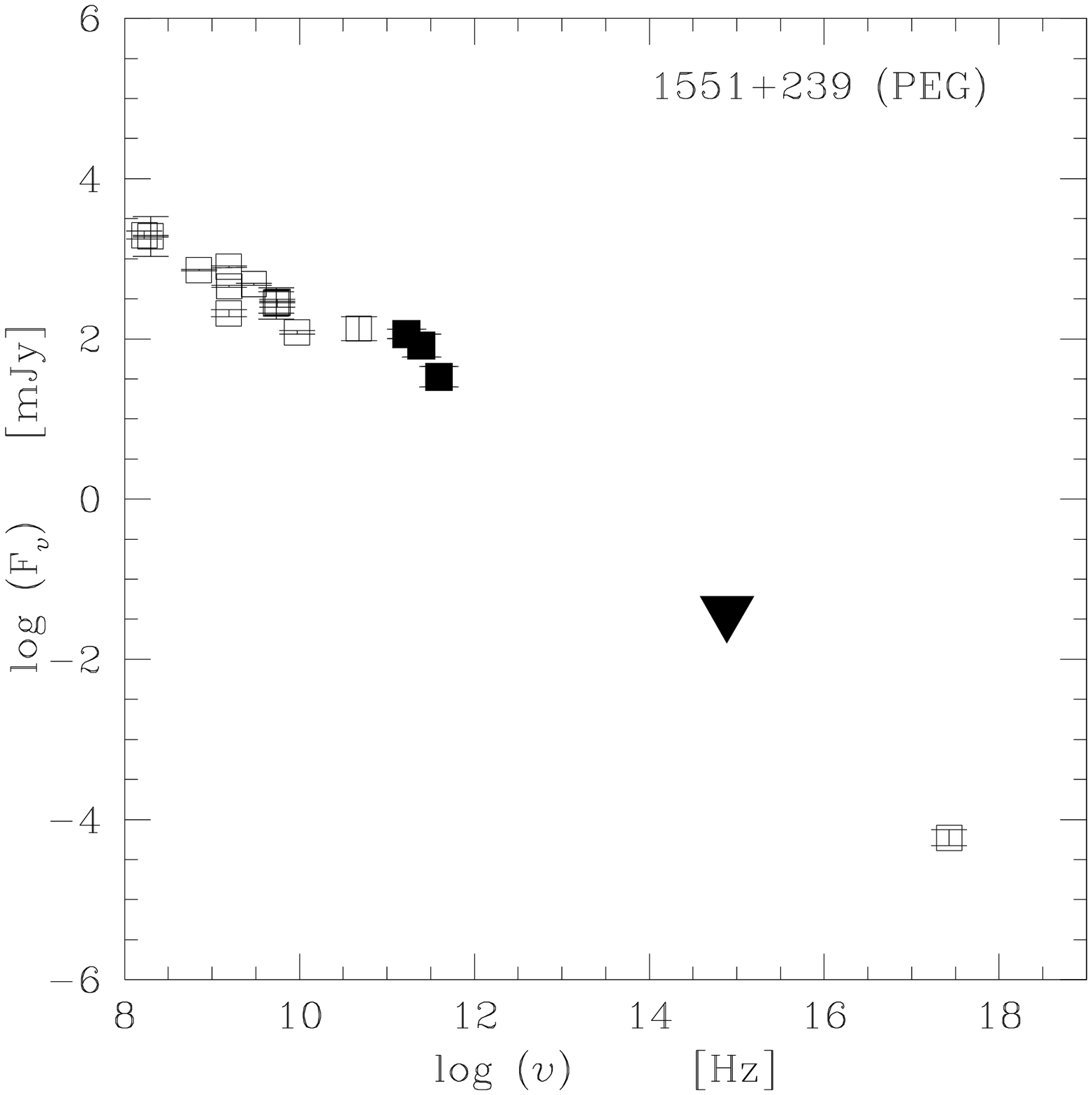}
\includegraphics*[width=5.5cm]{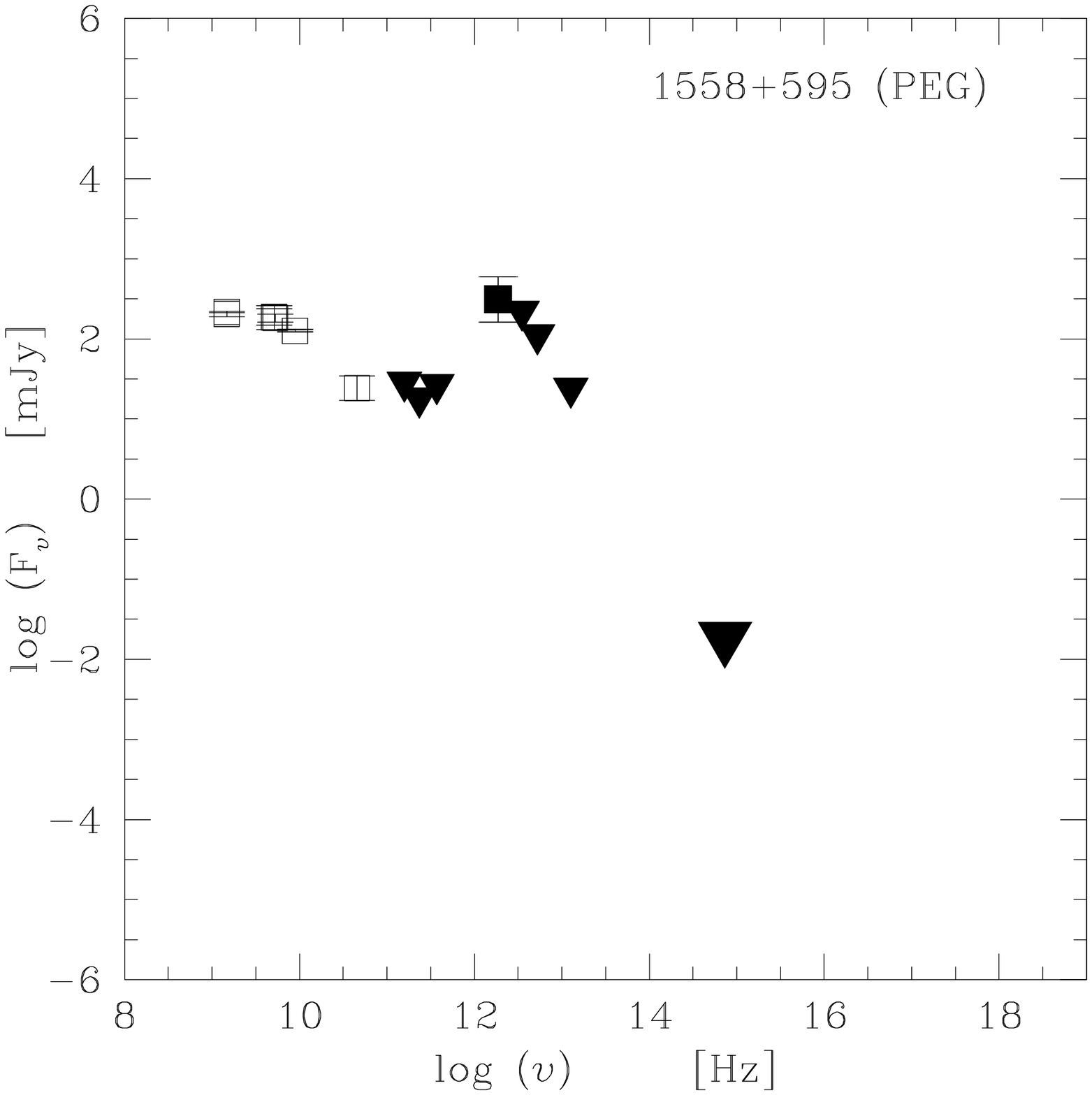}
\includegraphics*[width=5.5cm]{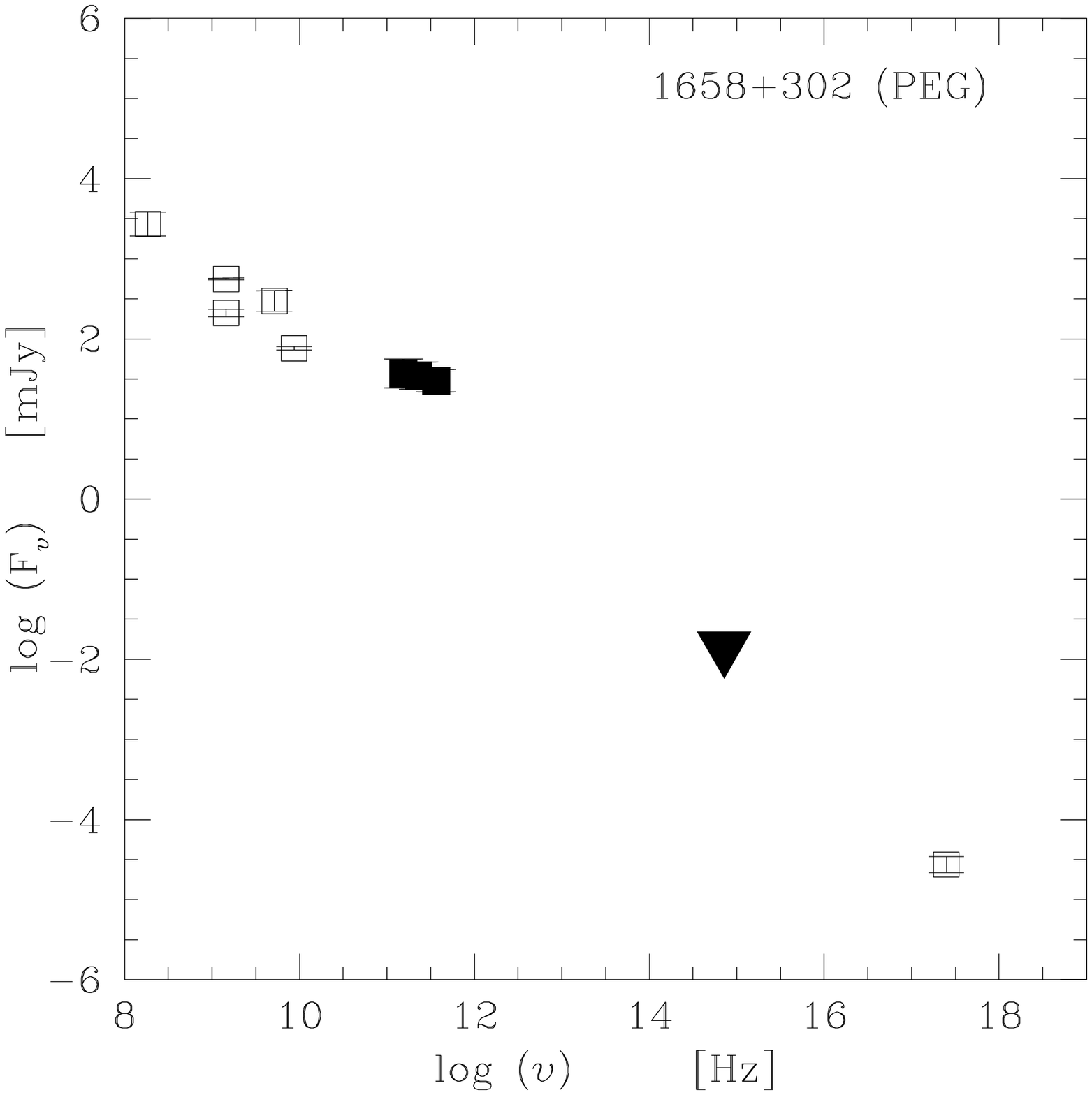}

\includegraphics[width=5.5cm]{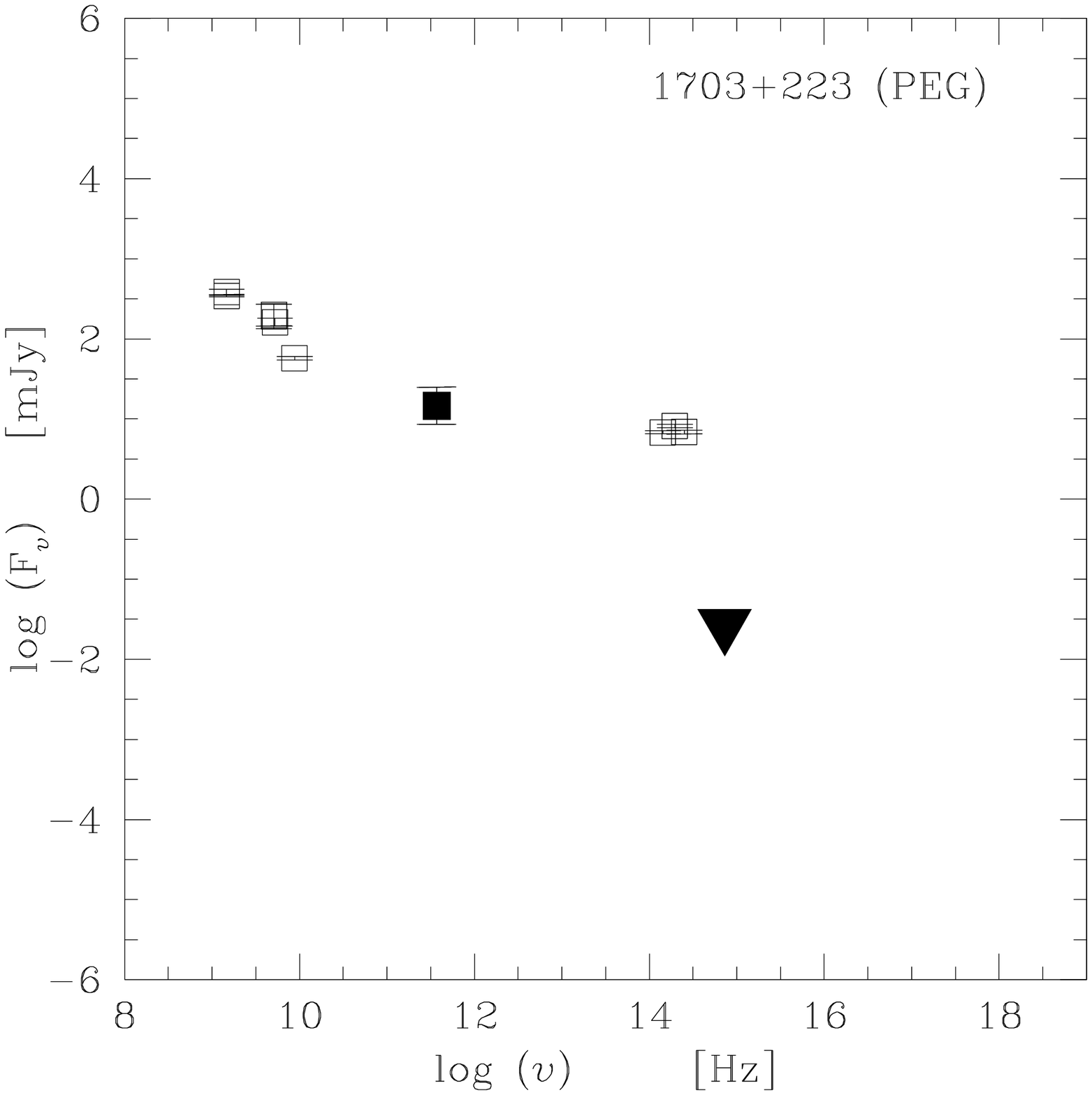}
\includegraphics*[width=5.5cm]{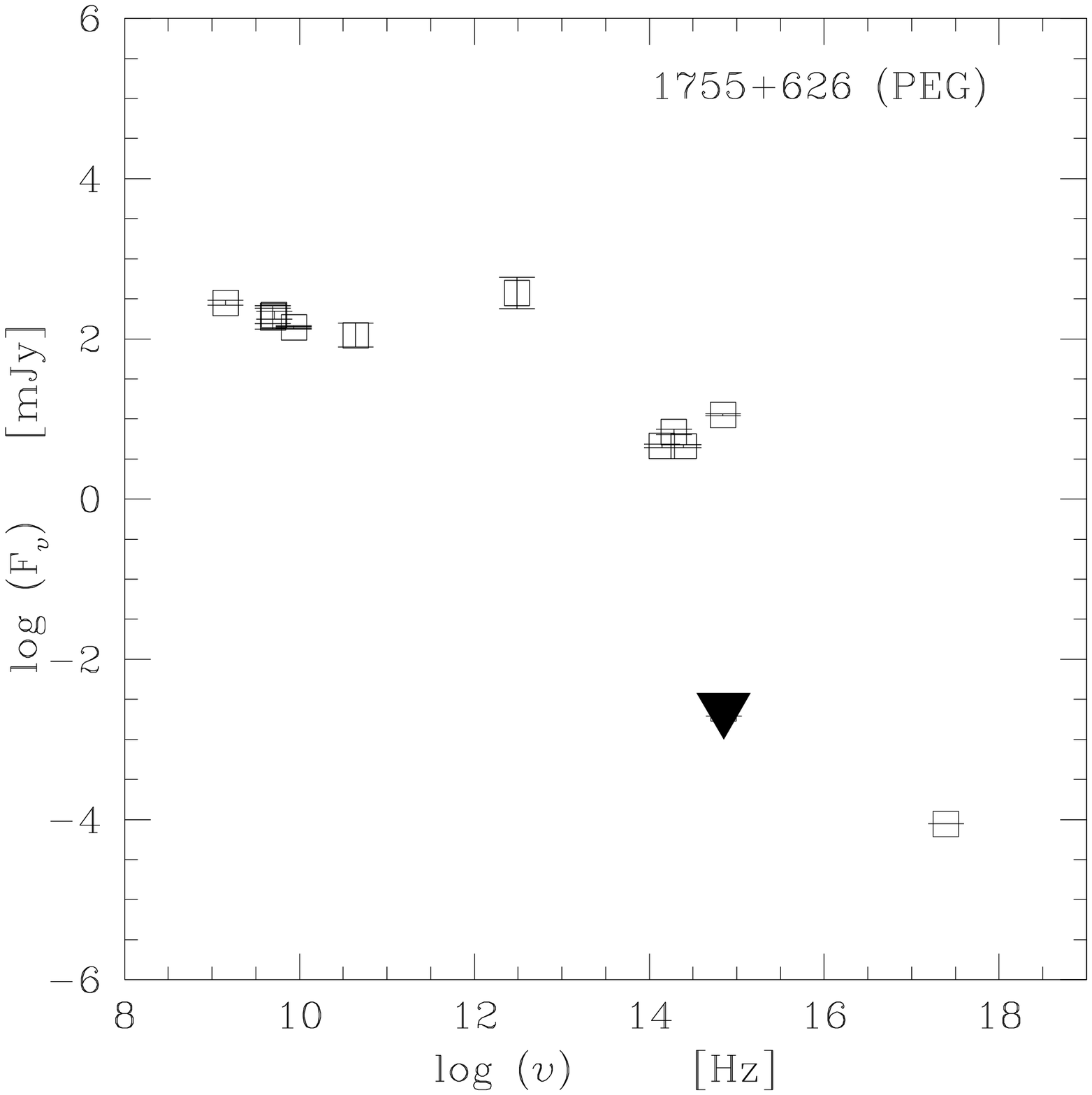}
\includegraphics*[width=5.5cm]{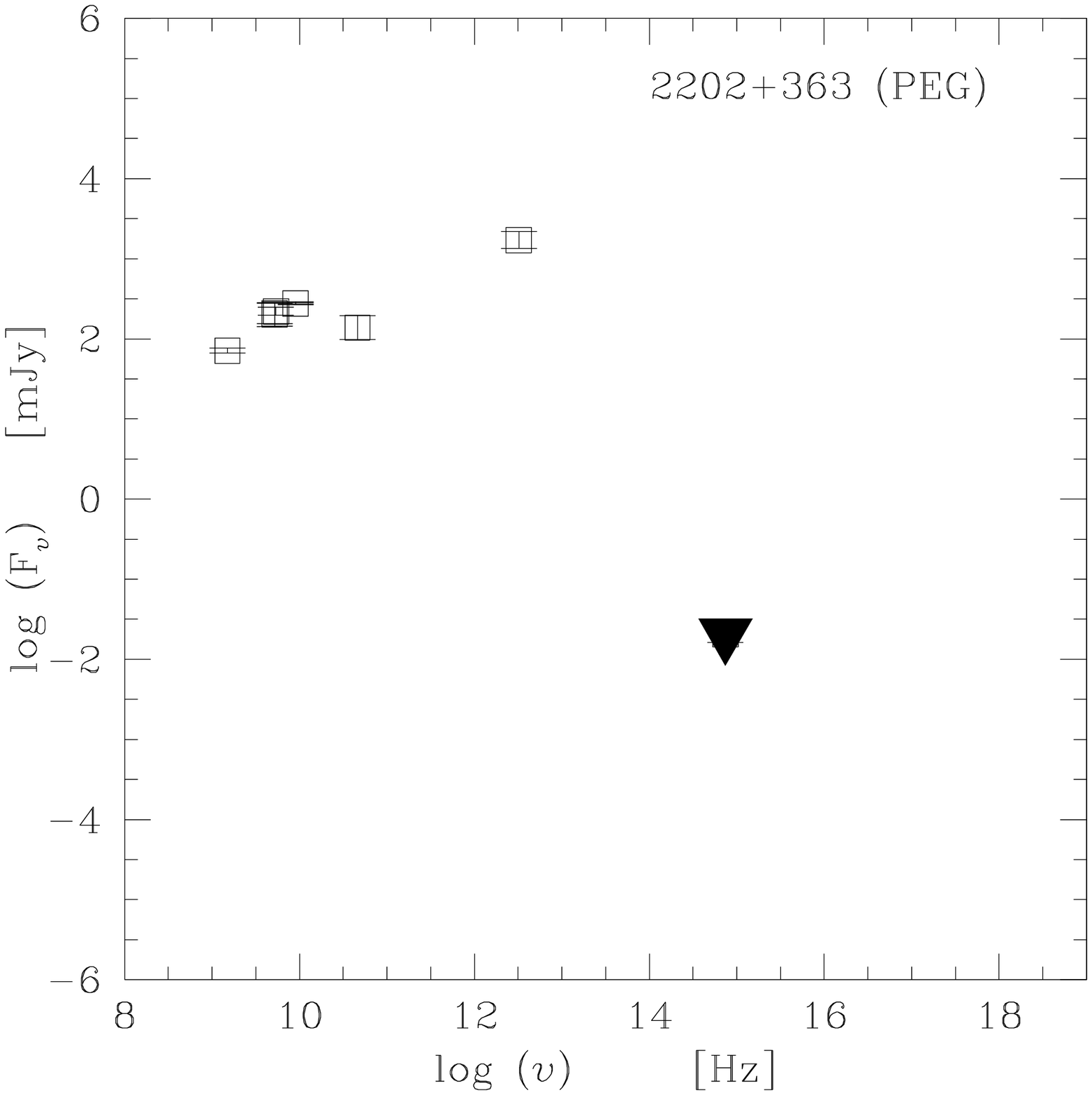}

\includegraphics[width=5.5cm]{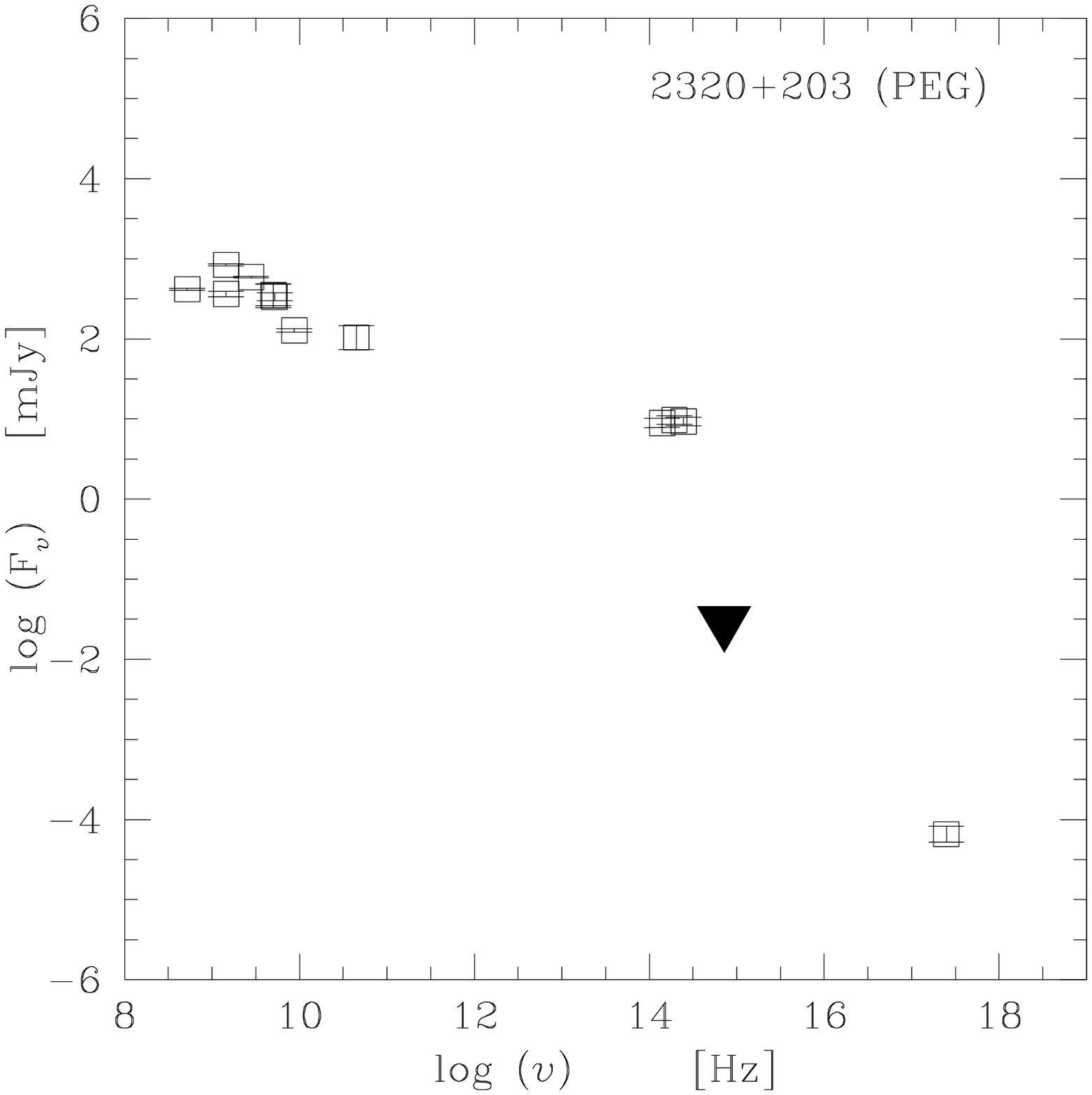}
\includegraphics*[width=5.5cm]{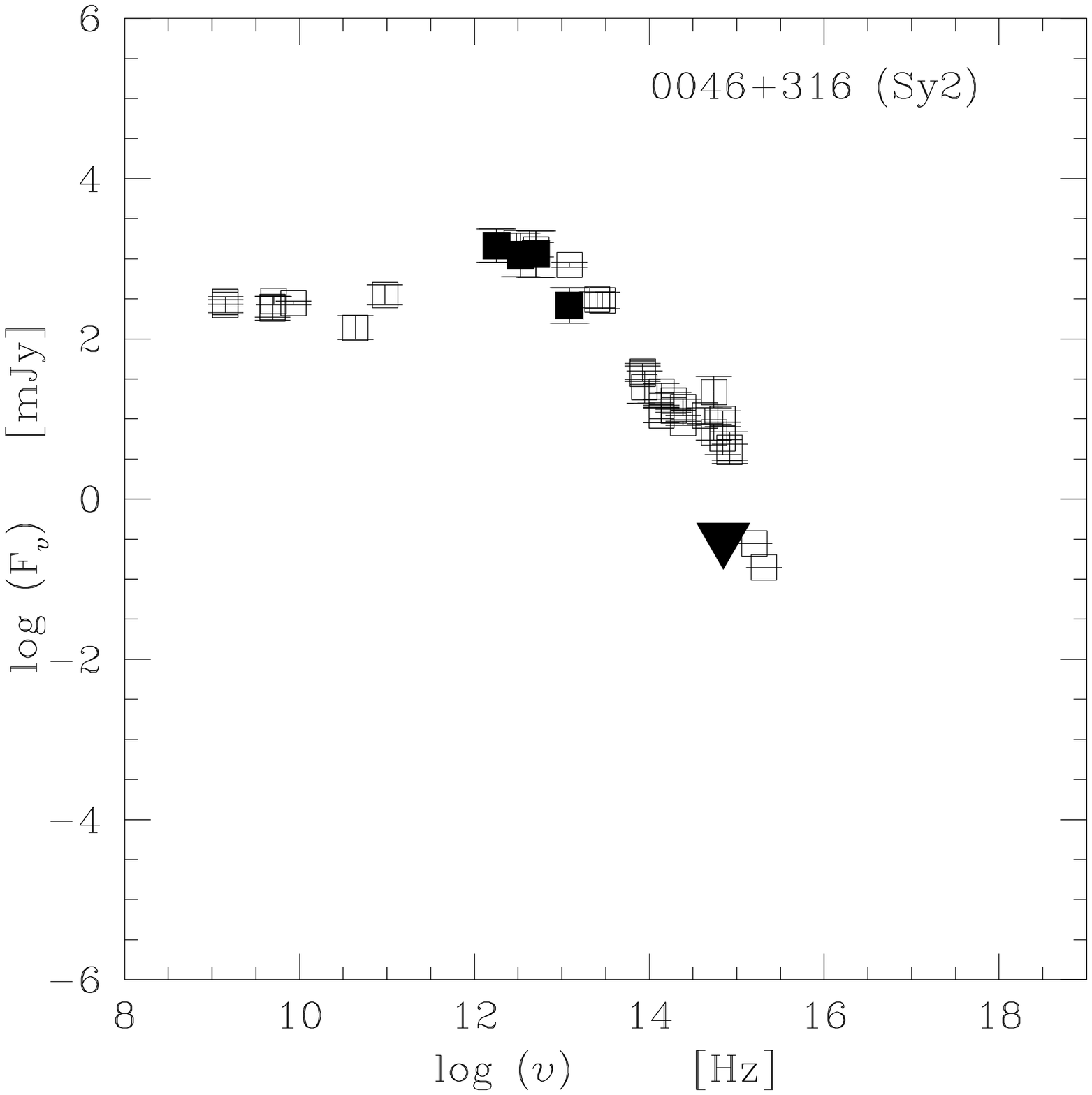}
\includegraphics*[width=5.5cm]{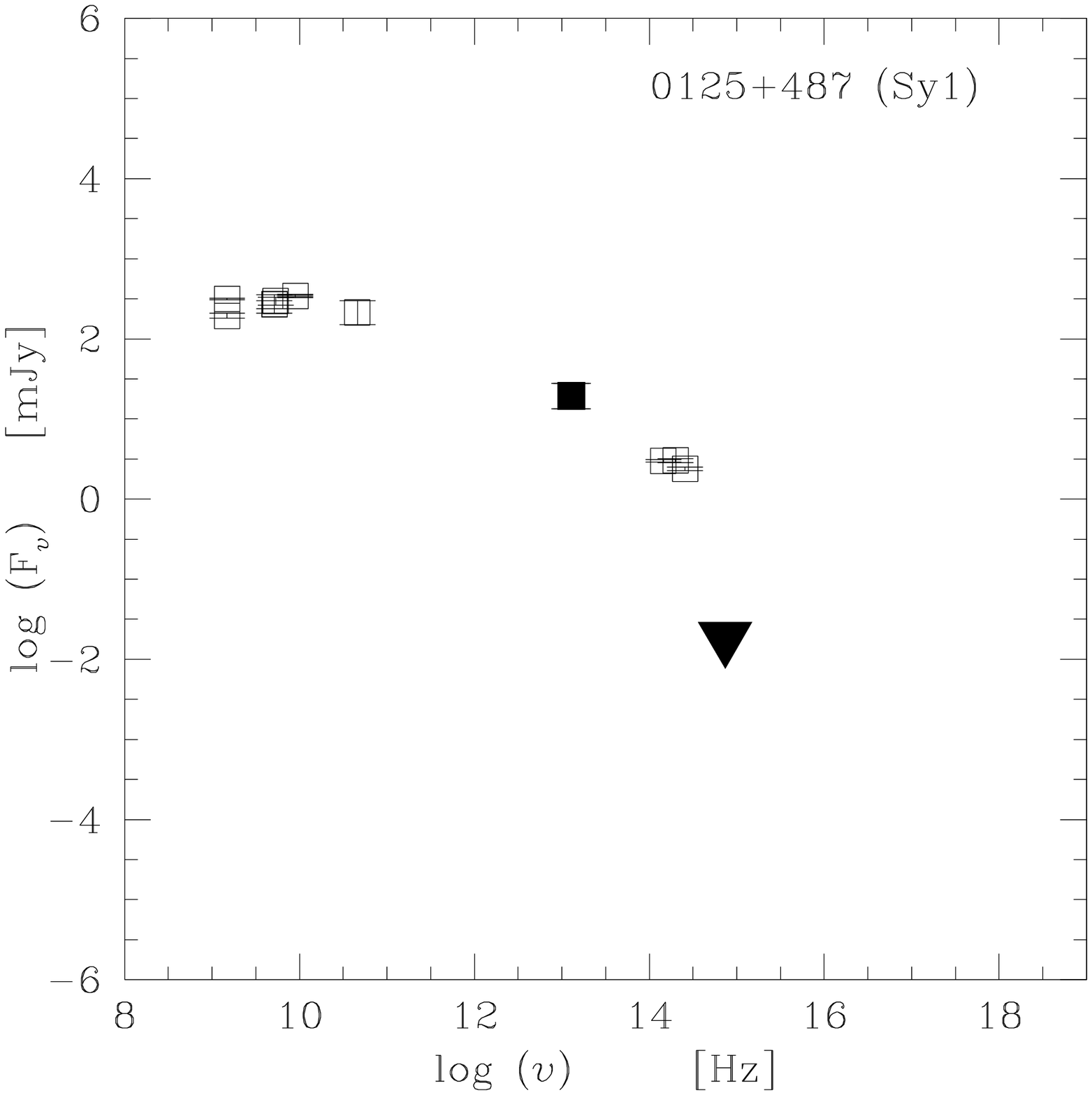}

\end{figure*}

\begin{figure*}

\includegraphics[width=5.5cm]{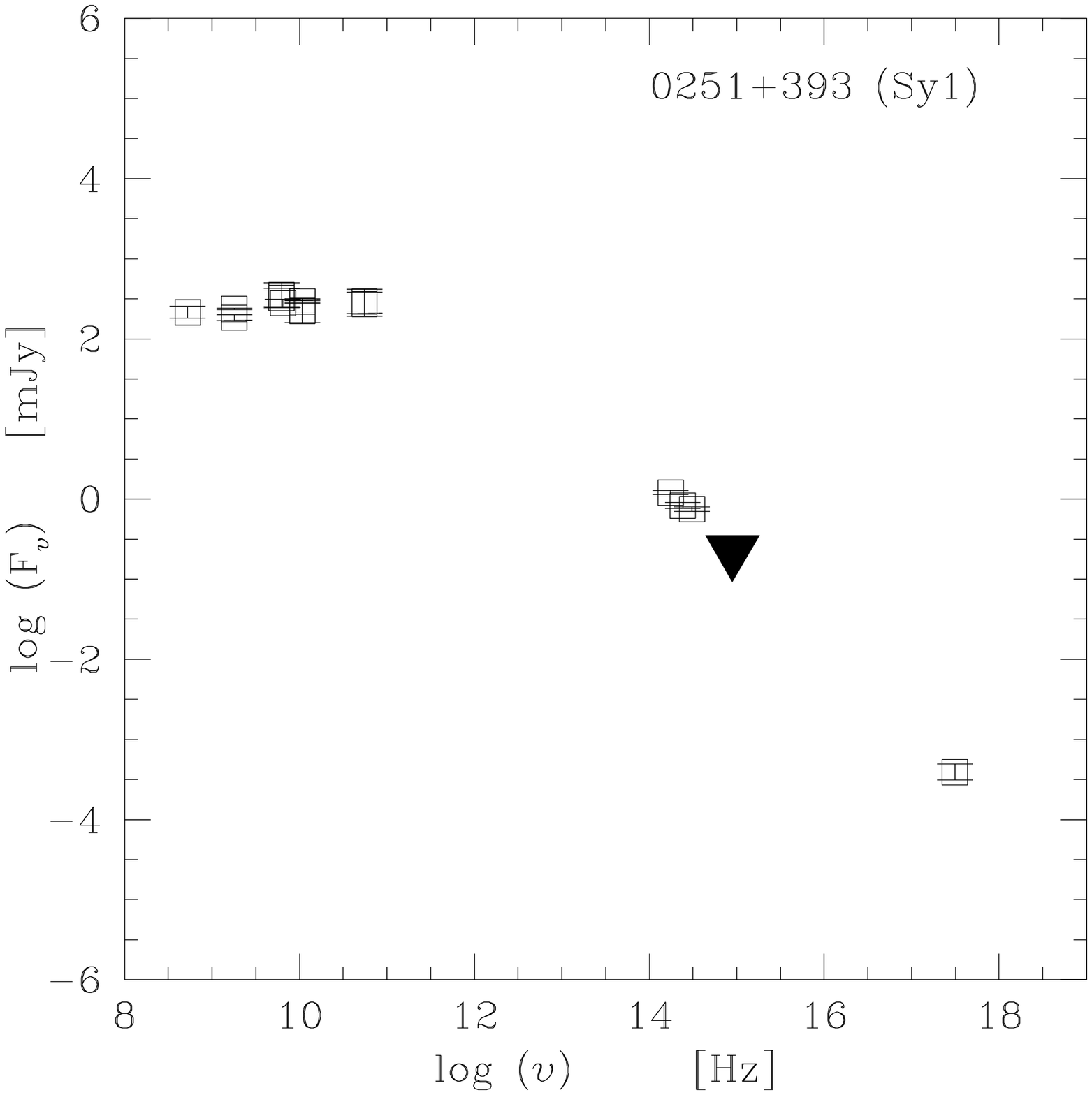}
\includegraphics*[width=5.5cm]{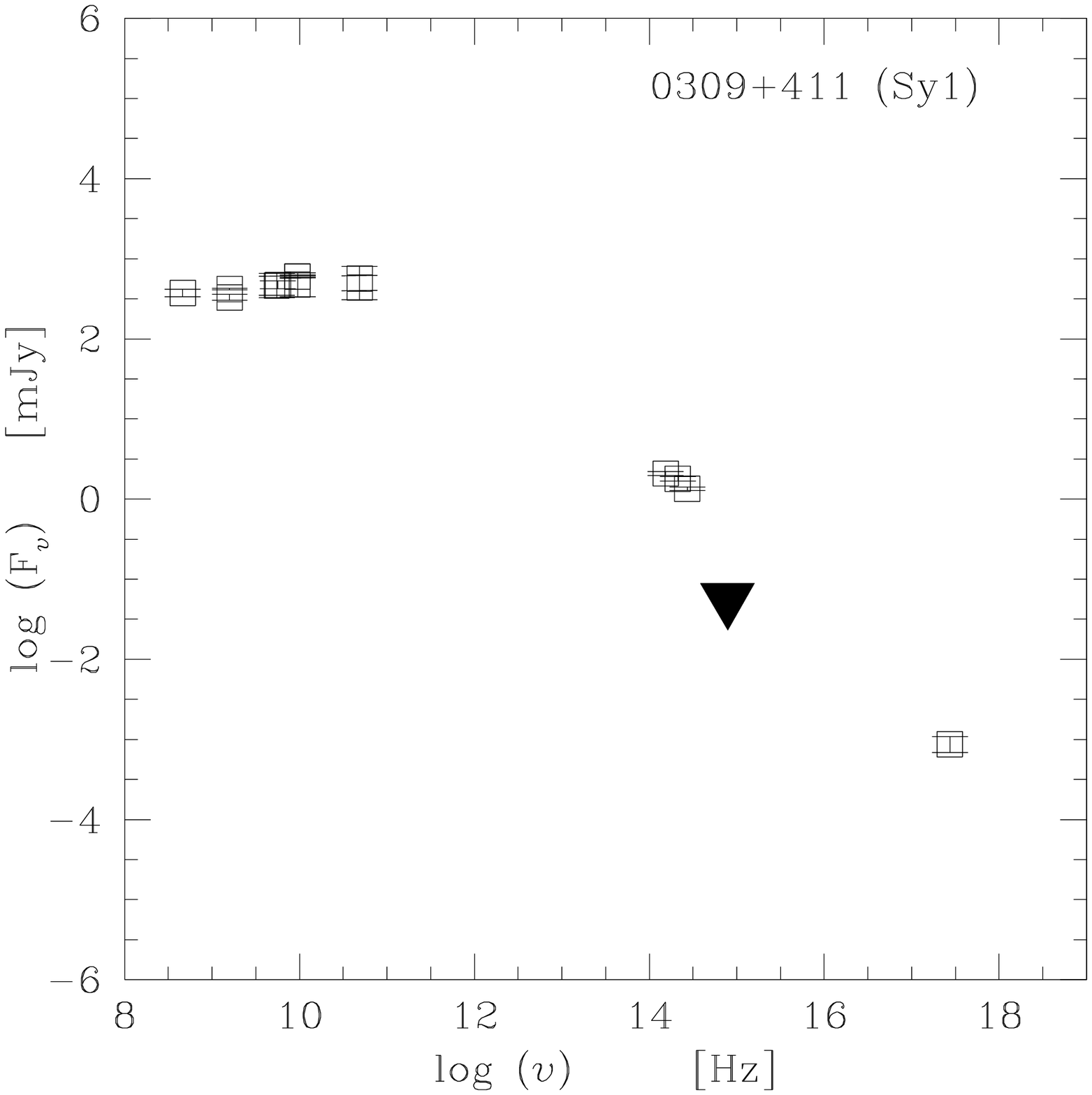}
\includegraphics*[width=5.5cm]{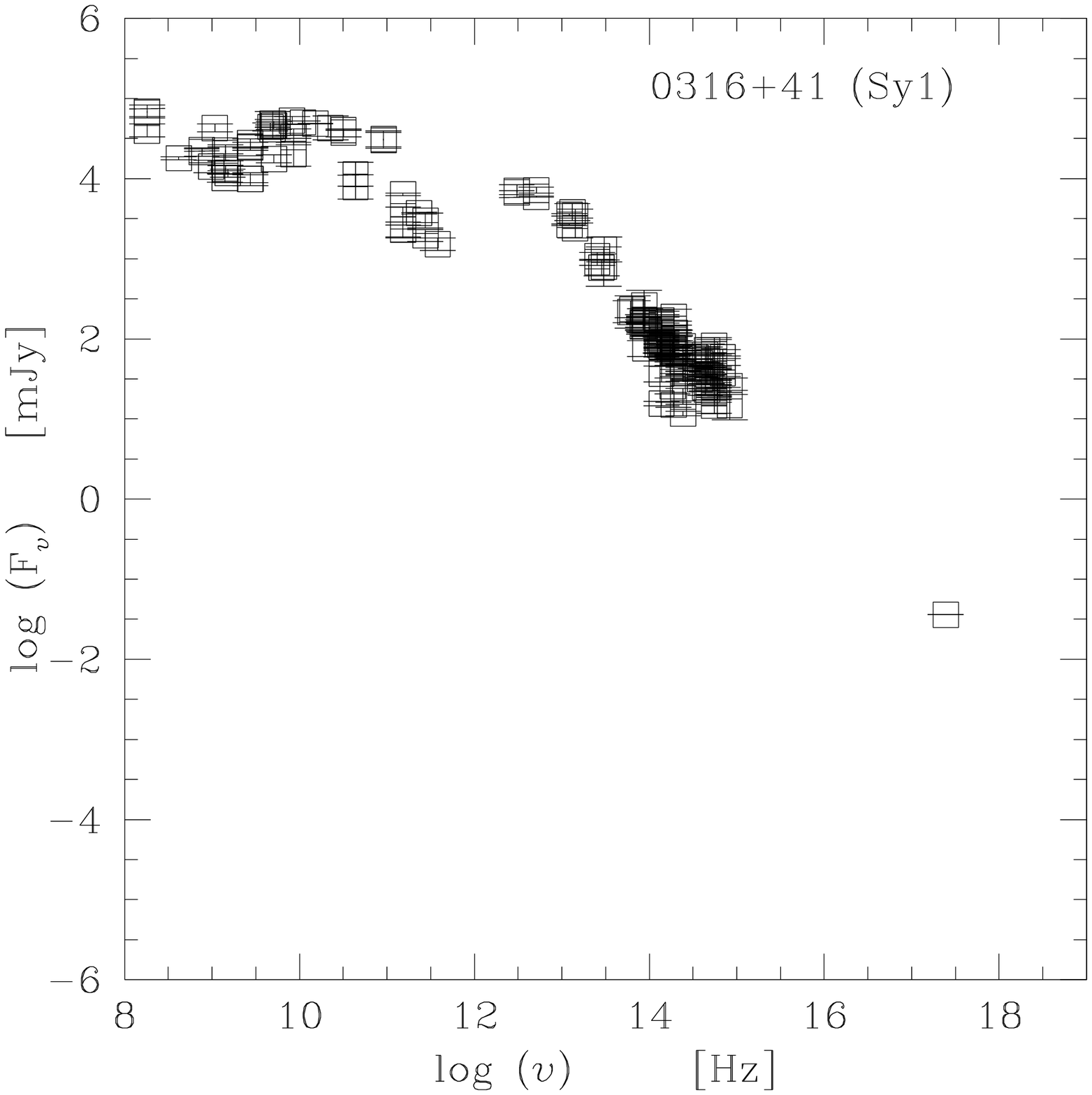}

\includegraphics[width=5.5cm]{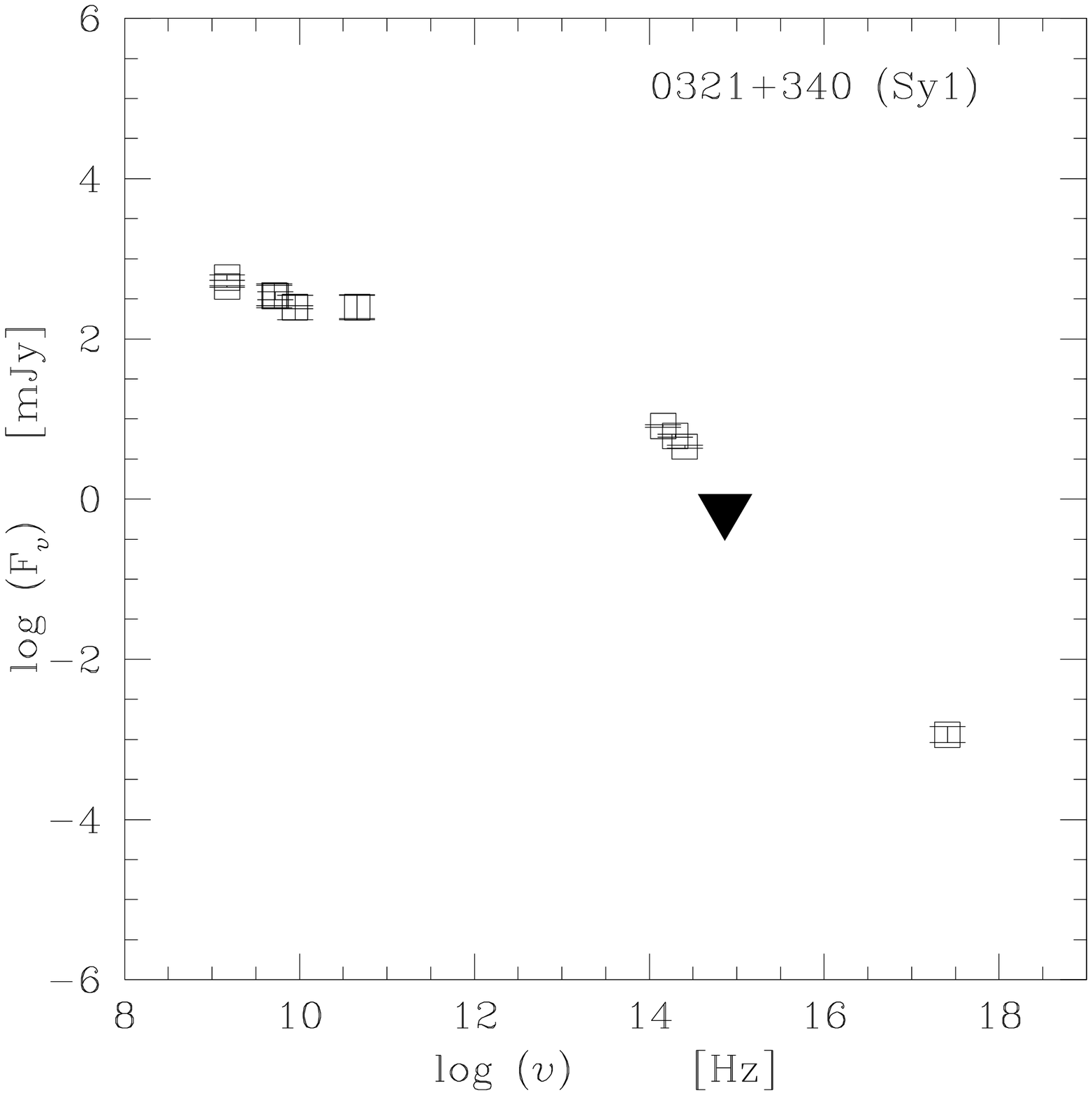}
\includegraphics*[width=5.5cm]{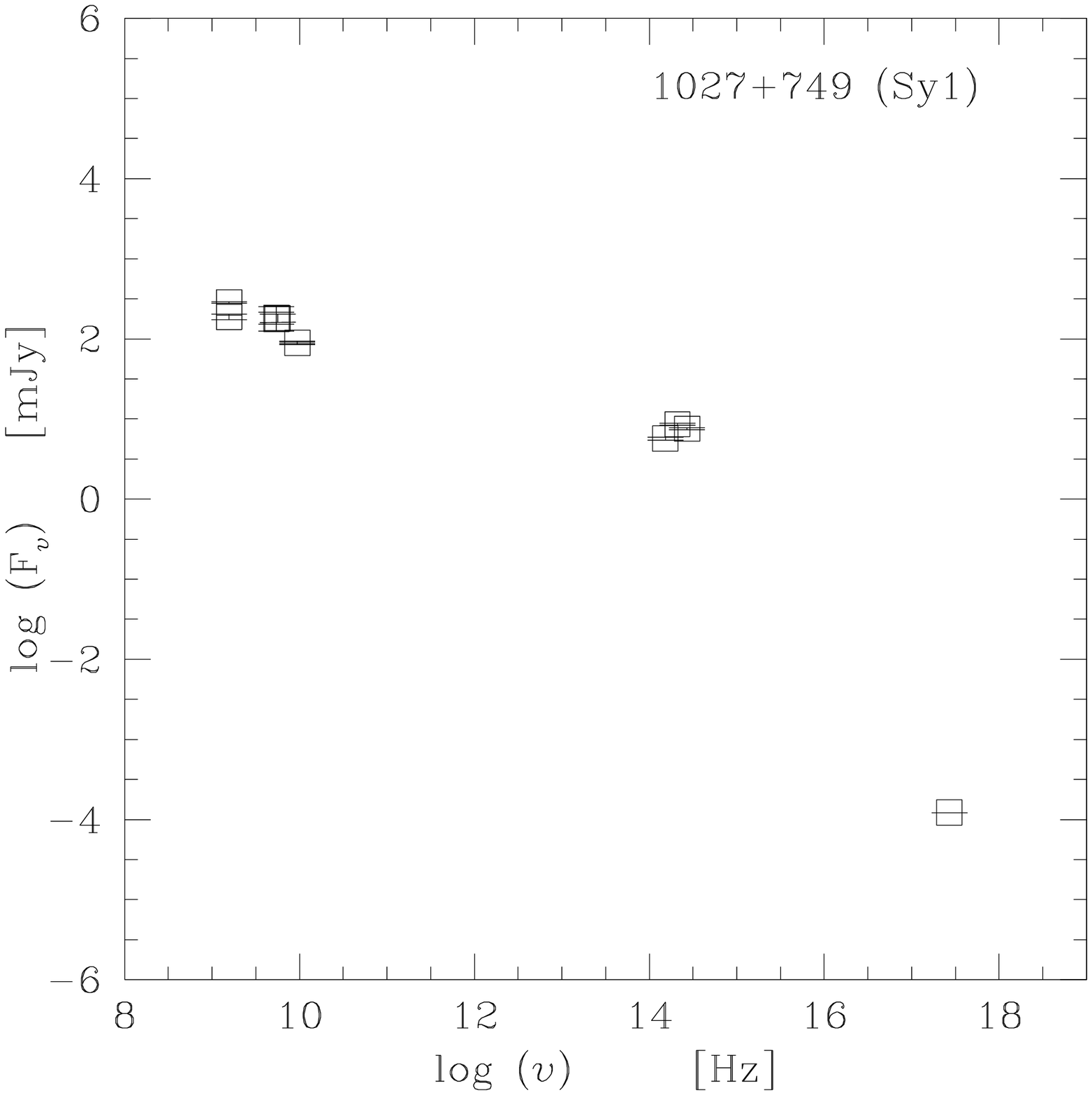}
\includegraphics*[width=5.5cm]{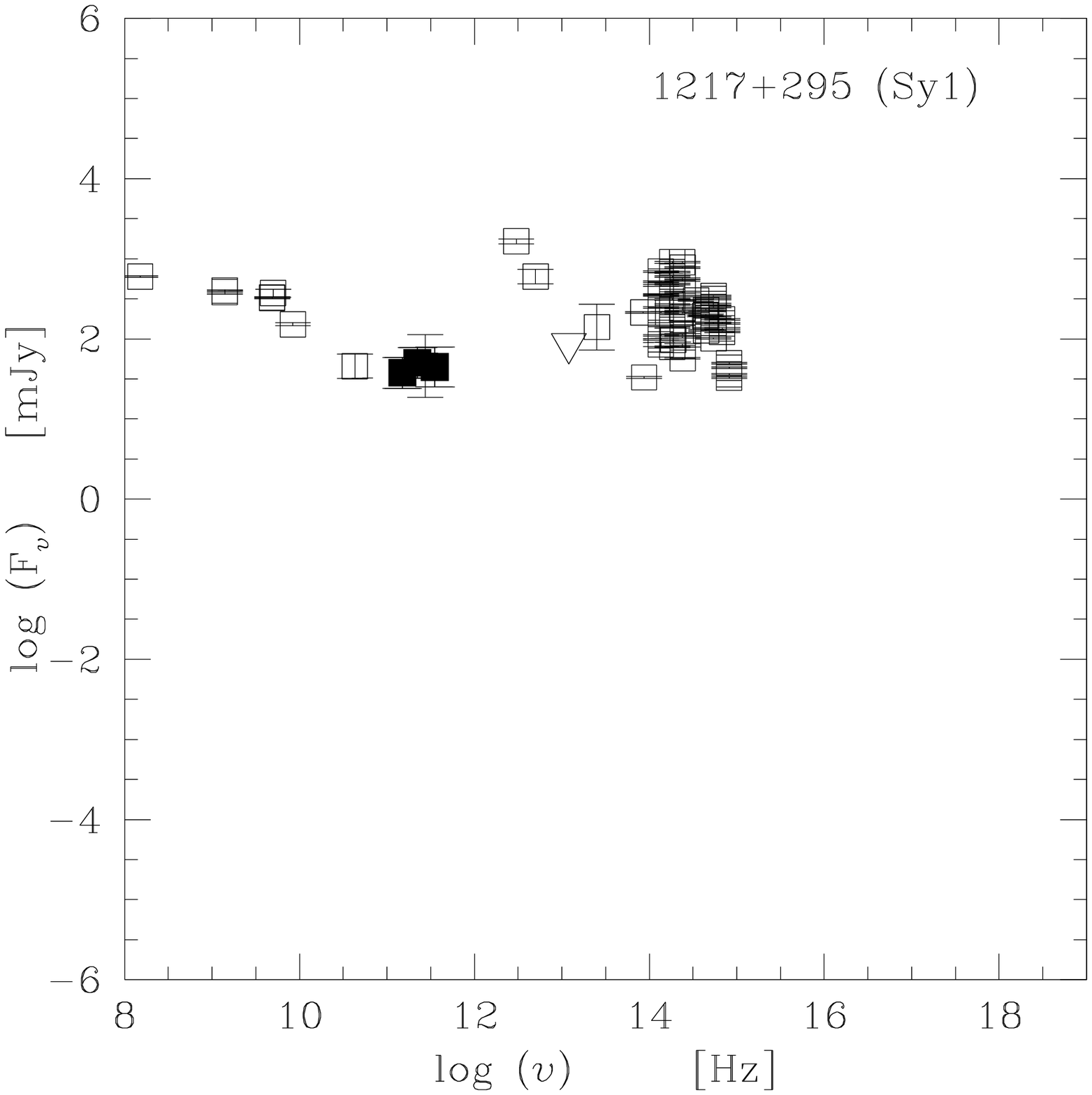}

\includegraphics[width=5.5cm]{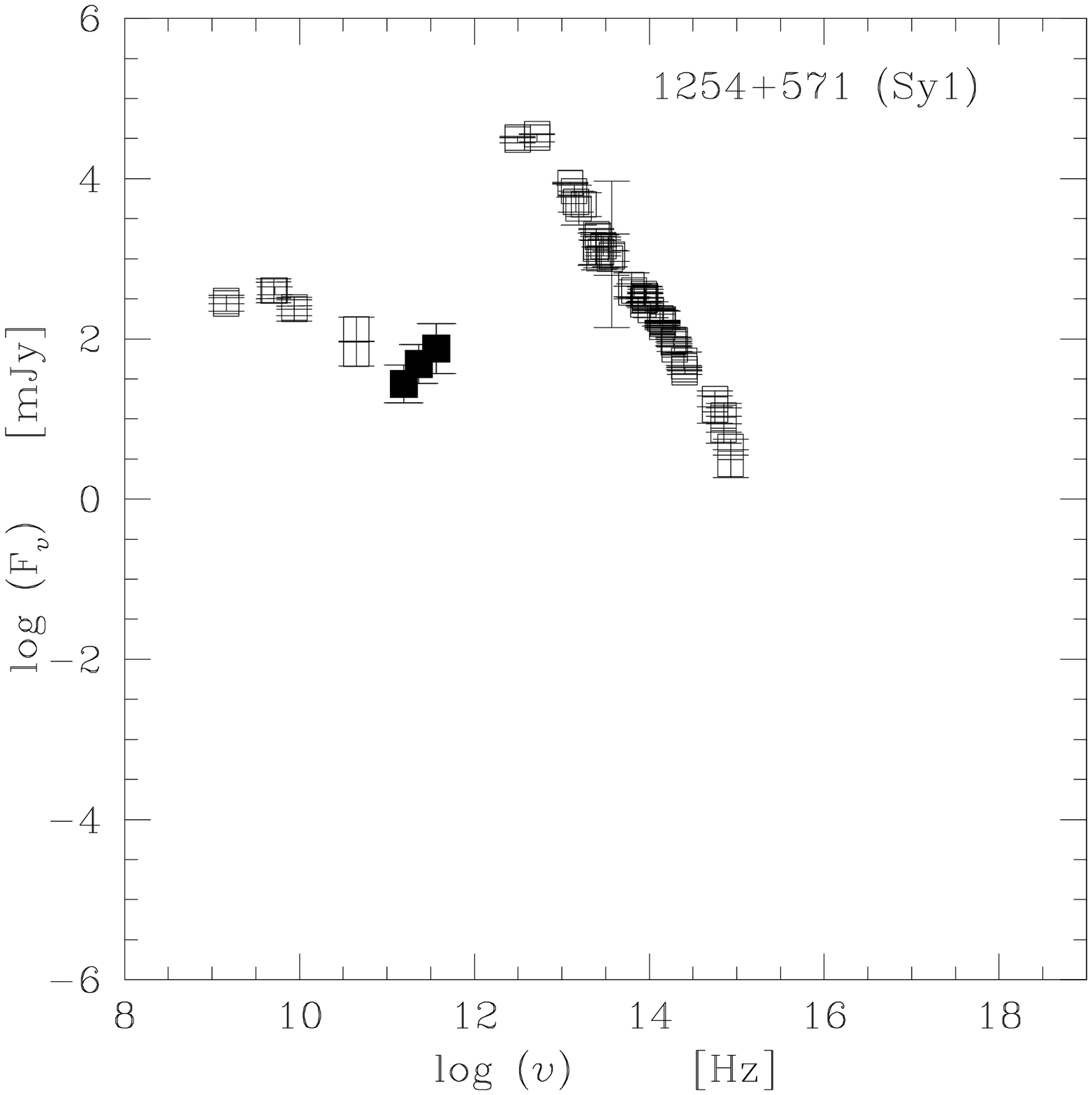}
\includegraphics*[width=5.5cm]{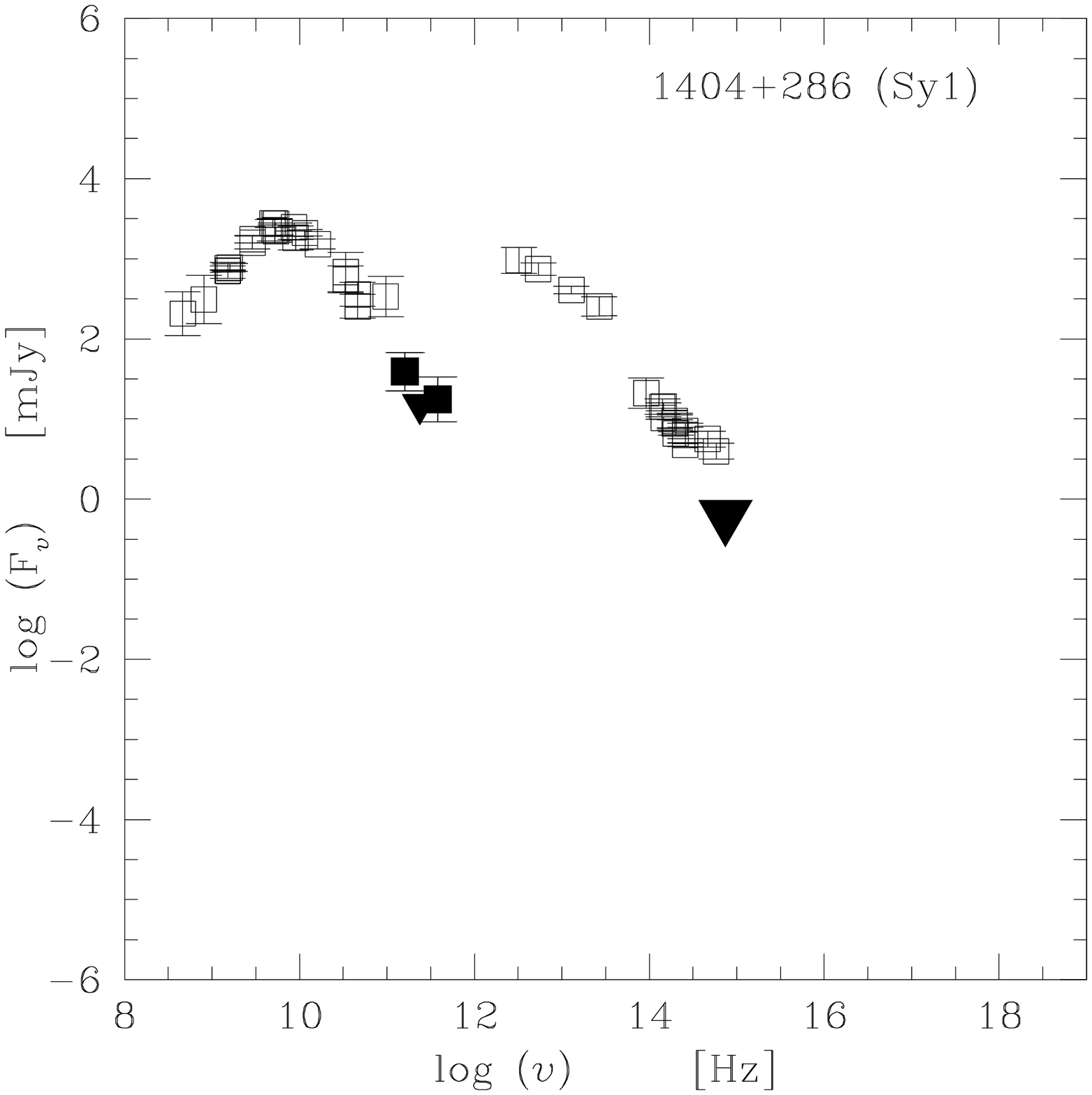}
\includegraphics*[width=5.5cm]{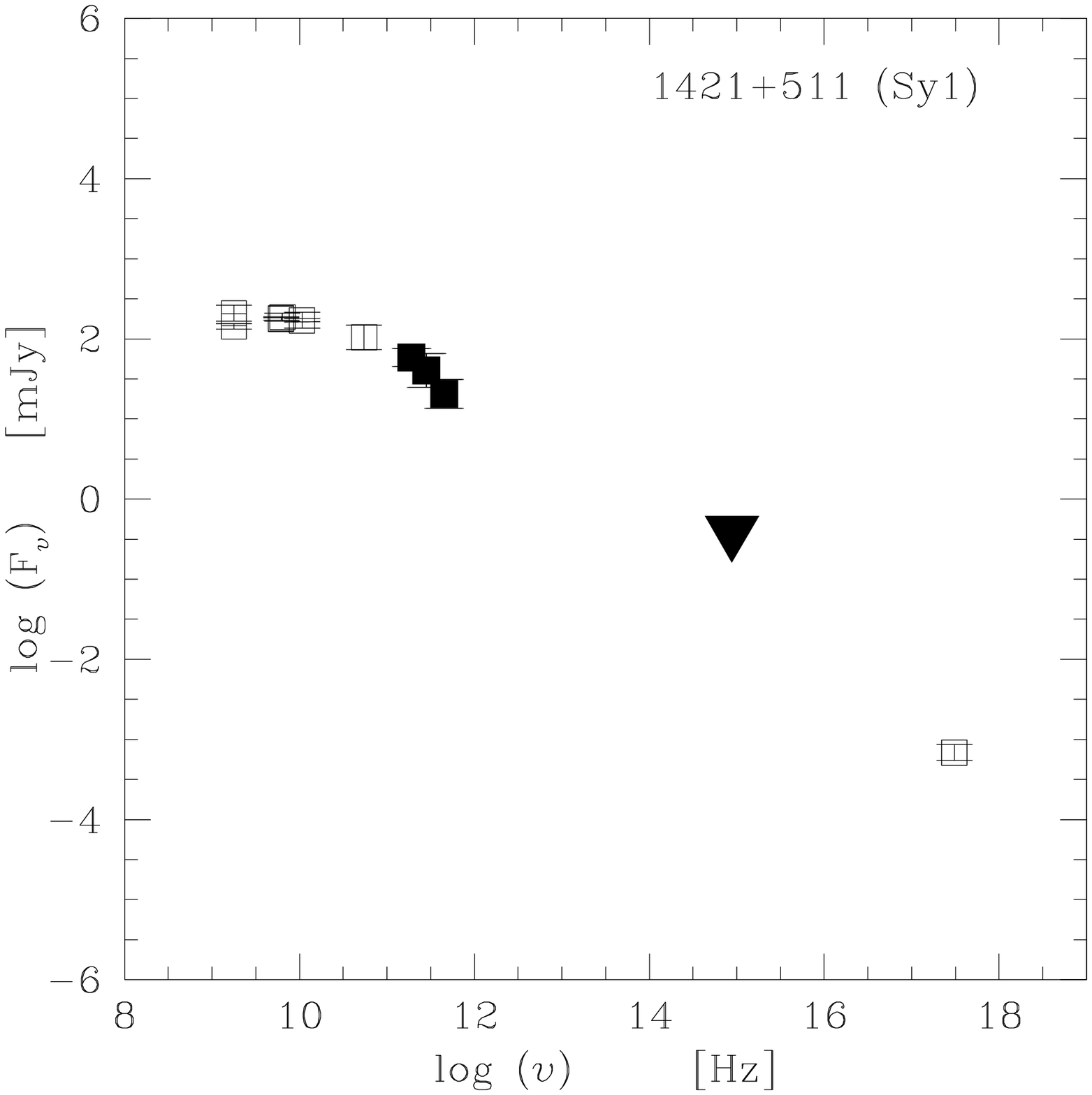}

\includegraphics[width=5.5cm]{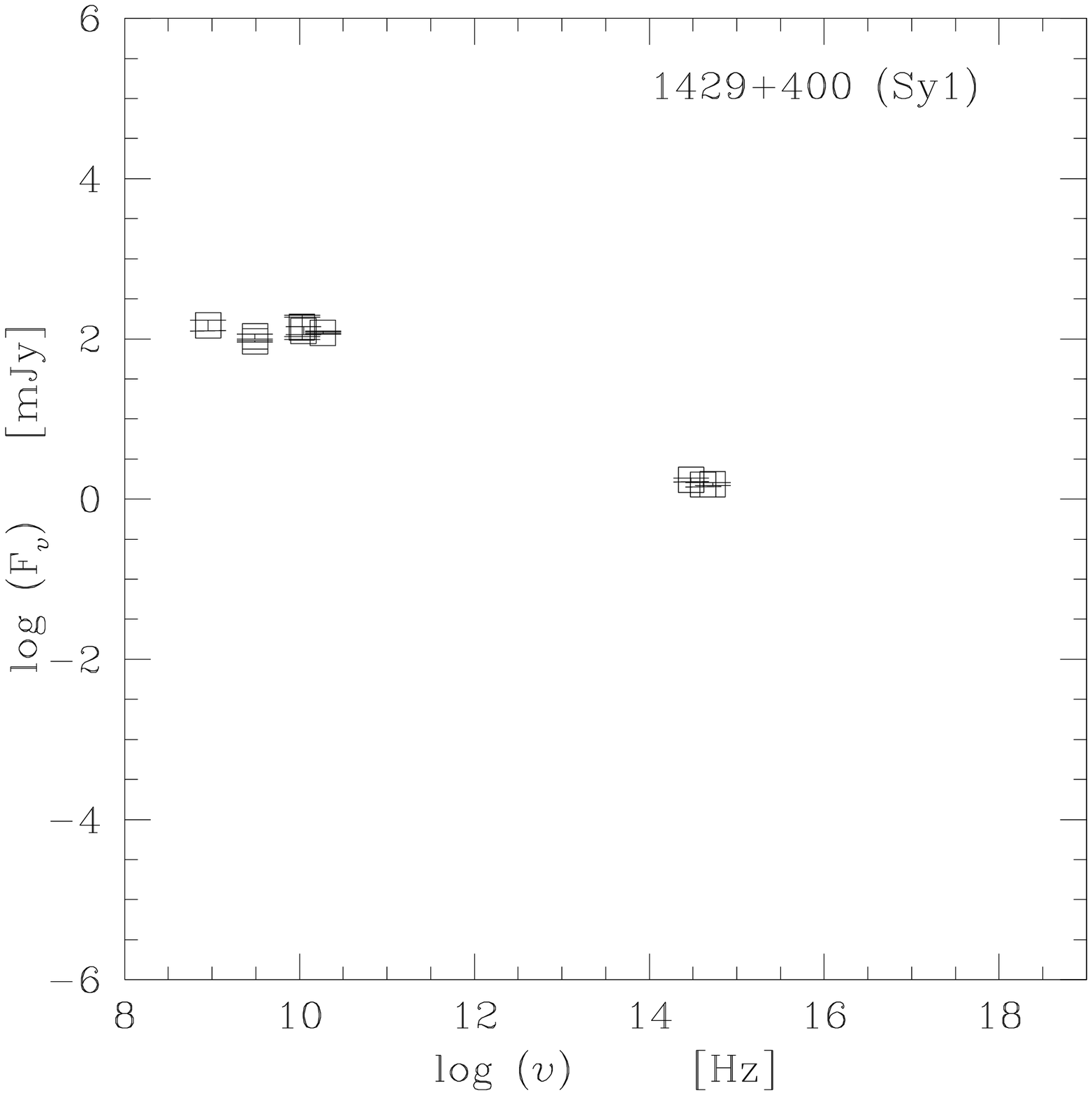}
\includegraphics*[width=5.5cm]{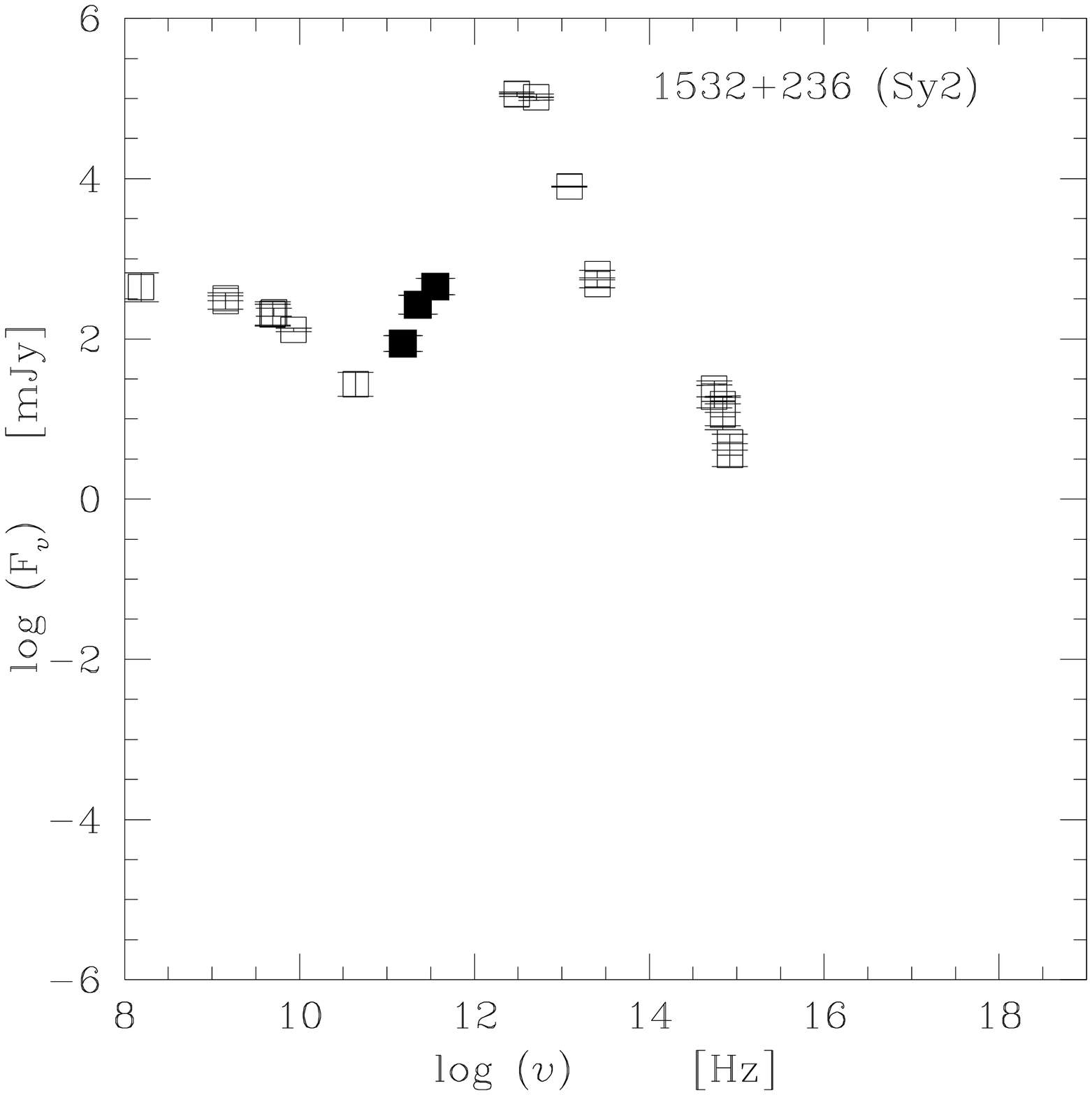}
\includegraphics*[width=5.5cm]{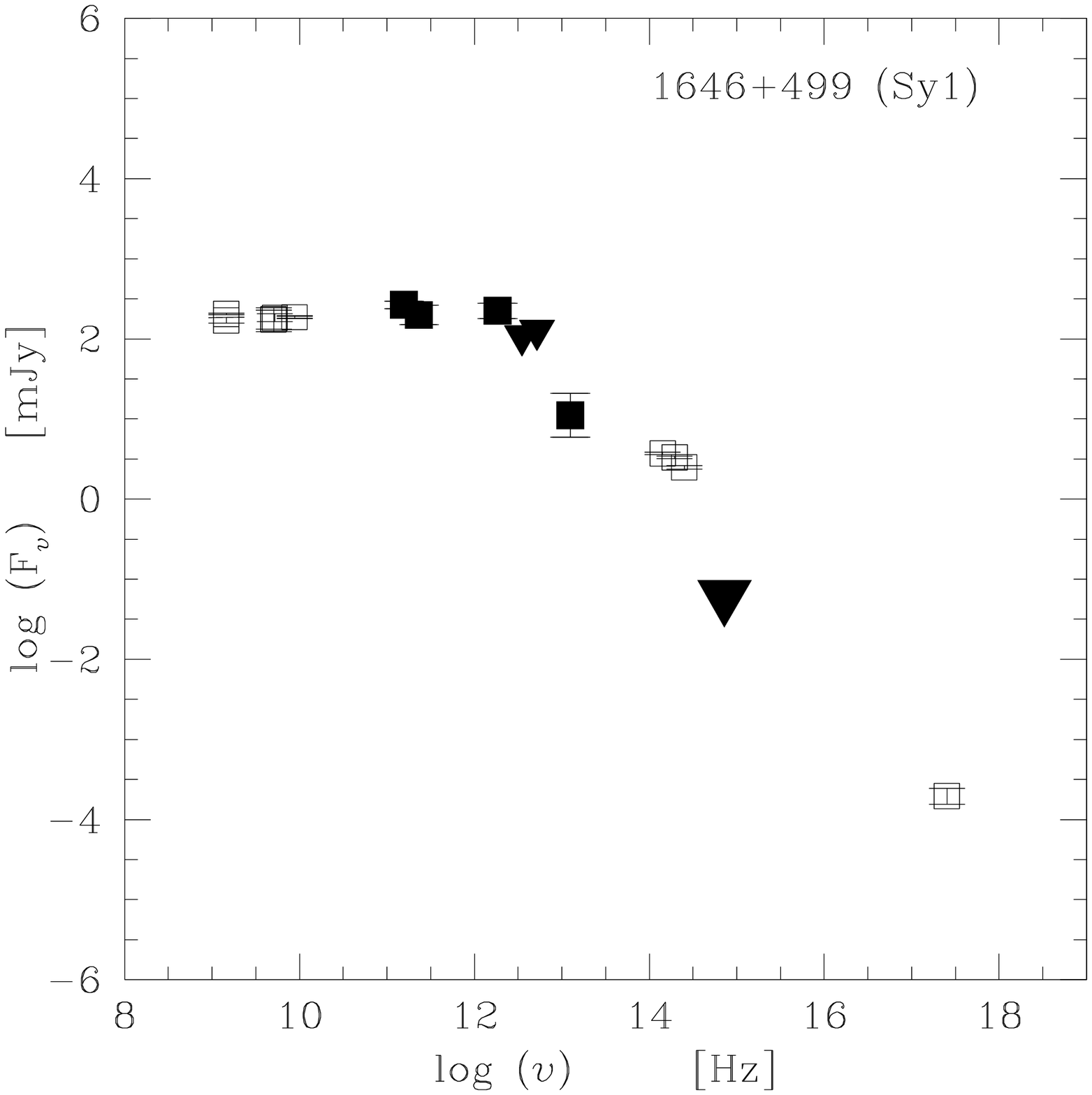}

\end{figure*}

\begin{figure*}

\includegraphics[width=5.5cm]{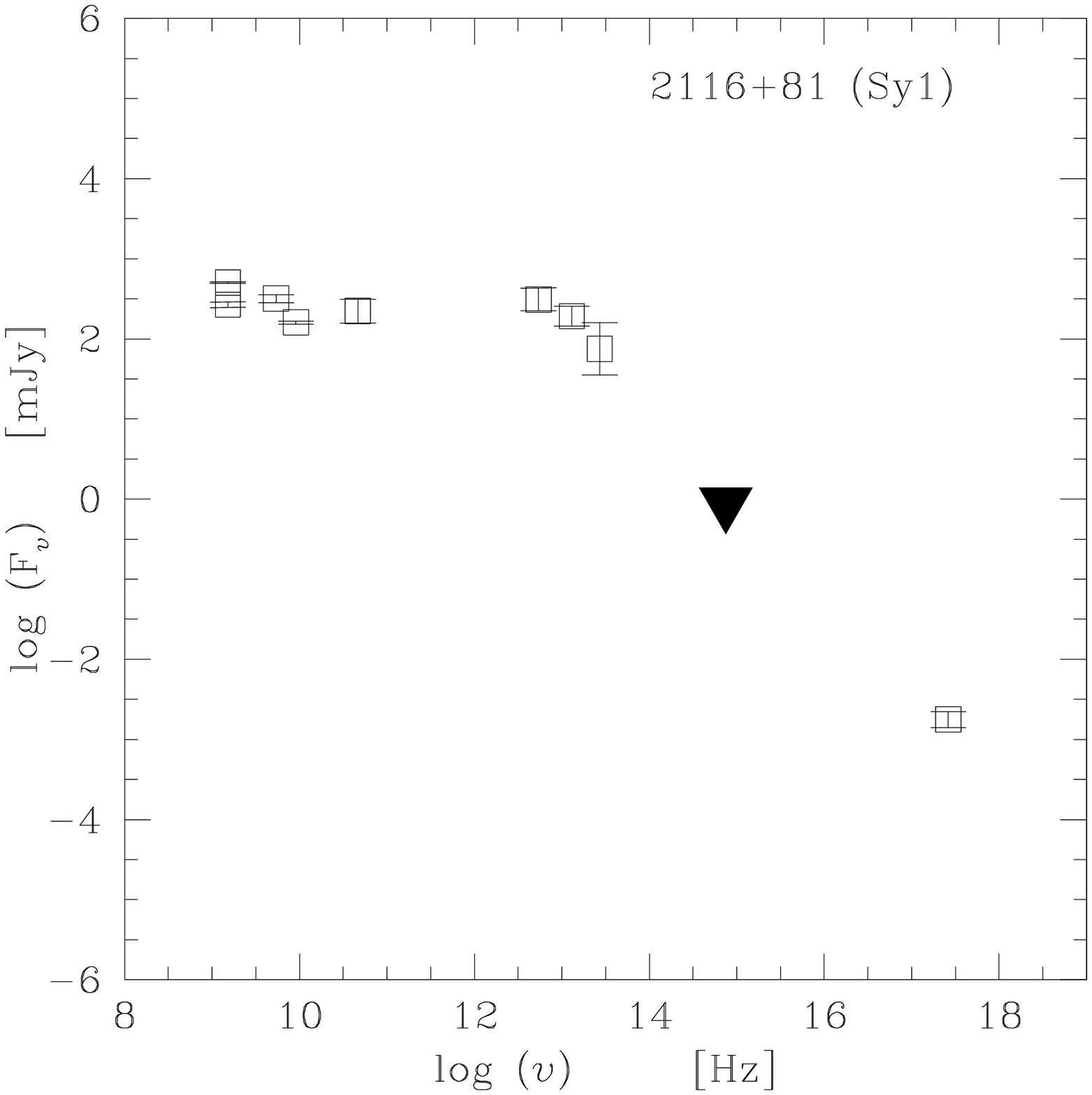}
\includegraphics*[width=5.5cm]{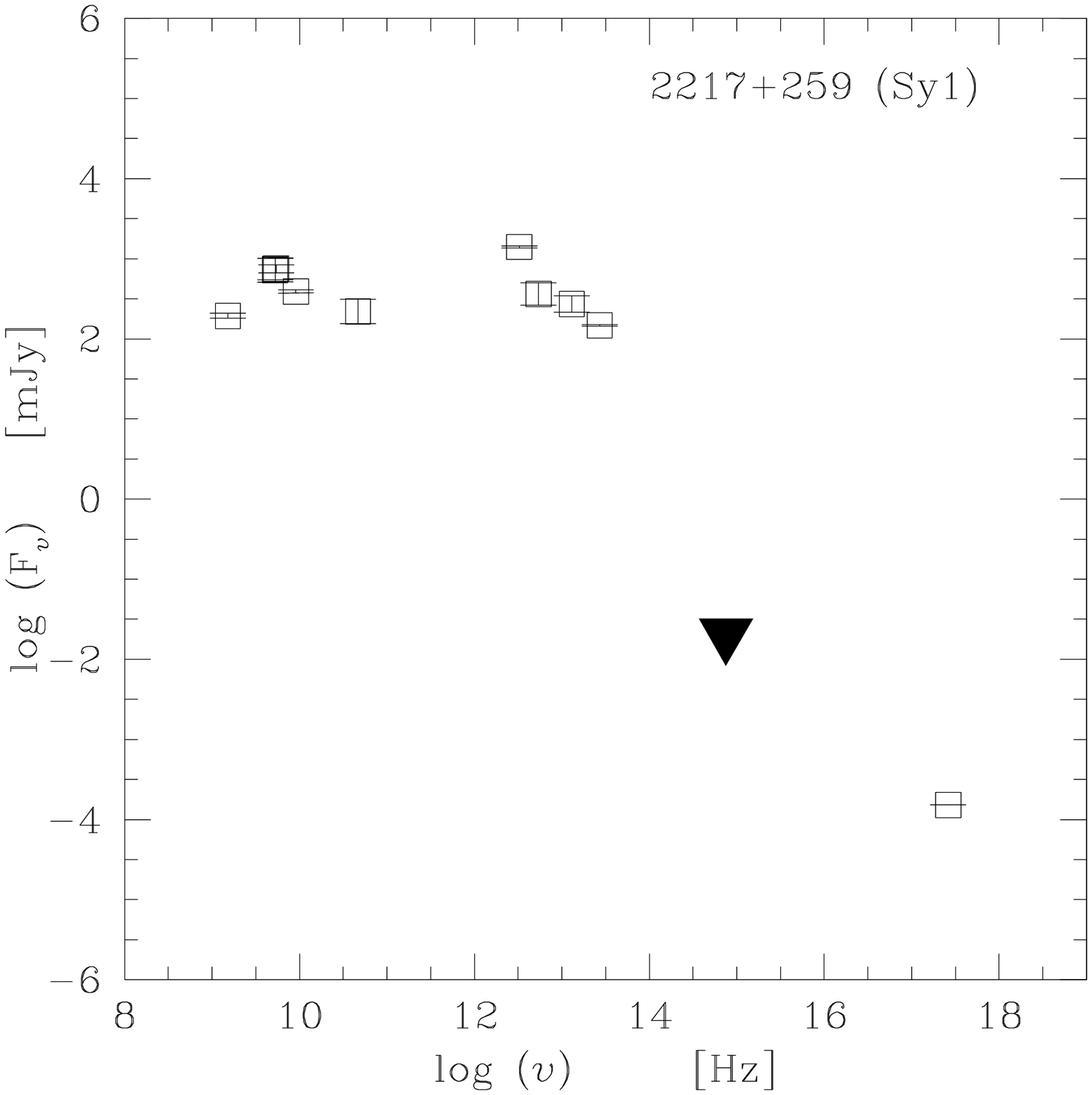}
\includegraphics*[width=5.5cm]{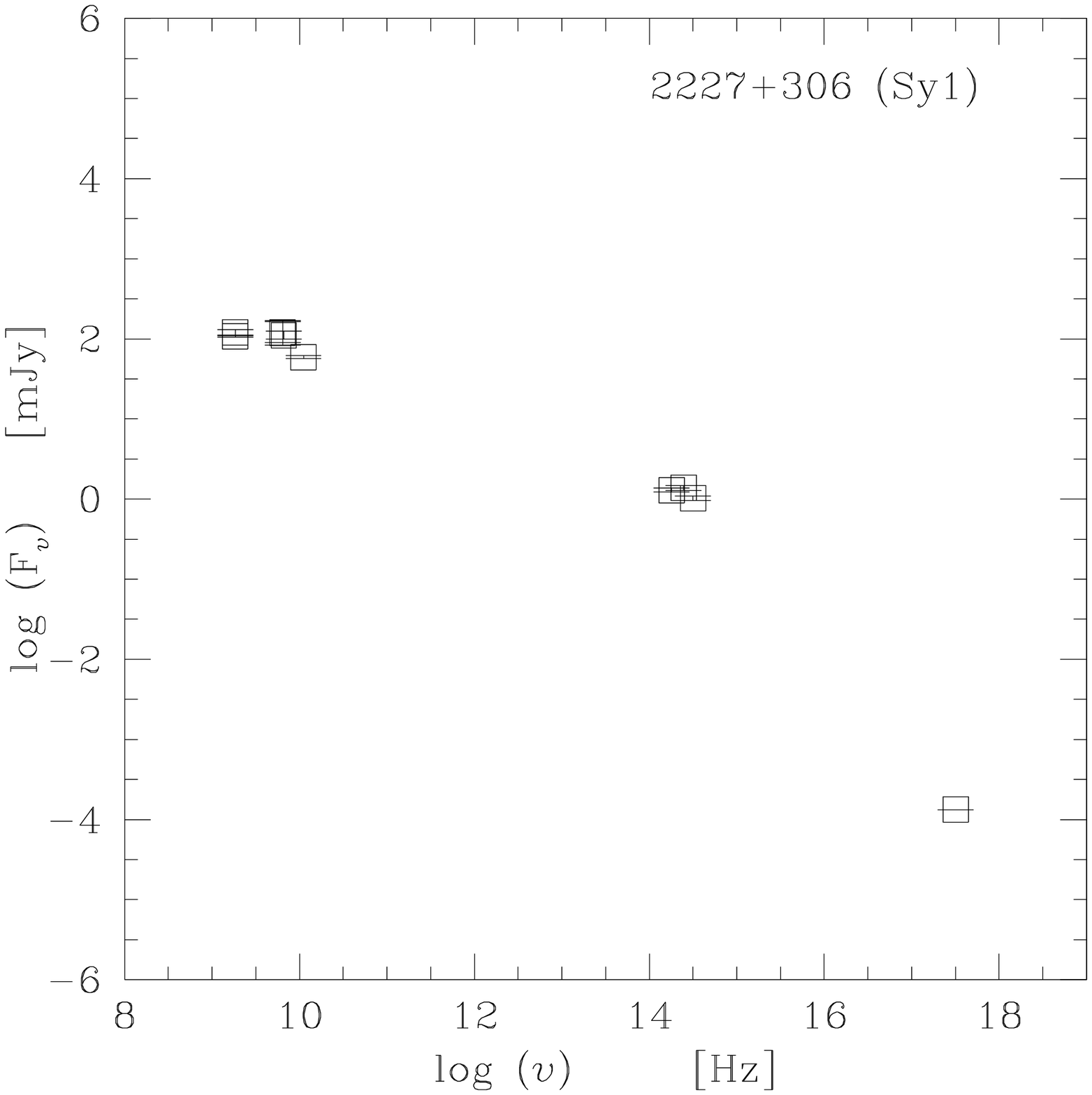}

\includegraphics[width=5.5cm]{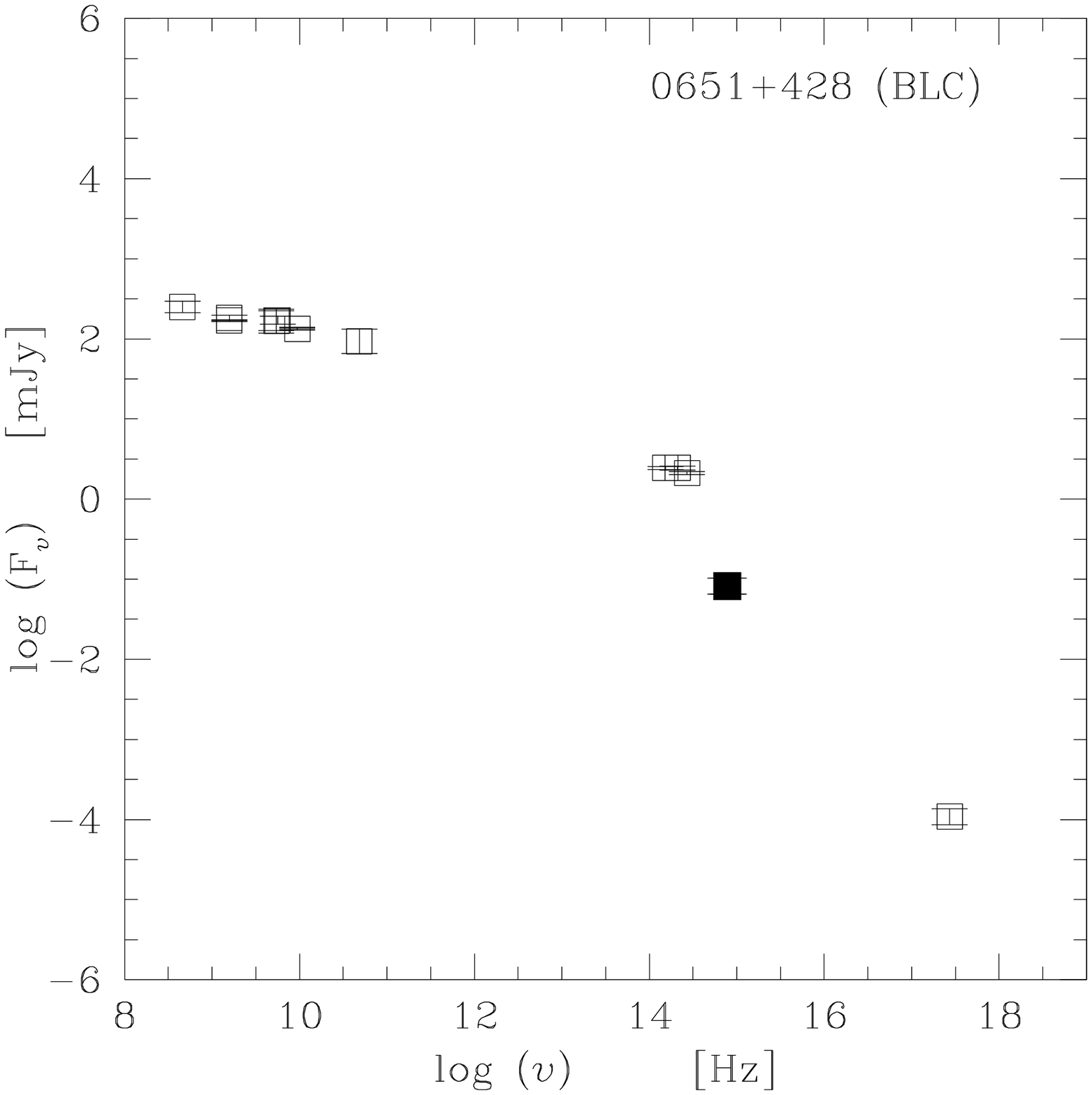}
\includegraphics*[width=5.5cm]{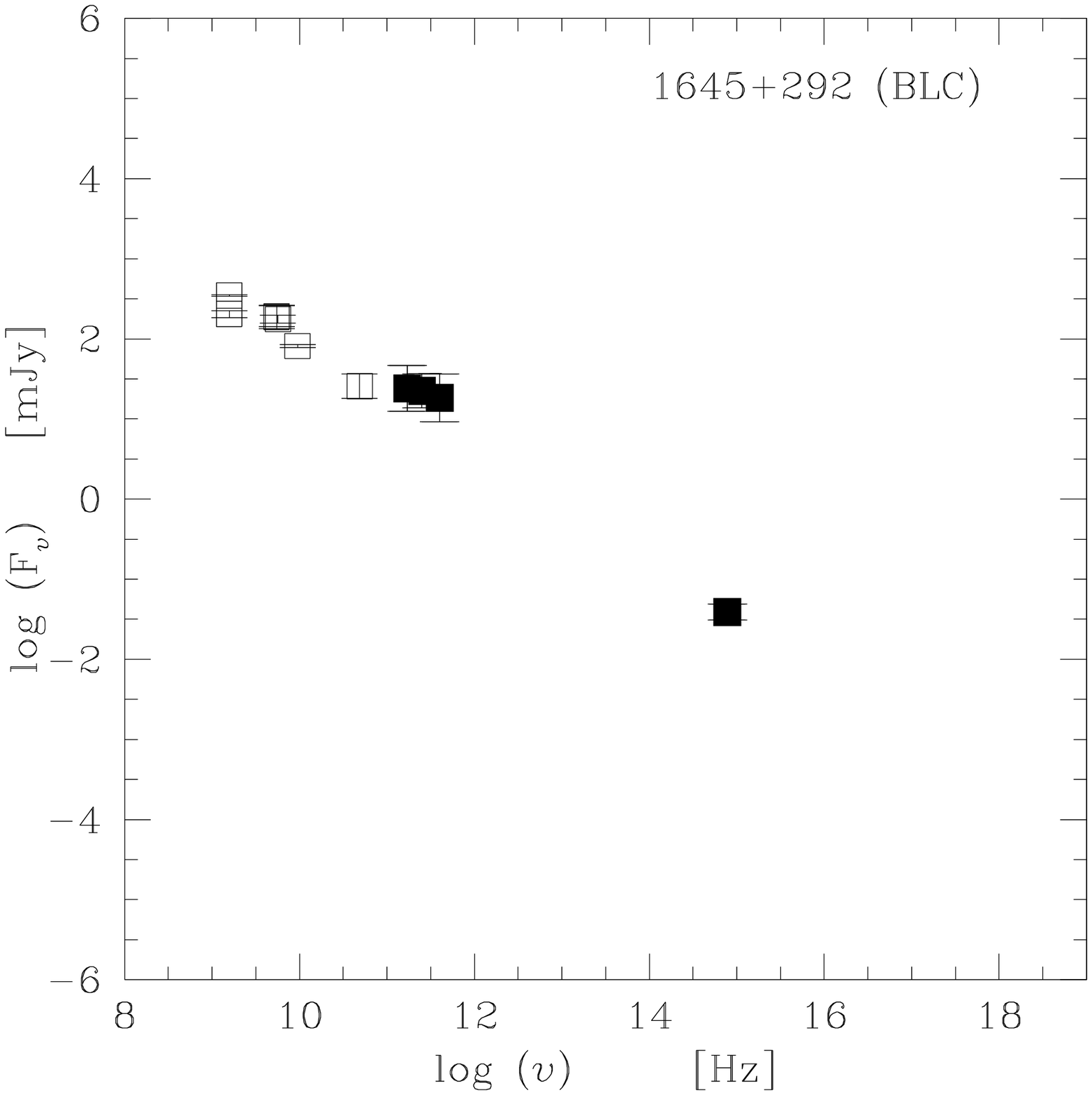}
\includegraphics*[width=5.5cm]{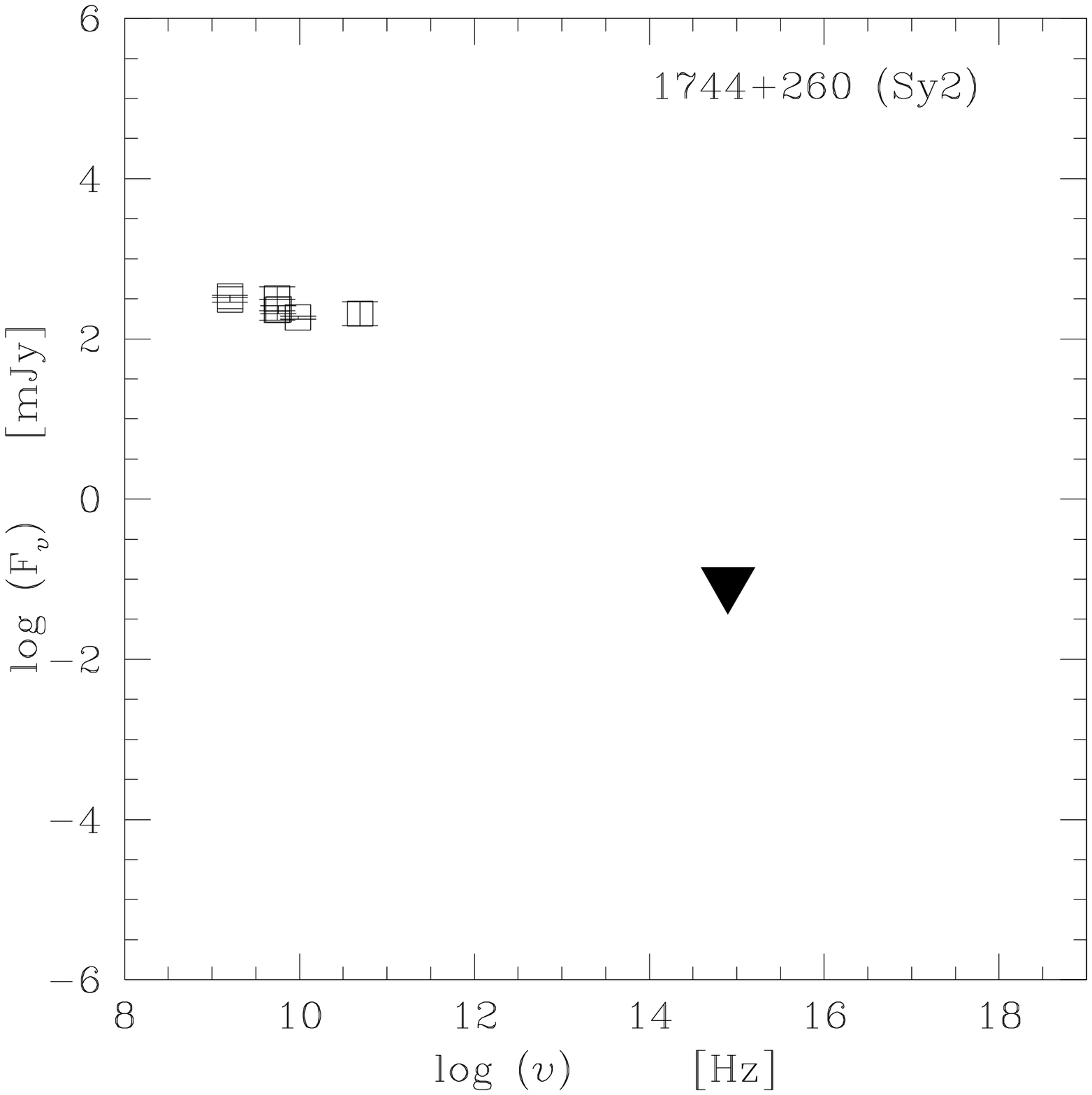}

\includegraphics[width=5.5cm]{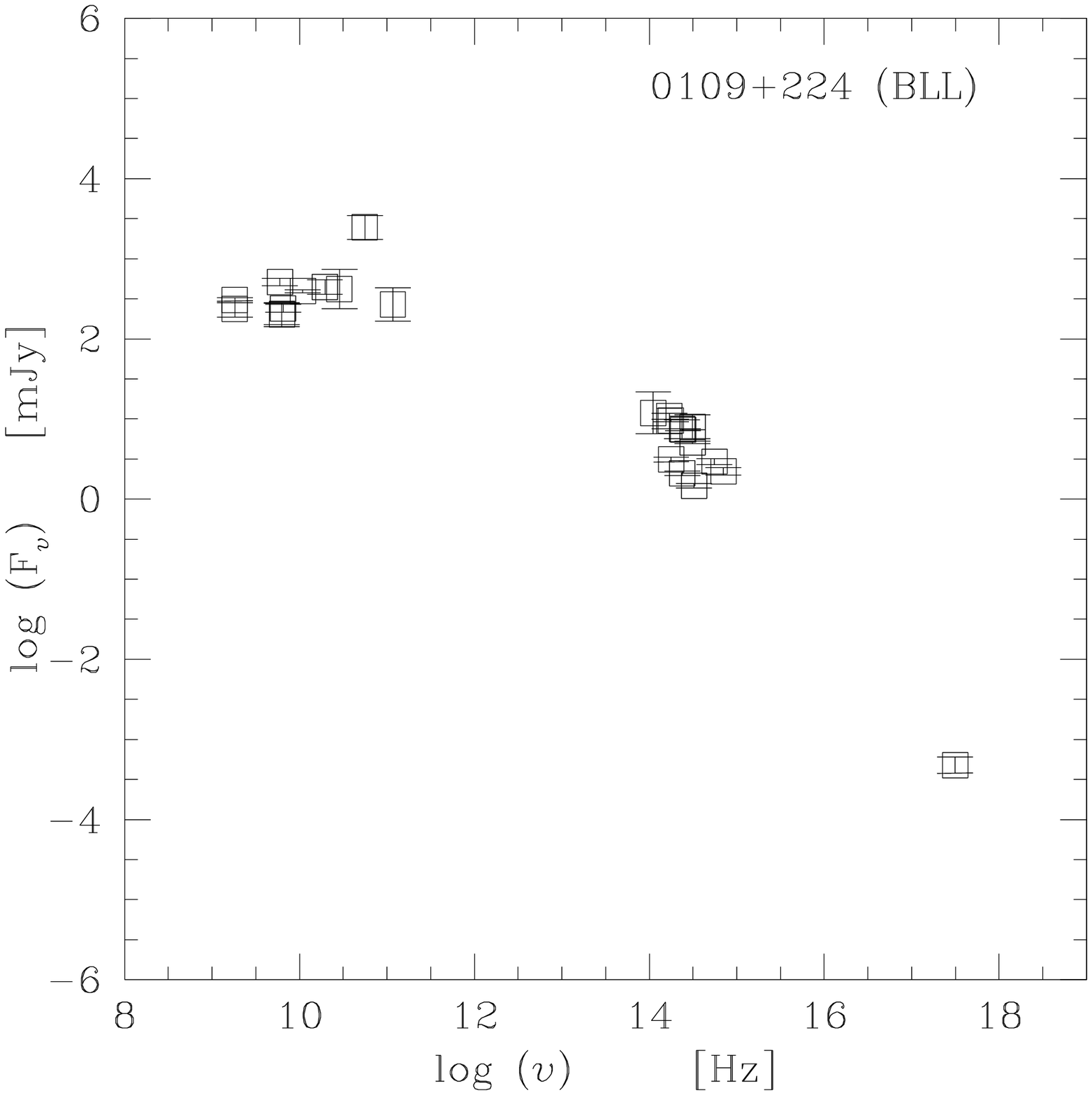}
\includegraphics*[width=5.5cm]{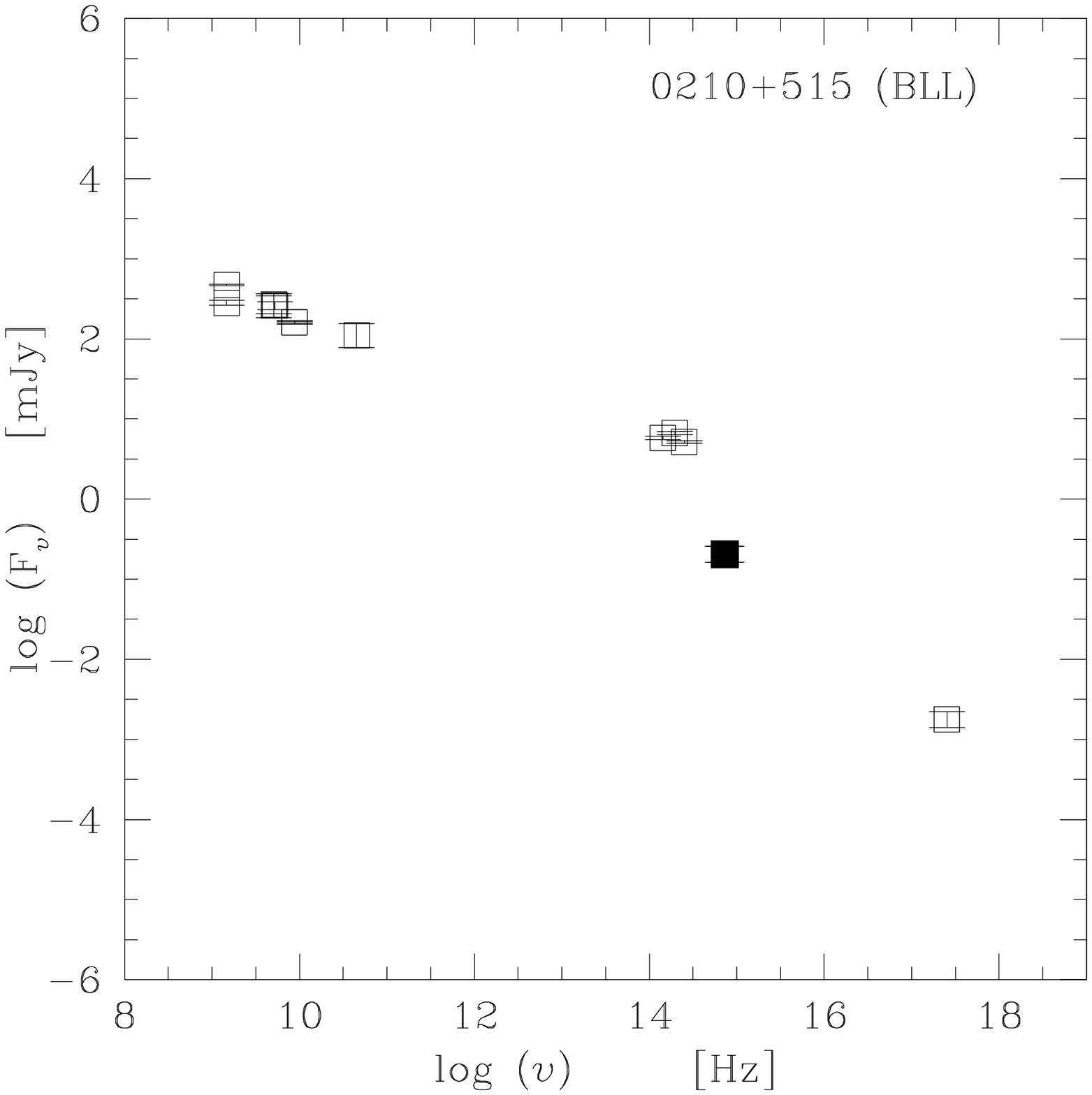}
\includegraphics*[width=5.5cm]{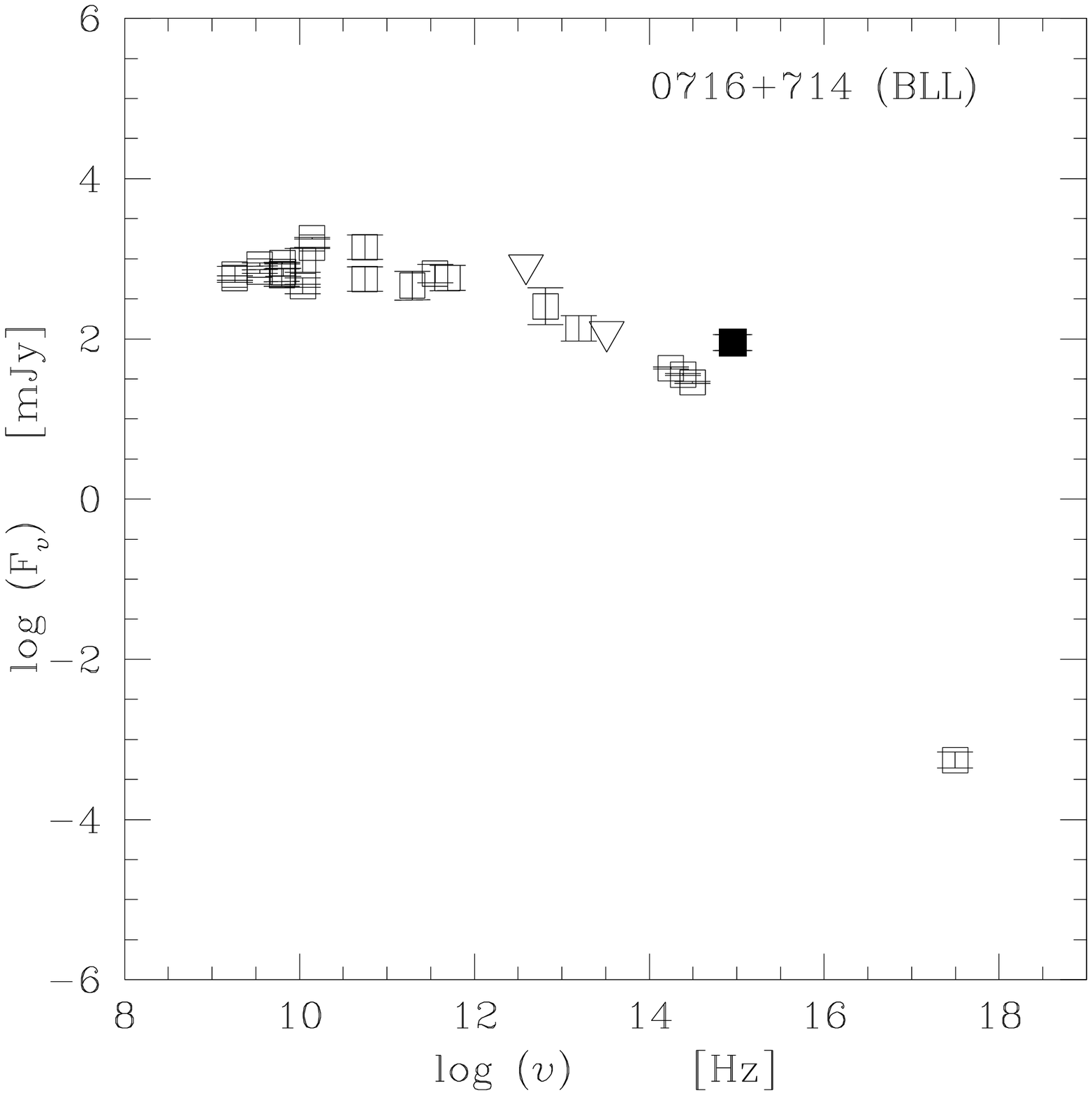}

\includegraphics[width=5.5cm]{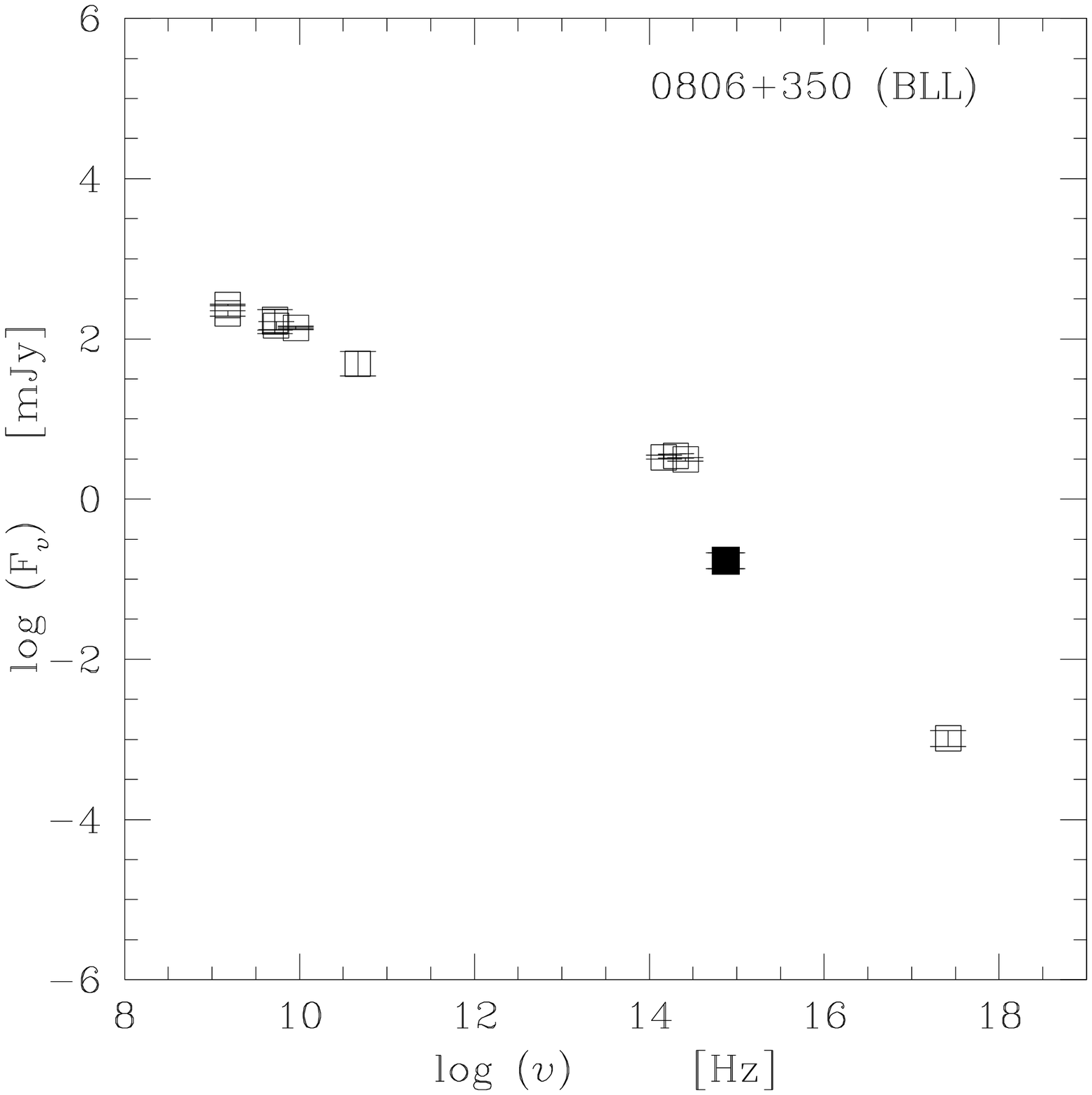}
\includegraphics*[width=5.5cm]{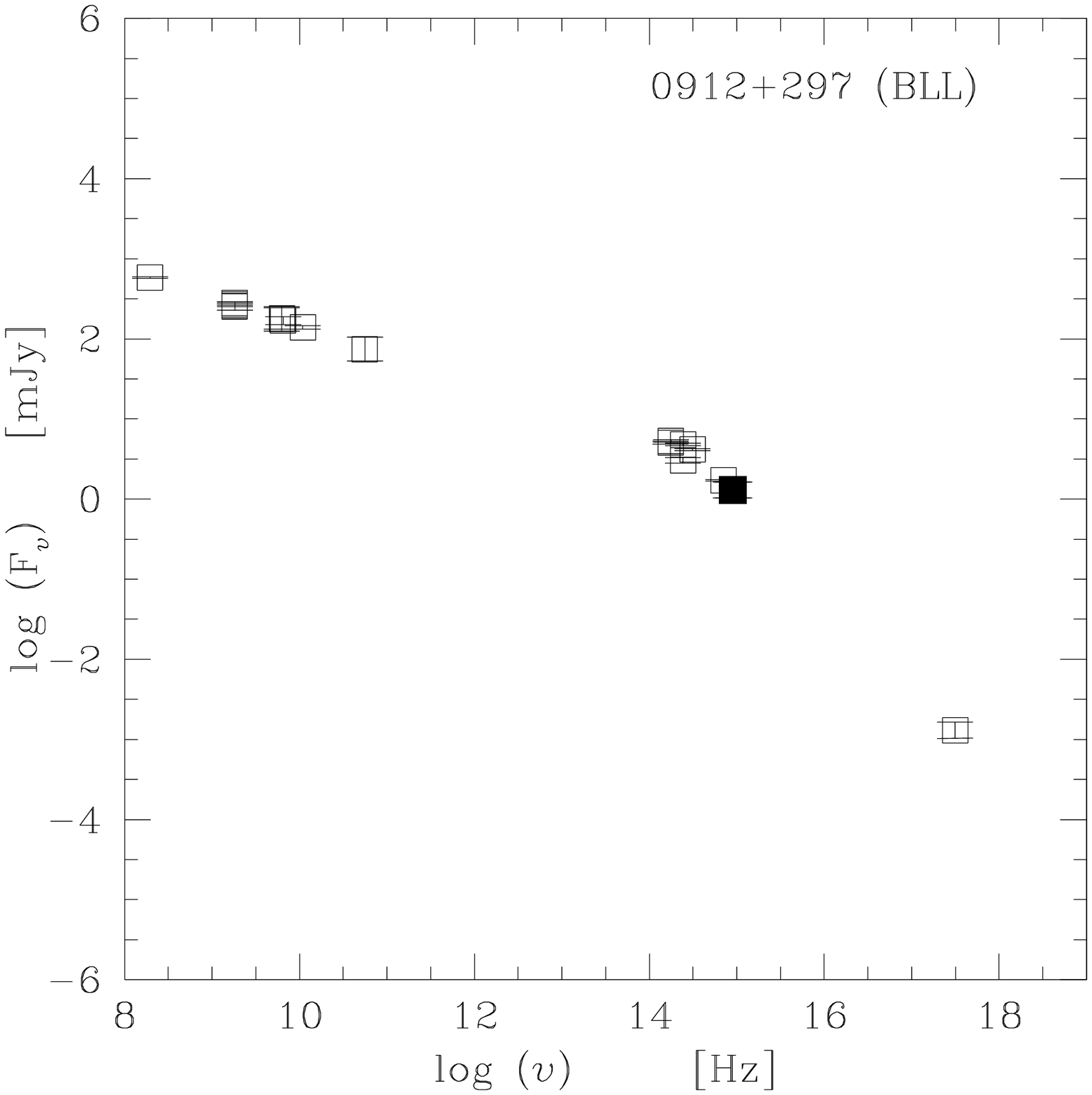}
\includegraphics*[width=5.5cm]{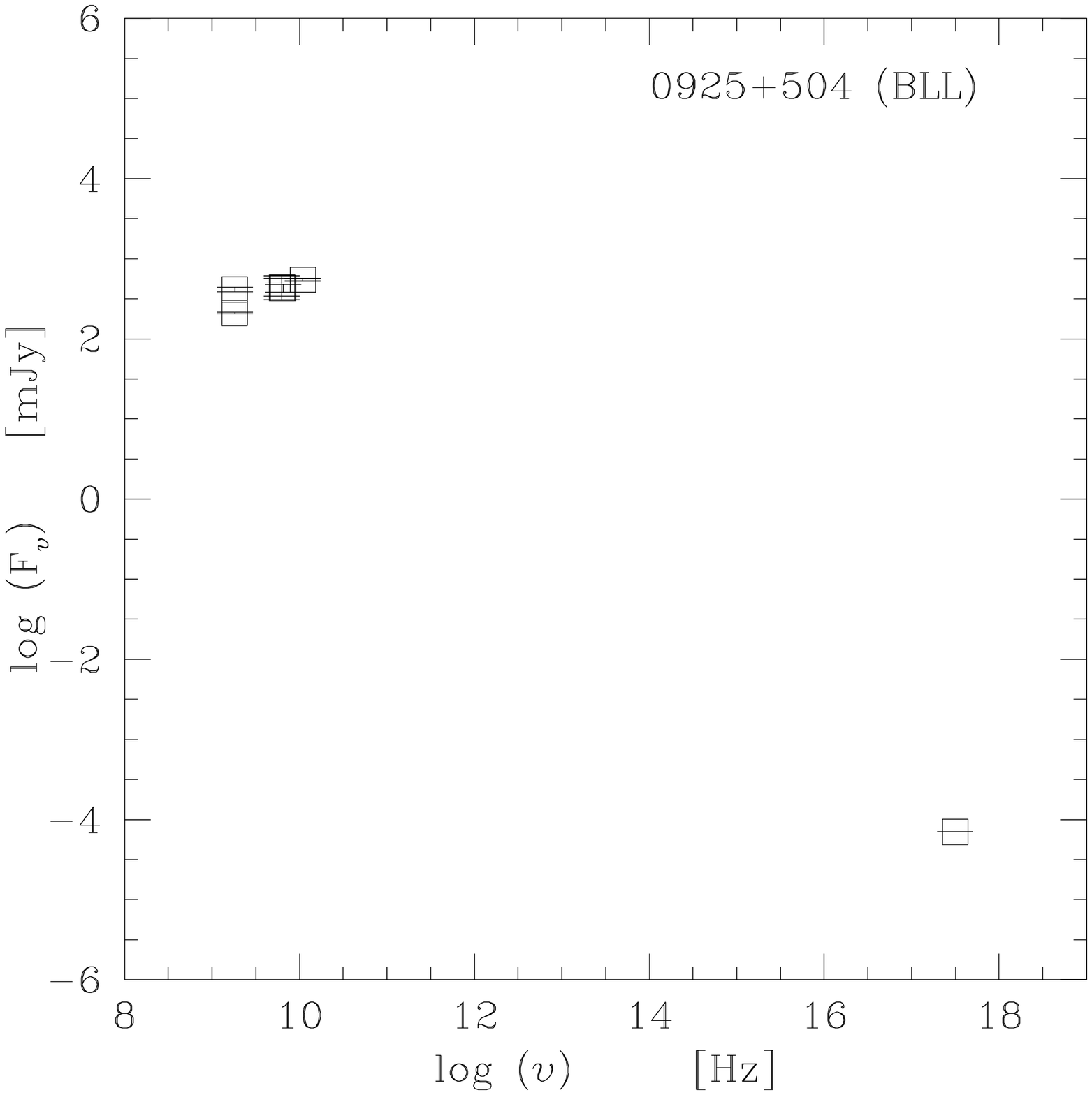}

\end{figure*}

\begin{figure*}

\includegraphics[width=5.5cm]{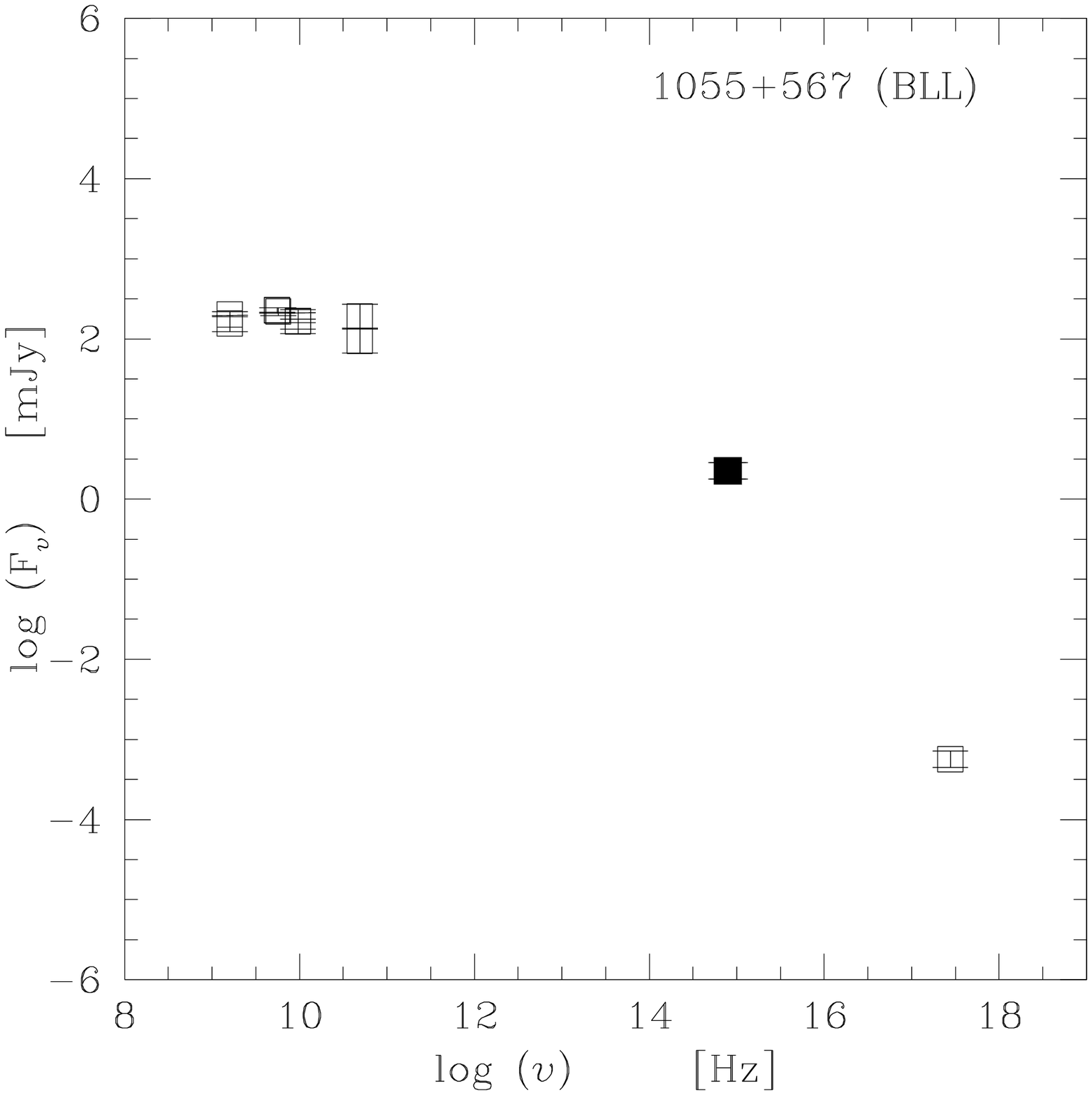}
\includegraphics*[width=5.5cm]{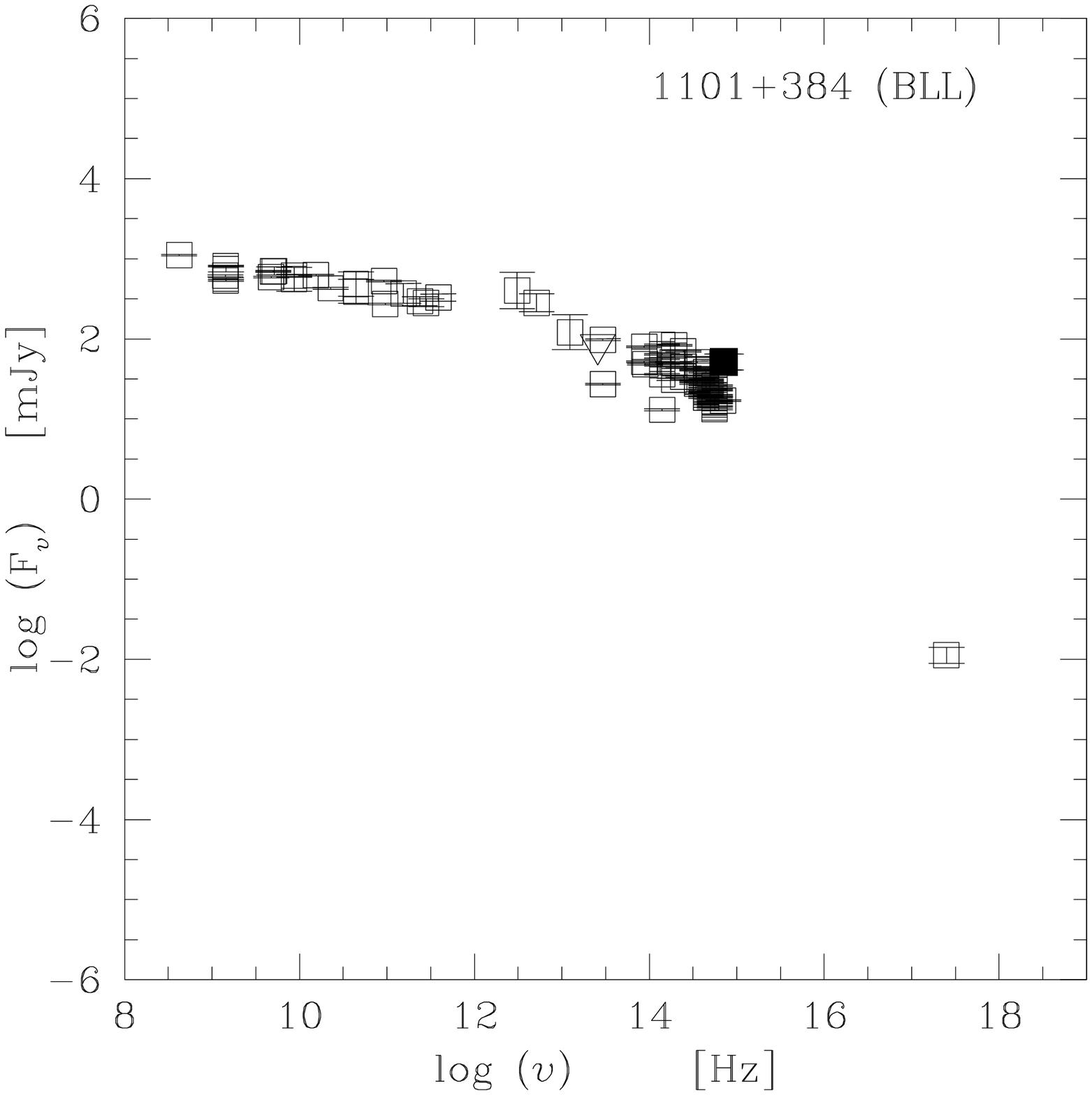}
\includegraphics*[width=5.5cm]{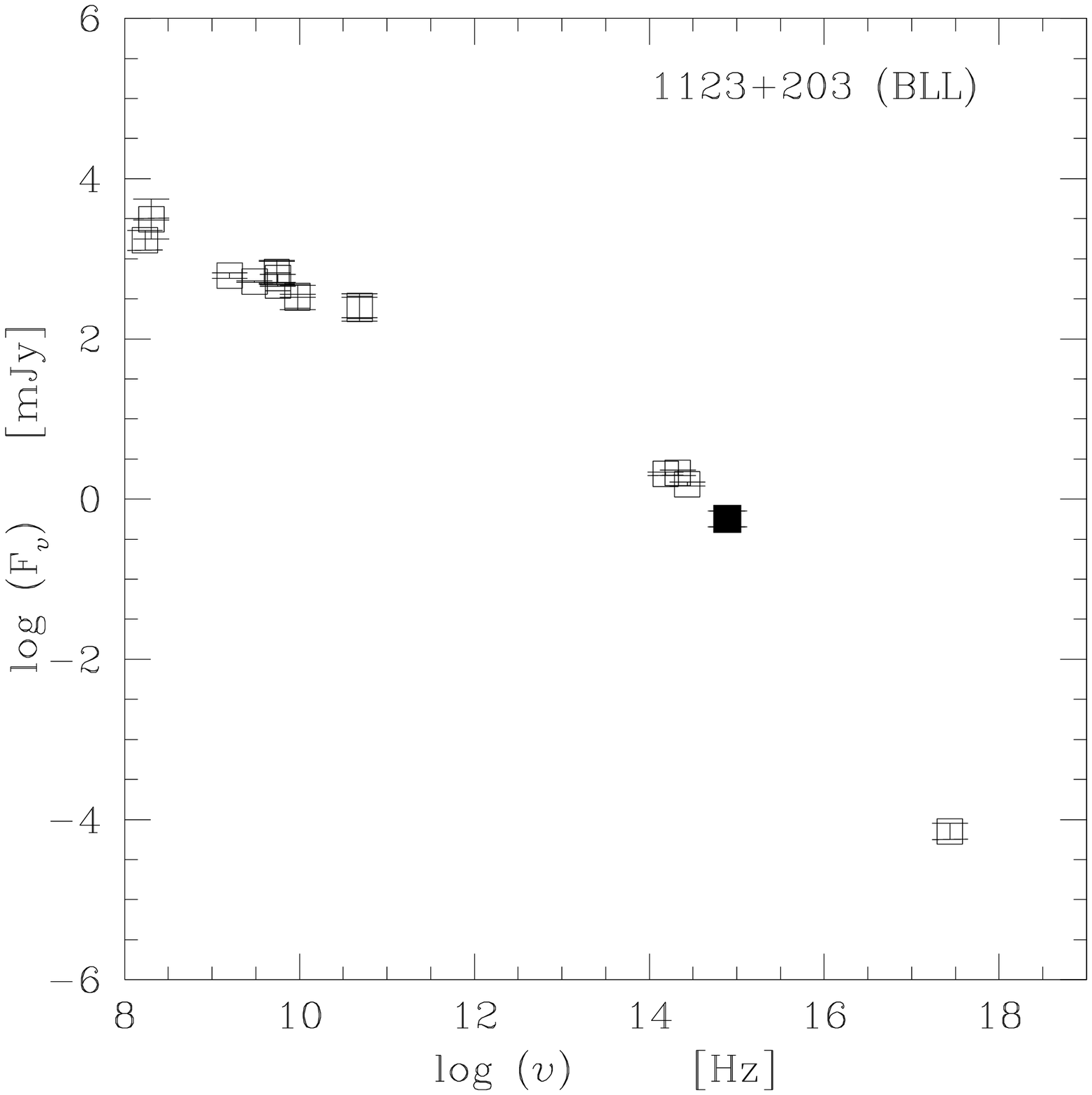}

\includegraphics[width=5.5cm]{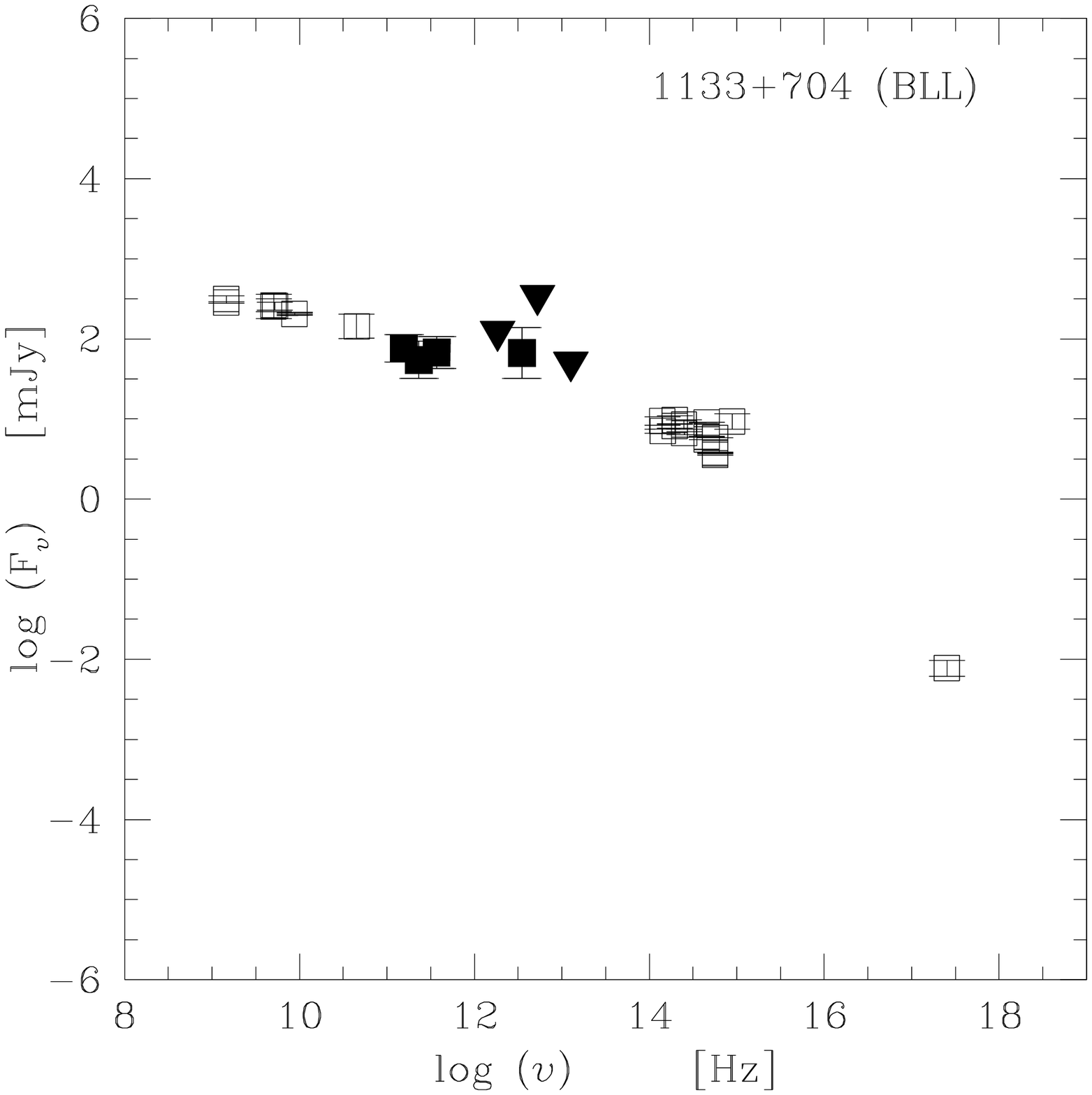}
\includegraphics*[width=5.5cm]{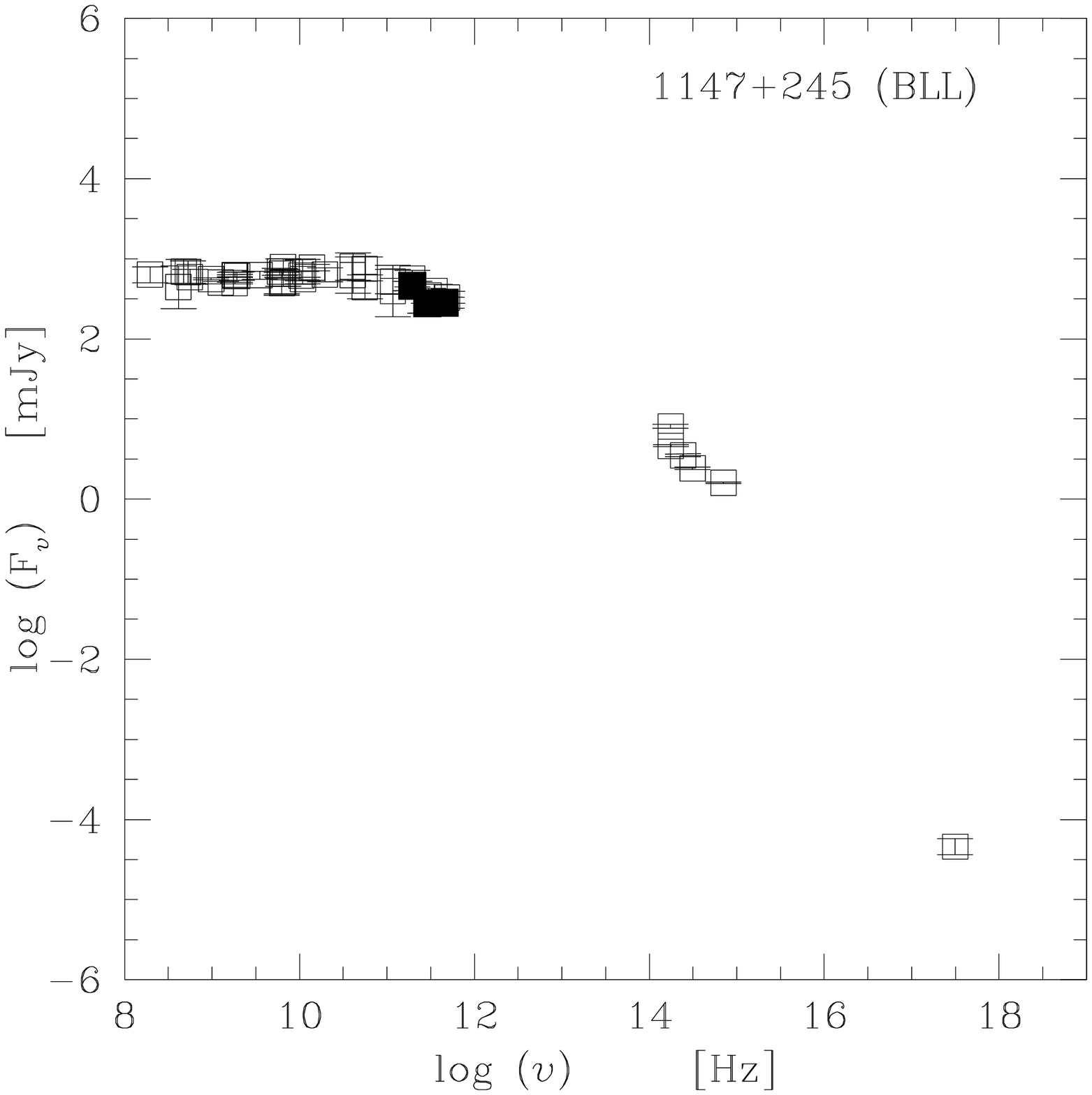}
\includegraphics*[width=5.5cm]{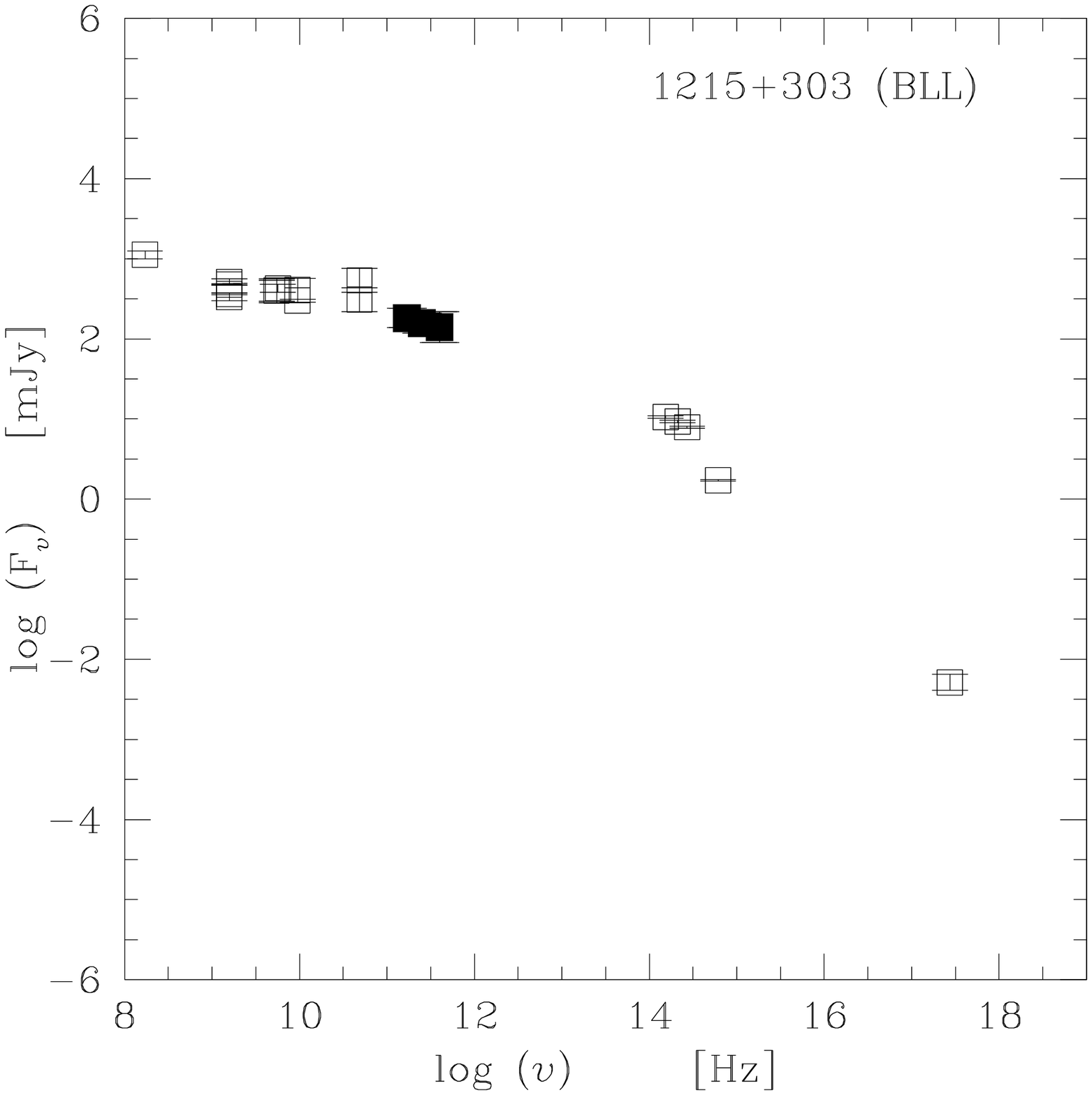}

\includegraphics[width=5.5cm]{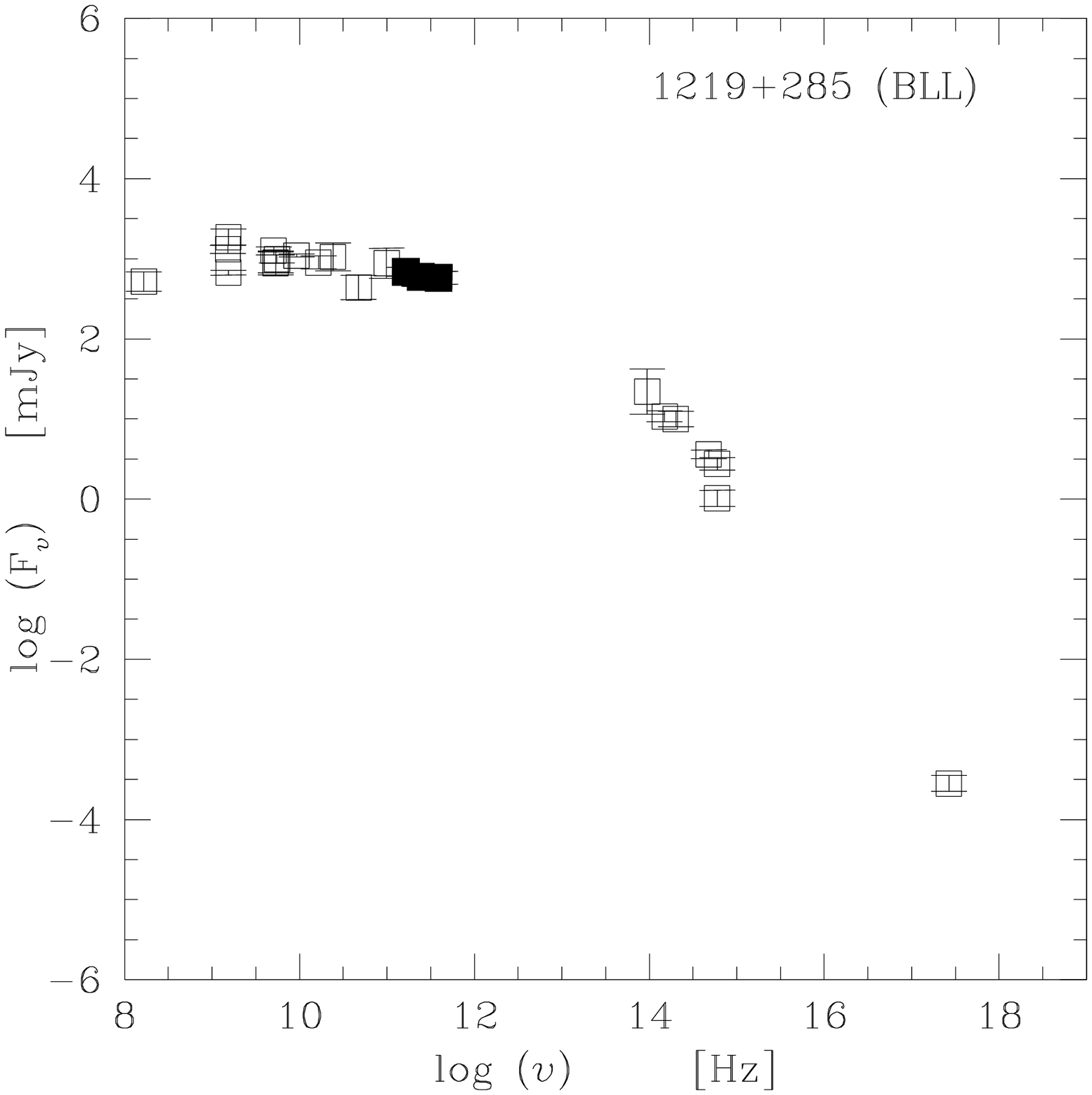}
\includegraphics*[width=5.5cm]{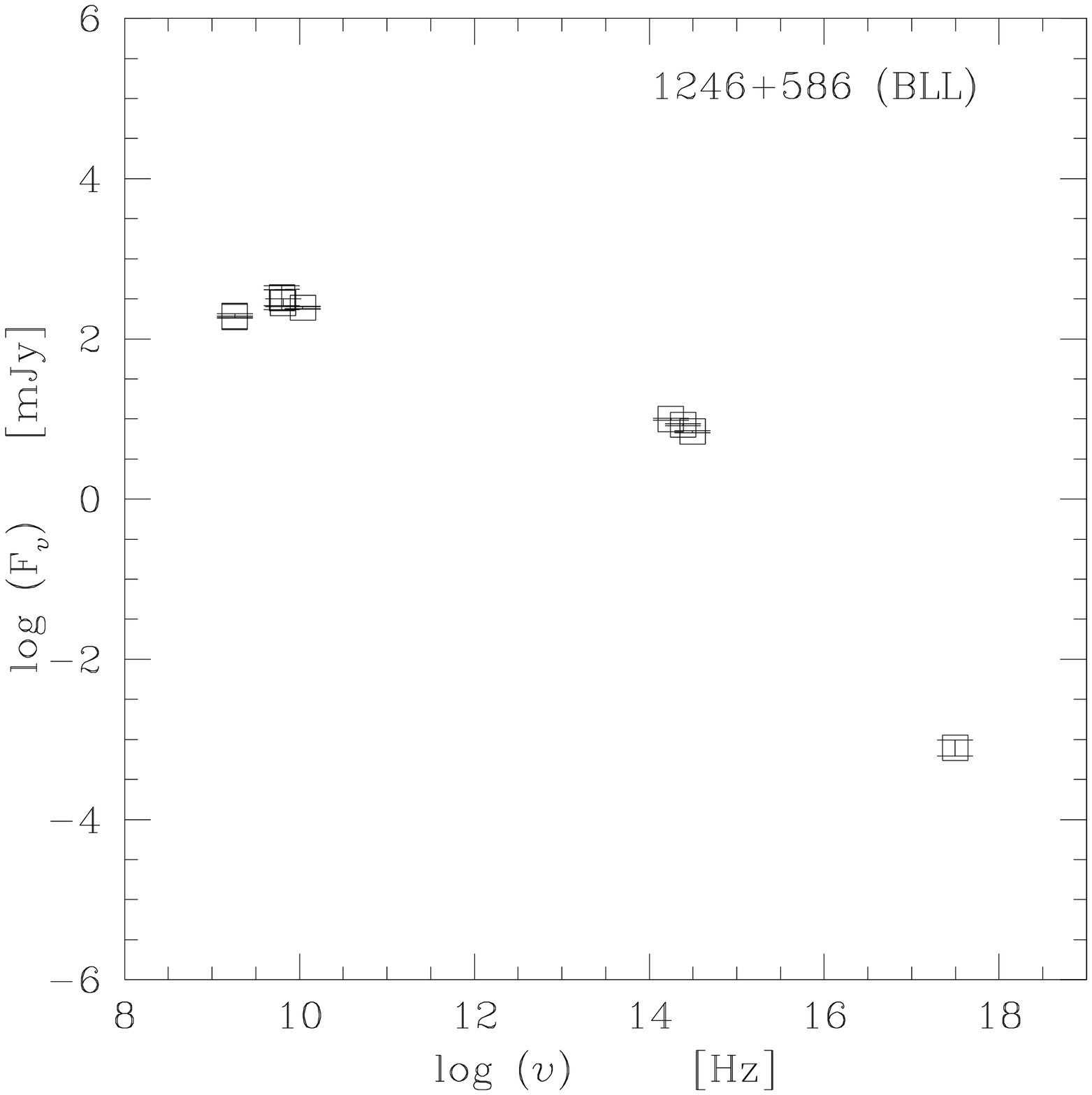}
\includegraphics*[width=5.5cm]{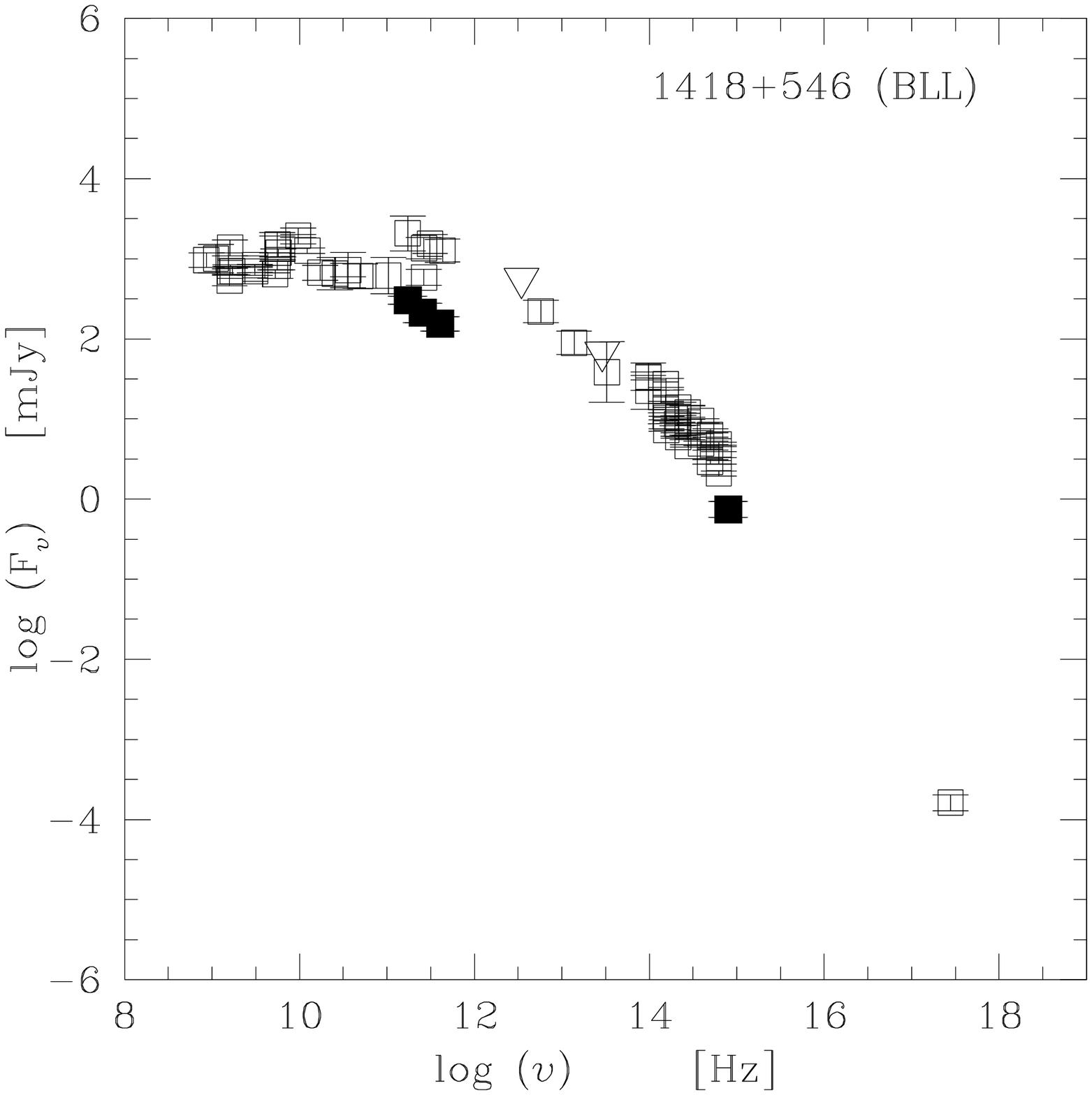}

\includegraphics[width=5.5cm]{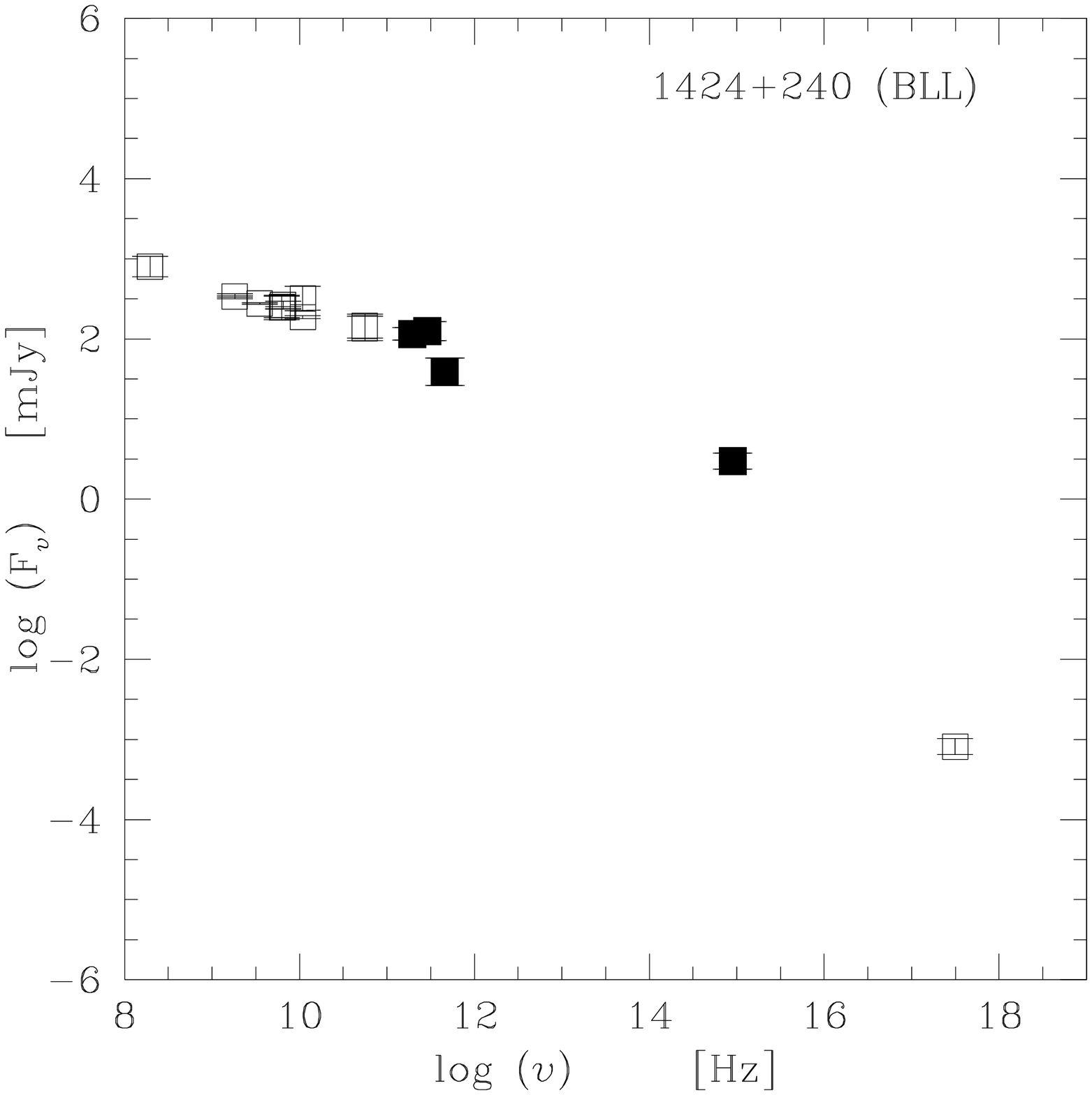}
\includegraphics*[width=5.5cm]{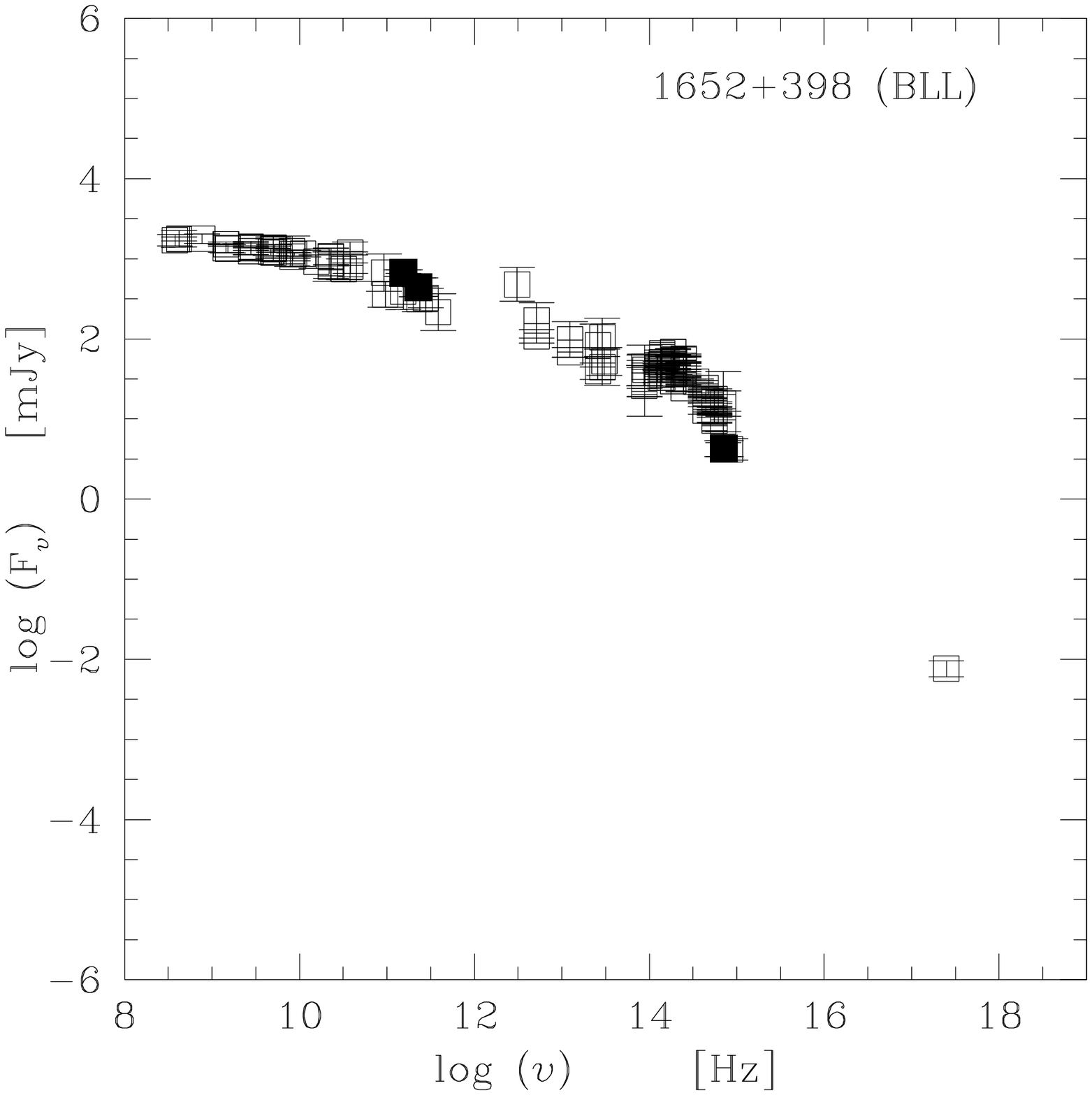}
\includegraphics*[width=5.5cm]{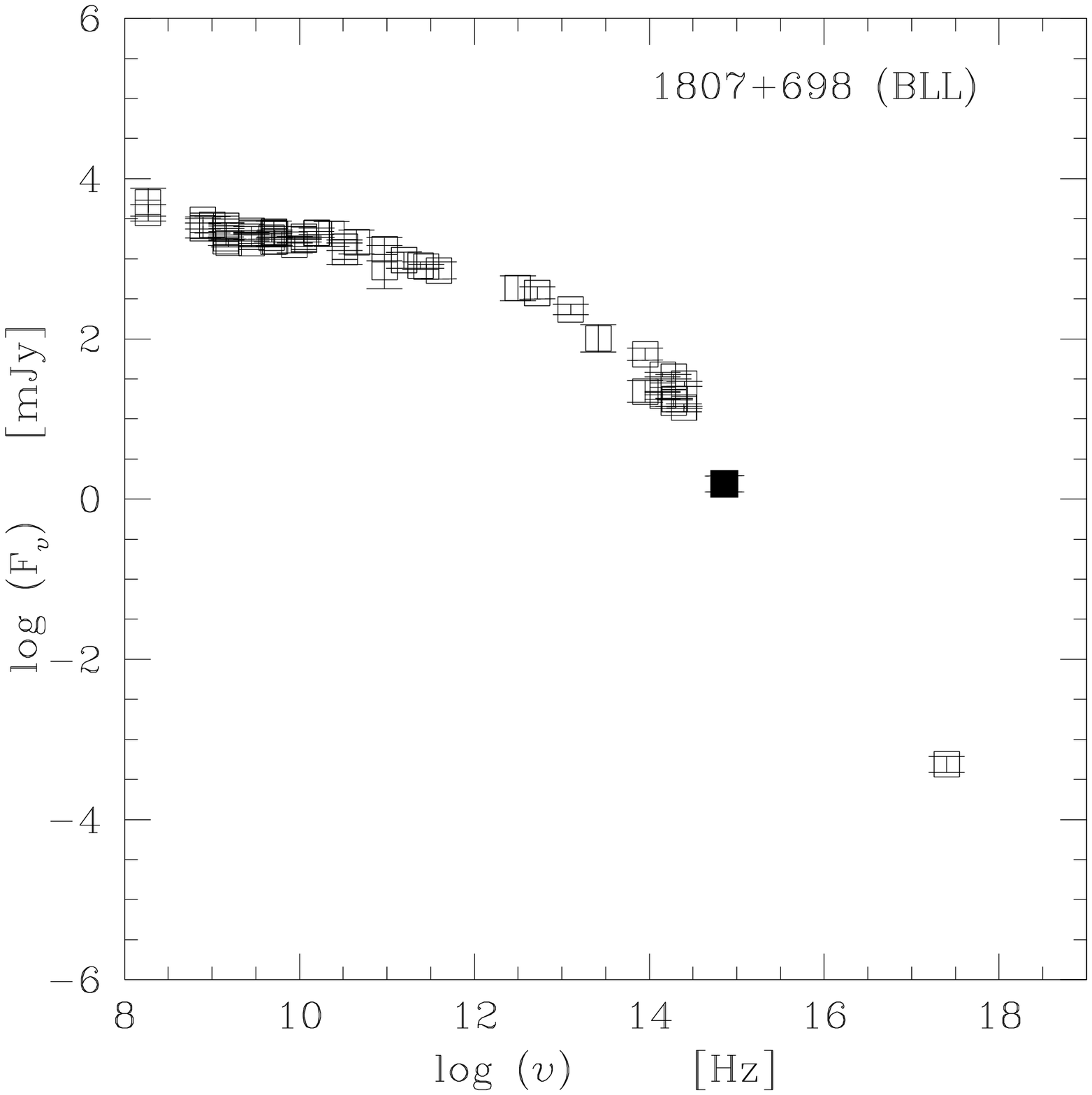}

\end{figure*}

\begin{figure*}

\includegraphics[width=5.5cm]{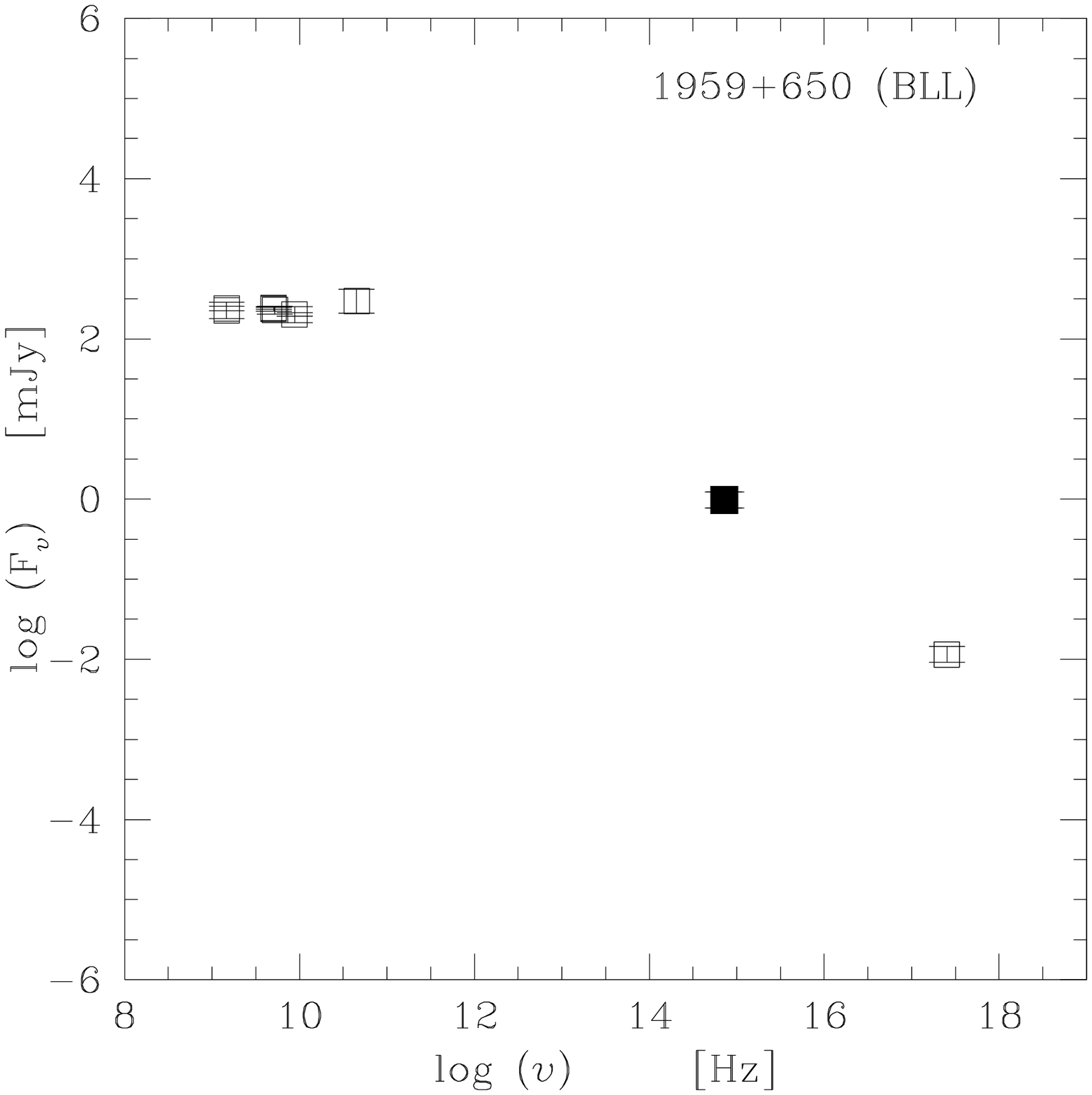}
\includegraphics*[width=5.5cm]{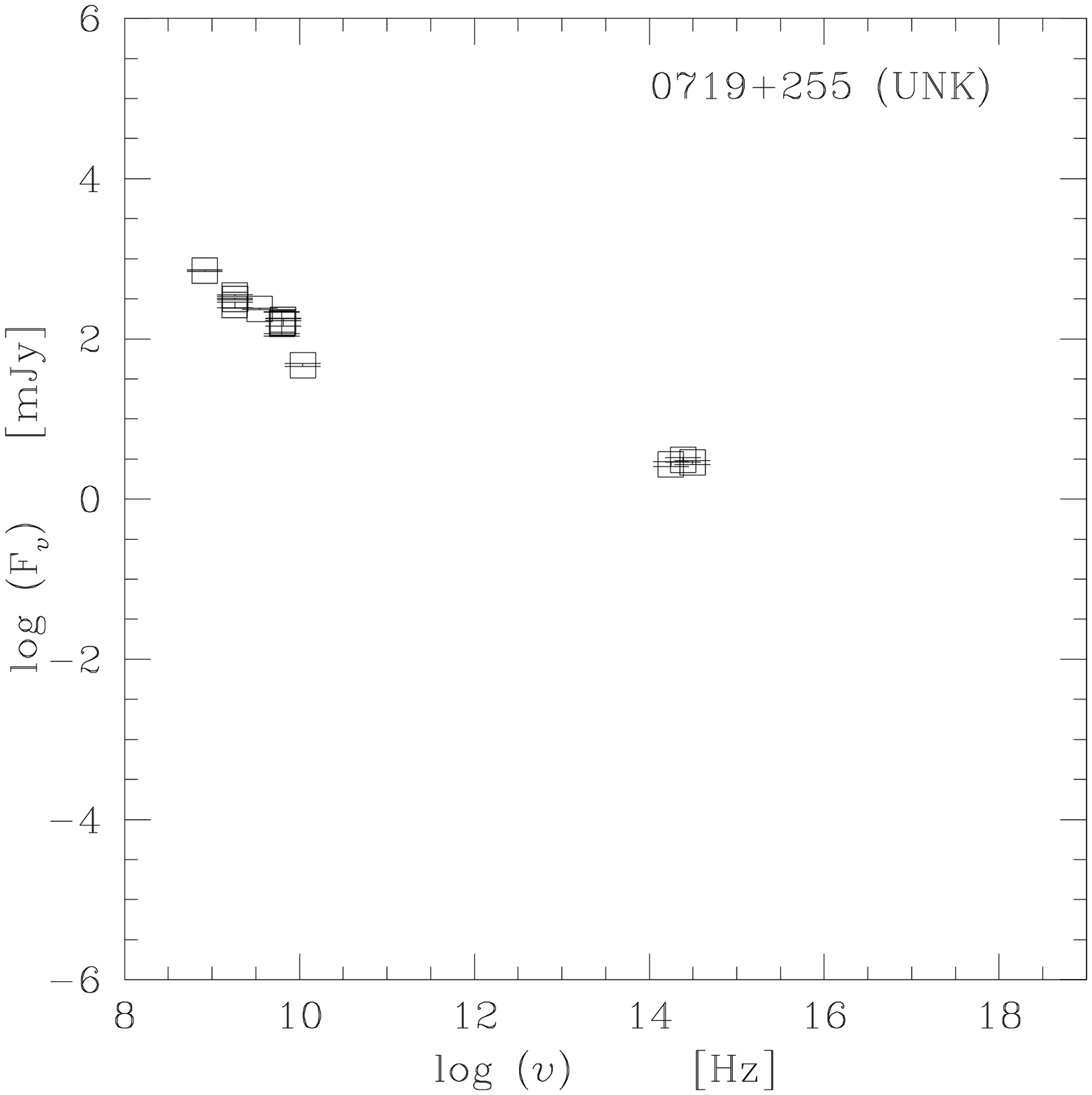}
\includegraphics*[width=5.5cm]{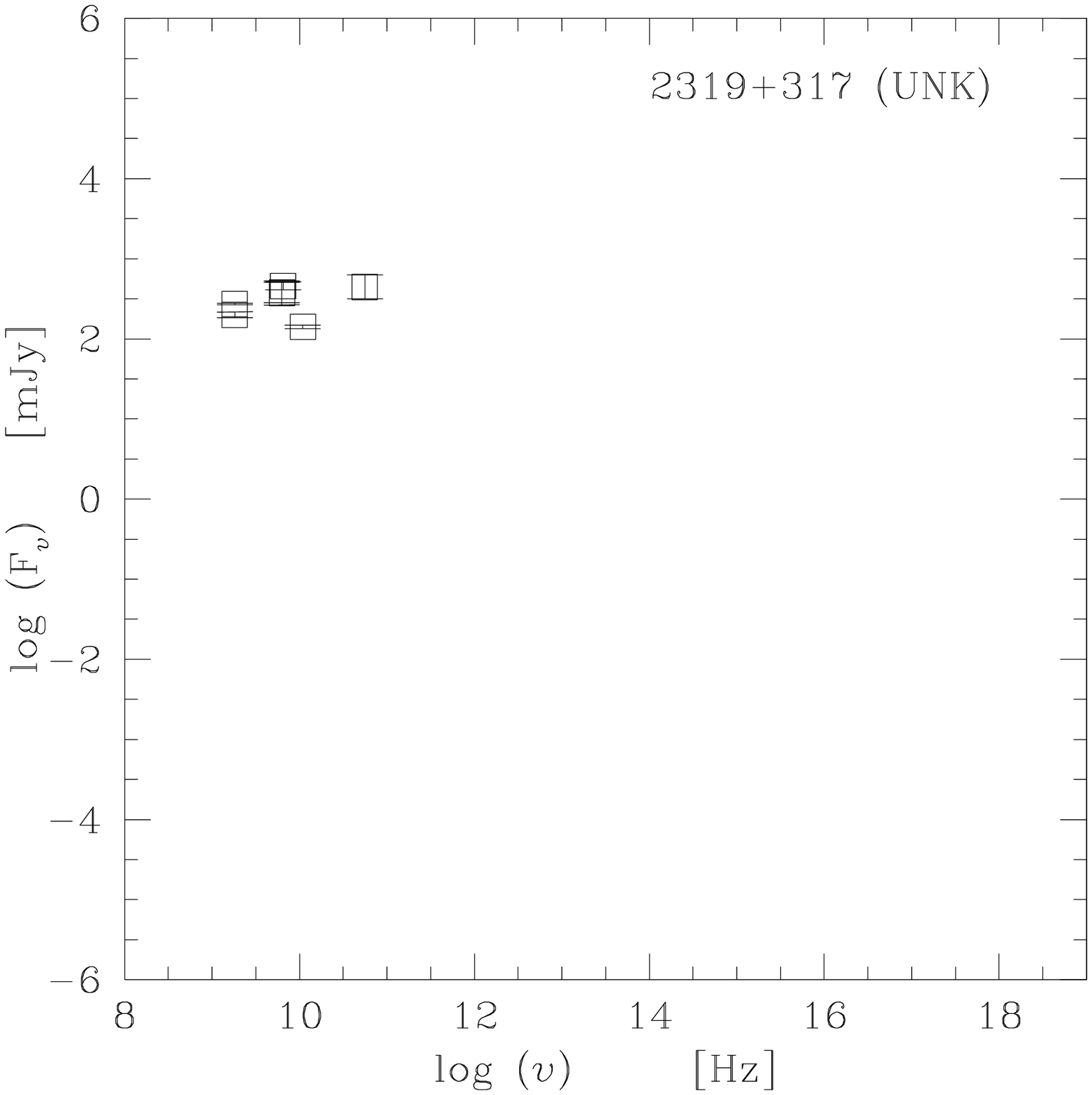}

\includegraphics[width=5.5cm]{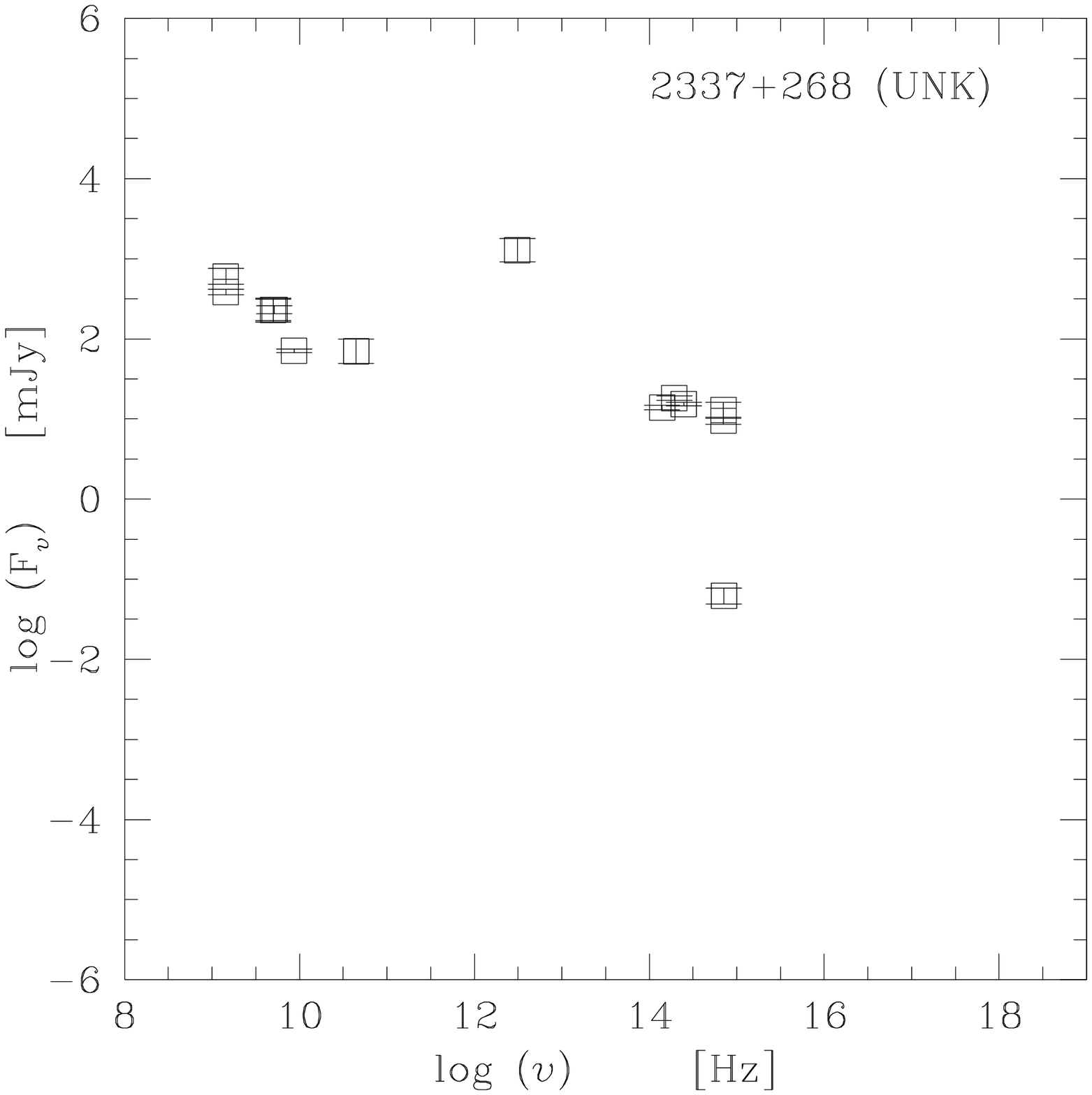}
\hspace{11cm}

\caption{For each object the SED in units of log (F$_\nu$) $vs$ log ($\nu$). 
The objects are ordered by optical classification, the first group being the PEGs,
then the Seyfert-like objects, then the BL Lac candidates, then the BL Lacs, and
finally 3 objects for which are missing optical spectra and are marked as UNK. 
Empty  symbols represent data found in the literature and in the NASA/IPAC
Extragalactic Database (NED). 2MASS JHK photometry and detections in the Bright {\it ROSAT} 
All-Sky Survey are included. Filled symbols represent the following data:
SCUBA data - 150, 222 and 353 GHz; ISOPHOT data -- 170, 90, 60 and 25 $\mu$m and
Optical data --  optical flux densities estimated through the 4000\AA\ break contrast, 
$F^{AGN}$ (see text). Inverted triangles represent upper limits. Note that in the
case of PEGs and Seyfert-like objects the estimated $F^{AGN}$ represents an upper limit, 
for this reason the data points are plotted as inverted big triangles.}
\label{seds}
\end{figure*}

\end{document}